\documentclass[preprint]{elsarticle}
\usepackage{lineno,hyperref}
\modulolinenumbers[5]

\usepackage[utf8]{inputenc}
\usepackage[T1]{fontenc}
\usepackage{geometry}
\usepackage{nameref}

\usepackage{physics}
\usepackage{amsfonts, amssymb, amsmath}
\usepackage{stmaryrd}
\usepackage{listings}
\usepackage{appendix}
\usepackage[makeroom]{cancel}
\usepackage{mathtools}
\makeatletter
\newsavebox\myboxA
\newsavebox\myboxB
\newlength\mylenA
\newcommand*\xoverline[2][0.75]{%
    \sbox{\myboxA}{$\m@th#2$}%
    \setbox\myboxB\null
    \ht\myboxB=\ht\myboxA%
    \dp\myboxB=\dp\myboxA%
    \wd\myboxB=#1\wd\myboxA
    \sbox\myboxB{$\m@th\overline{\copy\myboxB}$}
    \setlength\mylenA{\the\wd\myboxA}
    \addtolength\mylenA{-\the\wd\myboxB}%
    \ifdim\wd\myboxB<\wd\myboxA%
       \rlap{\hskip 0.5\mylenA\usebox\myboxB}{\usebox\myboxA}%
    \else
        \hskip -0.5\mylenA\rlap{\usebox\myboxA}{\hskip 0.5\mylenA\usebox\myboxB}%
    \fi}
\makeatother

\usepackage{amsthm}
 
\theoremstyle{definition}

\theoremstyle{remark}

\usepackage{graphicx}
\usepackage{url}
\usepackage{mathrsfs}

\usepackage{todonotes}

\usepackage[numbers]{natbib}

\usepackage{sidecap}
\sidecaptionvpos{figure}{t}
\begin{document}
\begin{frontmatter}

\let\WriteBookmarks\relax
\def\floatpagepagefraction{1}
\def\textpagefraction{.001}

\title{Cosine series quantum sampling method with applications in signal and image processing}                      

\author[1]{Kamil Wereszczy\'nski \corref{cor1}}
\author[1]{Agnieszka Michalczuk}
\author[1]{Damian P\k{e}szor}
\author[1]{Marcin Paszkuta}
\author[1]{Krzysztof Cyran}
\author[1]{Andrzej Pola\'nski}
\address[1]{Department of Graphics, Computer Vision and Digital Systems, Faculty of Automatic Control, Electronics and Computer Science, Silesian University of Technology, Akademicka 16, 44-100 Gliwice, Poland}

\cortext[cor1]{Corresponding author}
\tnotetext[2]{This work was [partially] supported by the research project for young scientists (RAU-6, 2020) of the Silesian University of Technology (Gliwice, Poland).}

\begin{abstract}
A novel family of \emph{Cosine series Quantum Sampling (QCoSamp)} operators appropriate for quantum computing is described. The development of quantum algorithms, analogous to classical algorithms, we apply to the harmonic analysis of signals. 
We show quantum sampling through measurements of a quantum system, and after operators of the family are applied, allow for input signal mapping with a Fourier series representation. 
Technical methodologies employed, facilitating the implementation of each QCoSamp algorithm to a quantum computer and application to the field of signal and image processing we also described. 
\end{abstract}

\begin{keyword}
\texttt{quantum computing} \sep quantum information theory \sep quantum operator \sep quantum sampling \sep Fourier sine-cosine series \sep signal processing \sep image processing

\end{keyword}
\end{frontmatter}
\section{Introduction}\label{sec-introduction}

In this paper we propose a novel family of  \emph{Cosine series Quantum Sampling (QCoSamp)} quantum operators for defining quantum versions of classical algorithms having Fourier series decomposition as their basis. In applying QCoSamp operators to quantum computation, this scenario includes a sequence of three steps, (i) Preparation of the quantum system. (ii) Evolution of the system, according to a suitably defined QCoSamp operator. (iii) Projection of the qubits states onto the measurement basis.
The preparation step (i) describes setting the initial state of each qubit to either $\ket{0}$ or $\ket{1}$ and involves qubits encoding input values and "auxiliary," so-called \textit{ancillae} qubits, which we generally set to $\ket{0}$. 
The \textit{unitary} QCoSamp evolution operator (ii) plays the role of the algorithm and is composed of "atomic" unitary operators called \textit{quantum gates}, physically implemented in quantum computers. 
Finally, measurement (iii) can be described as the collapse of the wave function from the quantum state to one of the measurement operator's eigenstates and is observable. This measurement depends on the initial state, evolution operator, and definition of the measurement basis within a certain probability. 
Quantum advantage (speedup) i.e., a reduction of computational complexity compared to conventional computation, is achieved by programming the evolution operator to produce a superposition of qubits and/or their entanglement or cluster (or multipartite) entanglement  \cite{cluster_state_Briegel_2001, zhang_cluster-state_2006, pysher_parallel_2011, irons_control_2017, ottaviani_multipartite_2019}. E.g., superposition of $Q$ qubits can lead to parallel processing of $2^Q$ values, which researchers could utilize for very efficient computation.
Numerous programming ideas have defined quantum computational algorithms, regarding quantum advantage, are extensively described in the literature \cite{montanaro_quantum_2016, gyongyosi_quantum_2019, quantum_zoo}. This work proposes a new methodology for creating algorithms based on QCoSamp) operators. It allows for quantum versions of many classical computational procedures that rely upon the decomposition of input signals into the Fourier series components. We show the implementation of QCoSamp algorithms on a quantum computer at the quantum gates level and present several applications in signal and image processing.

\subsection{State of the art}

To outline the background of the proposed methodology, we review several quantum algorithms focusing on two criteria. The first one is whether a quantum algorithm is "oracular" or not. In other words, whether its construction is an oracle (or a black box), a function to determine which configuration of the problem we define as desirable or "good" and that which are not desirable or "bad." As introduced to quantum information theory from Turing machines theory (e.g., Soar in \cite{soare_turing_2009}) in 1992 by Deutsch and Jozsa \cite{Deutsch_Jozsa} and Berthiaume and Brassard \cite{berthiaume_quantum_1992, berthiaume_oracle_1992}.
The second criterion describes how the output of a quantum algorithm is created and then utilized. More specifically, quantum algorithms are divided into two separate classes, \emph{deterministic} (\cite{cleve_quantum_1998}) which reach a singular, resultant output state (or a state belonging to a specified subset) with a probability equal to $1$ (or close to $1$), and \emph{non-deterministic} (\cite{knill_quantum_1996, de_wolf_characterization_2000}), \emph{quantum sampling} (e.g. \cite{wocjan_speed-up_2008, childs_weak_2007}) or \emph{state tomography} (Haah in \cite{haah_sample-optimal_2017}) which provide a distribution over the space of probable output states by iteration (shots) of the quantum algorithm. \\
The development of quantum computing began with oracular deterministic algorithms. Deutsch-Jozsa's algorithm \cite{Deutsch_Jozsa} supplied the solution to the "black box" problem of whether a binary function belongs to either a "balanced" or "constant" type by reaching the output state either $\ket{0}$ or $\ket{1}$.
Quantum algorithms developed by Brassard et al. in 1997 \cite{brassard_exact_1997} and Grover in 1998 \cite{Grover_algorithm}, return output state corresponding to the solution to the "inverse black box" problem of searching through the unindexed list. Grover's algorithm assumes that there is only one "good" solution and produces a homogeneous superposition of "bad" solutions. The \emph{amplitude amplification} algorithm by Brassard et al. \cite{brassard_quantum_2000} is a generalization of Grover's search algorithm that allows for many "good" solutions and does not produce homogeneous superposition for "bad" solutions. For Grover's algorithm, the number of steps is known and equal to $\sqrt{2^N}$ (where $N$ is the number of qubits). The number of steps in the amplitude amplification algorithm is equal to $1/\sqrt{a}$, where and $a$ is the probability that we will measure a "good" solution after evolution driven by the oracle, in the worst-case $a$ is unknown. In the same paper \cite{brassard_quantum_2000} the authors developed the algorithm to determine $a$, which is called the \emph{amplitude estimation} algorithm. Theese algorithms are extensively used in quantum programming research; a lot of work continues to be undertaken by many researchers concerned with further improvement of these algorithms, e.g., Grinko et al. \cite{grinko_iterative_2019} (iterative method of amplitude estimation), Suzuki et al., \cite{suzuki_amplitude_2020} (reduction of the quantum circuit size). Grover and Radhakrishnan \cite{grover_is_2005} described a faster algorithm that divides the entire domain of a search into blocks, then performing the amplitude amplification in each of these blocks in parallel and then comparing the results –\emph{quantum partial search}. Shor, \cite{shor_algorithms_1994} proposed a widely known algorithm, which searches for two prime divisors of $N$ in polynomial time. The basis of this built upon these two preceding algorithms: (1) \emph{phase estimation} described in details by Cleve et al. in \cite{cleve_quantum_1998}, known also as \textit{phase kickback} and (2) \emph{Quantum Fourier Transform} proposed by Coppersmith \cite{coppersmith_approximate_2002}, implemented for quantum circuits by Weinstein et al. \cite{qft_weinstein_implementation_2001}. Lee and Selby, \cite{lee_generalised_2016} proposed the generalized version of phase estimation; in their work, they extended the notion of phase estimation to the case of multiple qubits and proved that a generalized kick-back mechanism could obtain every phase transformation, and important to our present work, the superposition preserved by this controlled transformation (pp: 6-8, 13).\\
Examples of oracular but non determinitstic algorithms include the integration method proposed by Abrams and Williams, \cite{abrams_fast_1999}, where the value of the integral is approximated by iteration. The authors of \cite{abrams_fast_1999} defined oracular formulation for quantum integration and studied the relationship between algorithm complexity and computational error. Heinrich, \cite{heinrich_quantum_2002, heinrich_quantum_2003, heinrich_problem_2003} used quantum summation/integration for functions in Hilbert and Sobolev spaces. Papageorgiou et al., \cite{kais_quantum_2014} described the integration algorithm and introduced an integration method based on an amplitude amplification algorithm \cite{brassard_quantum_2000}. The papers referenced above present the theoretical issues of quantum integration, focusing on mathematical background and computational complexity. However, the authors omitted structural and implementation aspects of the oracle, which represented an abstract function. Therefore, the implementation aspects needed further development. Kane and Kutin, \cite{kane_quantum_2010} were seeking oracles reflecting specific properties of the polynomial, e.g., having a polynomial degree of $d-1$?.
Van Apeldoorn et al., \cite{van_apeldoorn_convex_2020} applied oracular, non-deterministic approach to a convex function optimization. They presented a theoretical background, diagrams of oracles, and the quantum circuit corresponding to the final solution. Gilyén et al. \cite{gilyen_optimizing_2019} made insights into the structures and properties of oracles. They distinguished three types of oracles: probability, binary, and phase. Their other contribution was the discovery and proof of inter-convertibility of oracles, which allowed for transforming of an oracle from one type to another.\\
Non-oracular,  non deterministic algorithms are an important, burgeoning area of quantum computation; they can define complete input-output algorithms and part of other quantum algorithms.  
The \emph{Quantum Fourier Transform} (QFT) algorithm is used widely in research as a part of other quantum algorithms of all types. Ruiz et al. in \cite{ruiz-perez_quantum_2017} after Draper \cite{draper_addition_2000} and Beauregard et al., \cite{beauregard_quantum_2003} propose the usage of QFT for introducing a number to a system of qubits with a technique they called \emph{distributed phase encoding} and then perform several arithmetic operations on these numbers, such as adding or mean and weighted mean computing.\\
In image processing, Beach et al., \cite{beach_quantum_2003} show encoding images in a quantum system and present sketches of the quantum image processing algorithms. Phuc Q. Le, \cite{le_flexible_2011} introduces the \emph{flexible representation of quantum images FRQI} method for encoding images basing on the rotation operator and shows the process of storing and retrieving images from the FRQI states. Yao et al. \cite{yao_quantum_2017QIP} use that representation for edge detection. Yan et al., \cite{yan_quantum_2017QIP} demonstrate an application of this method for multi-channel images and algorithms for image processing: watermarking, image encryption, steganography (hiding information in images). Zhou et al., \cite{zhou_global_2017} use this model of encoding images for translation operations of images.\\

\subsection{Motivation and Contribution}\label{sec-motivation}

The general motivation for introducing the QCoSamp family of operators is to provide a methodology supporting the possible creation of a wide range of quantum algorithms (unitary evolution operators). Construction of the QCoSamp operator enables solving, with the quantum computer, problems formulated by a sine-cosine transformation in harmonic analysis. QCoSamp operators support creating the quantum version of conventional algorithms by defining the corresponding non-deterministic quantum algorithm. Depending on the type of problem, it can be either oracular or non-oracular. \\     
Quantum algorithms currently created are at a low level of abstraction. The target (high rank) unitary operator, constructed by combining the closed set of atomic (low rank) operators called gates. There are frameworks provided by IBM (QComposer, QASM, QISKit) or Microsoft (F\#) dedicated to the creation of quantum algorithms in the graphical environment (Qcomposer), they use the assembler-like syntax (QASM) or commonly known classical programming languages like Python (QISKit) or C\# (F\#). The programmer has to express the problem to be solved in the Hilbert space unitary operator language and then decompose it to the available gates and not a trivial task. It requires a level of mathematical and physical competence, which is a barrier to the spread of quantum computation.
On the other hand, the rapid development of classical computation technologies, alongside hardware evolution, was made possible by reaching successive abstraction levels in programming languages, from the lowest level. The machine language or assembler, through low-level languages like C up to currently used expanded frameworks and libraries, where each framework or library covers the specific problem and enables the creation of large computer systems from these building blocks. We imagine that analogous development in quantum computation will take place, and we propose a tool for.\\
QCoSamp creates quantum algorithms possible at higher levels of abstraction; the quantum programmer must express the problem to be solved in terms of a Fourier series and then run the QCoSamp machinery described in this paper. 
The evolution of the prepared quantum initial state through QCoSamp generates a measurable output, i.e., probability distribution over the output qubit states. We demonstrate that the probability distribution over the output eigenstates obtained by quantum sampling can be an efficient solution to problems encountered in data analysis, such as integration and curve fitting and signal and image processing, kernel filtering, and feature extraction. 
We adapt the QCoSamp operator's structure to the specific background problem (e.g., one of the above listed), where parameters of the background problems can be defined either by suitable preparation of the input quantum state or by setting values of angles in the phase shift quantum gates. We name the resulting output states probability distribution \textit{quantum cosine sampled function (FCoSamp)} $\mu_N$. In section \ref{sec-mapping}, we show that there exists a mapping between FCoSamp and the Fourier sine-cosine series. Therefore, we can say that quantum computation's output approximates the function that can be mapped to the Fourier series. Therefore, QCoSamps operators can be used as a basis for quantum modeling of problems using the sine-cosine Fourier series.\\
Generation of the FCoSamp function involves the application of quantum operators, superposition, (cluster) entanglement, and interference. Superposition enables processing of the large bundle of values of those elements in the time required to process one such value on classical computers. Entanglement, together with (or by) an interference, allows for the projection of the demanded result from the superposed state (large) to the smallest possible dimension output, which makes launching the proper quantum sampling procedure possible, with a reasonable number of repetitions.
The methodology of construction for the FCoSamp function, encoding the arguments, frequencies, and phase shifts by the usage of superposition, interference, and entanglement phenomena employ some novel approaches to programming quantum algorithms.\\

As an example, in the case of quantum integration with FCoSamp, we know the integrated function, and therefore the frequencies and phase shifts of FCoSamp are also known; we want to sum up the values of FCoSamp for the large number of $x$. Therefore we superpose the $x$-s, which produce $2^{X+T}$ components of $X+T$ qubits, where $X$ qubits we dedicated to representing the $x$ number and $T$ qubits – for the FCoSamp construction. 
We add quantum phases to each component of the uniform superposition state. This quantum phase represents the value of the function to integrate for a number from its domain. Since there are $2^X$ coordinates, the superposition consists of $2^X$ values of the function to integrate for $x$-s spread out (not necessarily uniformly) on the range $[-\pi, \pi]$. 
At this point the measurement of the system will return a uniform probability distribution, because the phase does not influence the measurement (as $\big| e^{i\phi}\big| ^2=e^{i\phi}e^{-i\phi}=1)$. The interference phenomena, which we constructed with Hadamard transformation, serve as a bias for the measurement according to the phases that the given state contains, an analog of phase-shift detection by a Mach-Zehnder interferometer. 
However, suppose we apply the Hadamard transformation alone, to the taken superposed state; each component combines all added and subtracted elements of the superposed state after measurement, results in the sum of sequences of sine and cosine functions in the range $[-\pi, \pi]$ or its opposite number.
We obtain the integral function's final form, using the additional superposed qubit, called the \emph{ancilla}. This operation generates the state with alternate components equal to $1$ and phases representing the arguments of the integrated function and divided by a normalization factor making the components compliant with the postulates of QM. 
Then, with the tensor product of their identities and one Hadamard transformation, we construct interference limited to the two adjacent components: even and odd, which result in the components having phases added (for even) or subtracted (for odd) from $1$-s, which is the cluster entangled state. 
The quantum sampling of such a state and the summation of the results for even components is the value of the integral. Nevertheless, the number of eigenstates to be measured grows exponentially with the number of qubits encoding $x$, which makes reasonable accuracy of integration (e.g., in the 3-dimensional case and 32 qubits for argument in each dimension) impossible to sample due to the exponential growth in the number of demanded repetitions. For the given case, we obtain $7.92\cdot 10^{28}$ eigenstates for measurement, which require a time of computation a few orders of magnitude longer then the age of the Universe, even if we suppose the computer producing $10^9$ qubits per second, which is theoretically affordable using the quantum optics phenomena.

When developing non-deterministic quantum algorithms for efficient sampling, it is essential to properly scale (limit) the space of possible output states. One approach is by limitation of measurement basis; another method uses amplitude amplification algorithms, which do not limit the measurement basis but make a large part of it inaccessible. An example case where a measurement basis can be defined is quantum integration; we can only restrict the measurement basis to two eigenstates. The state obtained for $\ket{0}$ eigenstate is the projection on this state even coefficients, and is the sum of even states of the whole system. In contrast, the projection on the eigenstate $\ket{1}$ is the sum of odd states. The quantum sampling leads to obtaining approximation of the norm of states projected on the eigenstates belonging to measurement basis, which means that function sampled for $\ket{0}$ contains the sum of the values of the function to be integrated for the values spread out of the range $[-\pi, \pi]$, which we considered as the value of its integral. The opposite example is the curve-fitting problem. The output carries information the coefficient of the series $r_n, s_n$, so it has to contain all eigenstates representing the whole set of those coefficients. This space is too large for quantum sampling; hence we are forced to use the amplitude amplification algorithm (or similar) in such applications. 

We introduce the notion of balanced QCoSamps, which has a structure of a perfect binary tree (see Yuming Zou and Black in  \cite{perfect_binary-tree}), and the number of components is equal to the power of $2$. We discuss the advantages and disadvantages of balanced and not balanced QCoSamp. We prove (sec. \ref{sec-output-correctness}) that the inverses of lengths of all leaves in full binary tree sum to $1$. It demonstrates the correctness of the outputs of QCoSamp with an arbitrary number of components in the perspective of postulates for quantum mechanics.\\

To summarize, we present a method of generating quantum algorithms using quantum sampling, starting with describing the problem to be solved by the Fourier sine-cosine series. If the quantum programmer has this description, he can turn on the machinery described in this paper, especially in the sec. \ref{sec-theory-calculations} and obtain the algorithm ready for implementation on the quantum computer that will solve the given problem. Examples of problems allowing for quantum formulation by using QCoSamp, in the area of signal and image processing, are curve fitting \cite{abuelmaatti_simple_1993, guruswami_robust_2016}, regression \cite{bilodeau_fourier_1992, asrini_fourier_2014, pane_parametric_2014}, signal filtering \cite{chong-yung_chi_fourier_1997}, image kernel filtering \cite{chaudhury_fast_2011, ghosh_fast_2016}, image classification \cite{chii-horng_chen_statistical_1999}, etc. 
The QCoSamp operator uses canonical quantum gates: Hadamard, permutation, phase shifts, X and Z-Pauli and their controlled versions whilst also introducing the new quantum programming methods and concepts for the purpose of QCoSamp construction: \textit{twice permuted controlled operator}, \textit{ordering the coordinates}, \textit{comparison of the states}, \textit{encoding of constant data}, \textit{forging reference probability}. These concepts are useful for building the QCoSamp and using quantum algorithm implementation with QCoSamp. We present that the QCoSamp architecture is a full binary tree defined by Black in \cite{binary-tree, full_binary-tree}. Therefore QCoSamp can have the arbitrary number of components, in contrast to the example of known classical FFT algorithms solving Discrete Fourier Transform \cite{cooley_algorithm_1965, bruun_z-transform_1978, murakami_real-valued_1994, FFT-prime-numbers, rader_discrete_1968, bluestein_linear_1970}.\\
Along with the method of creation of the QCoSamps operators, we present the architecture (internal and external structure) of the QCoSamp and the explicit mapping with a two-step reconstruction of the approximated function from the FCoSamp. This mapping is the theoretical background allowing its use as a model of the Fourier series. Furthermore, we describe some novel modifications of existing quantum programming techniques necessary for creating the QCoSamp and using this operator for practical applications. We present that the proposed family of QCoSamps operators opens new research opportunities in quantum signal and image processing, such as quantum convolution filtering or quantum wavelet features for object description and recognition.

\subsection{Notation and nomenclature}

In this paper, we will use the Dirac notation concerning any issues connected with quantum science. We will use the following qubit notation convention:

\begin{itemize}
   \item Latin capital letters denote the qubit counts.
	\item Latin capital letters written as the ket-vectors stand for tensor products of qubits, e.g., $\ket{Q}$ means the tensor product of $Q$ qubits.
	\item Latin lower-case letters with subscripts, written as ket-vectors, where necessary, stand for qubit tensor products additionally indexed by a proper symbol. E.g., we often write down: $\ket{Q}=\ket{q_1...q_Q}$.
	\item Latin lower-case letters with subscripts denote the value of the qubit indicated by the index. This value is \textit{not} the eigenvalue but can be one either $1$ or $0$. 
	\item The argument of operators are in round brackets. If they appear, the operator doesn't touch the unlisted qubits in such brackets and avoid writing the tensor product with the unit operator $\mathbf{1}$. For example 
	$
	\mathcal{A}(C)\ket{BCD}=
	\big[
	\mathbf{1}^{\otimes B}\otimes 
	\mathcal{A}\otimes
	\mathbf{1}^{\otimes D}
	\big]
	\ket{BCD}
	$.  We say that operator $\mathcal{A}(C)$ \textit{touches} the qubits in state $\ket{C}$ only.
	\item For operator $\mathcal{A}=\mathcal{U}\circ\mathcal{B}$ we will define the operator $\mathcal{A}_{\mathcal{U}^-}=\mathcal{U}^\dagger\circ \mathcal{B}=\mathcal{B}$. In practice, $\mathcal{B}$ will have its complicated internal structure, and $\mathcal{U}$ will be very simple as compared to $\mathcal{B}$. Most often, it will be the Hadamard transformation creating the interference just before measurement. From the implementation point of view, the operator $\mathcal{A}_{\mathcal{U}^-}$ is the operator $\mathcal{A}$ without the ending operator $\mathcal{U}$.
	\item The probability of measuring the eigenstate $\ket{E}$ equal to $\ket{e}$ will be denoted with Fraktur "p": $\mathfrak{p}(\ket{E}=\ket{e})$ or, shortly $\mathfrak{p}\ket{E=e}$, omitting the round brackets for simplification of notation.
\end{itemize}

In this paper we will use the following nomenclature:
\begin{enumerate}
	\item QCoSamp is the name of the proposed method and the operators generated by this method.
	\item FCoSam is the function that arises from the system's quantum sampling evolving with one of the QCoSamp operators.
	\item Argument of the function is the variable denoted by $x$ and represents the argument of the modeled Fourier sine-cosine series. Depending on the application, there could be one or many arguments encoded in one of three ways (see sec. \ref{sec-selection-of-parameters} for details). 
	\item Parameters of the FCoSamp are variables that it depends on; they can be constant or changed (optimized) during the evolution process, depending on the application. There are two types of FCoSamp parameters:
	\begin{enumerate}
		\item \emph{FCoSamp frequencies}, which are, generally speaking, $n$-s in the eq. \ref{eq-fsn-approx} and \ref{eq-base-model} which could be understood as the numbers of consecutive harmonics (component) in both series.
		\item \emph{FCoSamp phases}, which are coefficients $r_n, s_n$ in mentioned equations and mapped to the Fourier series coefficients for the same component number $n$
	\end{enumerate}
	Notice that if we talk about frequencies, component (harmonic) numbers, phase shifts, we think about these quantities concerning the FCoSamp function. We emphasize it here because the same name of quantities arises on the operator description – e.g., the phase shifts $e^{i\varphi}$. In the case of an operator description, we will always talk about the "quantum phase," "quantum frequency," etc. 
	\item Reference values are the quantities that we compare FCoSamp (or its elements) to, during the computation process, or even after quantum sampling. 
	\item Arguments and parameters constitutes together \emph{elements} of the FCoSamp function.
\end{enumerate}

\subsection{Paper organization}

The structure of this paper is as follows.
In section \ref{sec-material-methods} and appendix \ref{appx-sec-mapping}, we define and prove the mentioned mapping, and, in section \ref{sec-quantum-programming-technique}, we describe the novel or adopted and extended programming techniques that are useful for the creation and usage of the QCoSamp. The conventional techniques are presented in the appendices \ref{sec-gates}-\ref{sec-oracle}. In section \ref{sec-theory-calculations}, we describe the QCoSamp in detail and present the method of building such operators from scratch. In section \ref{sec-results}, we present the experimental proof of the concept of QCoSamp obtained from the simulated and real quantum computers. We briefly describe the possible practical application in signal and image processing, and we outline the new and interesting research possibilities connected with the QCoSamp family. Section \ref{sec-discussion} is the short recapitulation of the paper.

\section{Methods}\label{sec-material-methods}

In this section, we present the general scheme of dealing with operators generated with QCoSamp. We give the mapping from the space of the coefficients of the sampled output probability distribution to the coefficients of the Fourier sine-cosine series. Next, we describe the novel or modified techniques used for the QCoSamp operator building or needed for its application. The conventional unmodified methods used in context QCoSamp, presented with notation consistent with the rest of the paper, are placed in the appendices \ref{sec-gates} to \ref{sec-oracle}.

\subsection{General quantum computation scheme using QCoSamp operators}\label{sec-geenral-scheme-of-using-QCoSamp}

The general scheme of quantum computation using the QCoSamp evolution operator consists of quantum computation and the part associated with interpretation:

\begin{enumerate}
	\item \textbf{Quantum computation by quantum sampling} is the iterative evolution (repeated sufficient times) of, once prepared, initial quantum states with given QCoSamp, ending with their with measurement. The measurement of one repetition gives one specific eigenstate at the output. Nevertheless, it is dependent on the algorithm and how many eigenstates there are possible to obtain. E.g., In the curve-fitting problem (sec. \ref{sec-app-signal-processing}), there is $2^{N\cdot (R+S)}$ potentially eigenstates possible to obtain, but in practice, their number is significantly limited due to the amplitude amplification algorithm (ibidem). We repeat the evolution - measurement procedure with the same input sufficient times and note each repetition's output eigenstate (or rather eigenvalue connected with this state). The histogram for the appearance of eigenstates is the output.
	\item \textbf{Interpretation part} is the generation of the solution of the problem modeled with the Fourier series corresponding to the applied QCoSamp, based upon the eigenstate appearance histogram. There is no general approach for the Fourier series generation, nor is there for the problem's solution, since it depends on the question posited to be solved. We must know the meaning of each output eigenstate from the perspective of the problem. In some cases (e.g., curve fitting, computing the value of the function), we construct the explicit series within only two reconstruction steps (sec. \ref{sec-reconstruction}). In other cases, such as integration, image filtering, and object recognition, the series works under the hood; therefore, output quantities are connected with the problem itself, not the Fourier series.
In integration, the probability for obtaining the state $\ket{0}$ after reconstruction is the value of the integral of the given function; in classification, the probability for the given eigenstate is the probability that the given object is in the class represented by this eigenstate.

\end{enumerate}
According to this scheme, we can specify steps that are performed to make computation using QCoSamps:
\begin{enumerate}\label{enum-steps-of-qcosam-computations}
	\item Create the \textbf{model of the problem} to be solved with the Fourier sine-cosine series:
	\begin{enumerate}
		\item Describe the problem with the Fourier sine-cosine series.
		\item\label{enumerate-parameters-encoding} Choose the data input method and parameters encoding the problem (see sec. \ref{sec-selection-of-parameters} for details).
		\item Set the measurement basis's dimension, making sure that it is as small as possible (but not smaller).
		\item\label{enumerate-dictionary} Specify the dictionary of the output eigenstates defining what each means in the problem domain to be solved.
	\end{enumerate}
	
	\item \textbf{Build the QCoSamp} corresponding to the series used in solving the problem with the instructions described in sec. \ref{sec-theory-calculations}.
	\item \textbf{Prepare the initial state} of the quantum system:
	\begin{enumerate}
		\item Set the \textit{ancillae} to $\ket{0}$.
		\item Set the qubits corresponding to the parameters and input data of the problem to the appropriate states according to the selected encoding method in point \ref{enumerate-parameters-encoding}.
	\end{enumerate}
	\item Proceed with \textbf{quantum sampling} of the same initial state. Make a normalized histogram of the results where one bin represents one of the eigenstates, and its height is the number of appearances of this state. 
	\item \textbf{Reconstruct the results} (see sec. \ref{sec-reconstruction}). In cases where the output eigenstates do not represent the series's quantities directly (e.g., object recognition), it is not possible to make a reconstruction. Because the output contains values of the function and other quantities, such as the probability of an image containing a given class of object; this step is then omitted.
	\item \textbf{Interpret} reconstructed results according to the problem and the dictionary of the eigenstates from the point \ref{enumerate-dictionary}.
\end{enumerate}

\subsection{Representation of Fouries series by Quantum cosine Sampled Function}\label{sec-mapping}

The real-value function $f$, that is the absolute integrable of the interval $[-\pi, \pi]$ can be expanded to the Fourier sine-cosine series form:
\begin{equation}\label{eq-fs-infty}
f(x)=\sum_{n=0}^\infty \bigg(\lambda_n cos(nx)+\gamma_n sin(nx) \bigg),
\end{equation}
with expansion coefficients $\lambda_n, \gamma_n$.
In practice one uses finite approximation given by the function as follows:
\begin{equation}\label{eq-fsn-approx}
f_N(x)=\sum_{n=0}^N \bigg( 
\lambda_n cos(nx)+\gamma_n sin(nx)
\bigg)
\end{equation}
The above function has many practical applications as mentioned in the introduction (see \cite{rao2014discrete, abuelmaatti_simple_1993, guruswami_robust_2016, bilodeau_fourier_1992, asrini_fourier_2014, pane_parametric_2014, chong-yung_chi_fourier_1997, chaudhury_fast_2011, ghosh_fast_2016, chii-horng_chen_statistical_1999}.
The essence of our approach is re-parametrization of the above function in the form that has the Quantum Cosine Sampled Function FCoSamp:
\begin{equation}\label{eq-base-model}
\mu_N(x)=
\sum_{n=1}^N
\bigg(
\frac{1+cos(nx+r_n)}{L}+
\frac{1+cos(nx+s_n)}{L}
\bigg)=
\sum_{n=1}^N\nu_n(x).
\end{equation}
In the above equation, where $N$ is a positive integer and a count of the sum of components, $L$ is the normalization factor arising from quantum postulates, the form of the above function $\mu_N$ comes from the QCoSamp operator architecture, we discuss in detail in sec. \ref{sec-theory-calculations}. In section \ref{sec-reconstruction} and appendix \ref{appx-sec-mapping} we prove that Fourier series in eq. \ref{eq-fsn-approx} can be represented, up to their scale factors, by the function FCoSamp, which is the result of quantum sampling of the state prepared using the QCoSamp operator and is an experimentally determined approximation of the function in the eq. \ref{eq-base-model}. The quantum algorithm, solving the given problem described by Fourier cosine series and is mapped to/from by the function given above, (eq. \ref{eq-base-model}) approximated by the FCoSamp function and constructed by using superposition, interference, and (cluster) entanglement for the elements of the $\mu_N$ function and dependent on the problem, already mentioned in sec. \ref{sec-motivation} and described in detail in the remaining part of the paper. 
To close, in considering representation of the Fourier series by the FCoSamp function, we describe the reconstruction of the Fourier series (or other quantities connected with it) in the next subsection (sec. \ref{sec-reconstruction}) and present below (eq. \ref{eq-mapping-FCoSamp-Fourier}) the final equation for realizing a mapping of such a representation, which is described in more detail in \ref{appx-sec-mapping}:
\begin{align}
\lambda_n &= cos(r_n)+cos(s_n)\nonumber\\
\gamma_n &= -sin(r_n)-sin(s_n),\label{eq-mapping-FCoSamp-Fourier}\\
\end{align}
where $n$ is the number of components.

\subsection{Reconstruction of Fourier series from quantum cosine sampled function}\label{sec-reconstruction}

During the generation of a particular FCoSamp, we should point out two important facts:
\begin{enumerate}
	\item In light of the facts described previously, and in \ref{appx-sec-mapping}, not every function is proper and therefore represented by QCoSamp – only the coefficients that are in the ball $\lambda^2+\gamma^2\leq 2$ we could be map to the phase shifts of FCoSamp. Therefore if we model a function with any coefficient outside this area, FCoSamp has to be multiplied by a factor of $(1/2) max_k(\lambda_k^2+\gamma_k^2)$.
	\item the second reconstruction has to be made due to a quantum normalization coefficient, the value of which is discussed in section \ref{sec-normalization-factor} "\nameref{sec-normalization-factor}". Where the need for this mapping can be imagined as decreasing the radius of the circle where $\lambda, \gamma$ exists. If we have this normalization factor equal to $1/(4M_n)$ for $n$-th component (see sec. \ref{sec-normalization-factor}), then we multiply by this coefficient of the resulting function.
\end{enumerate}

Therefore, the reconstruction is made according to the formula:
\begin{align}
\varrho_n&=2M_n max_k(\lambda_k^2+\gamma_k^2)=2^{\delta_n+1} max_k(\lambda_k^2+\gamma_k^2)\label{eq-reconstruction-FCoSamp-Fourier},
\end{align}
where $n$ is the number of a component, $\varrho$ is the reconstruction factor for some $n$-th component, $\delta_n$ is the depth of the component in the QCoSamp architecture (see sec. \ref{sec-base-model-architecture}) and $M_n=2^\delta_n$ (see sec. \ref{sec-base-model-architecture}, \ref{sec-normalization-factor}).
So, the resulting Fourier series is equal to:
\begin{equation}
f_N(x)=\sum_{n=1}^N \varrho_n\nu_n(x).
\end{equation}
Suppose the QCoSamp operator is balanced (ibidem). In that case, the reconstruction process simplifies because the depth of components is equal to the height of the QCoSamp, so $2^{\delta_n}=N$, the above equation reduces:
\begin{equation}
\varrho=2N max_k(\lambda_k^2+\gamma_k^2).
\end{equation}
It means that we do not need to multiply coefficients for each component separately and then compute the value of the function; instead, we can compute:
\begin{equation}\label{eq-fourier-reconstr-for-balanced}
f_N(x)=\varrho\mu_N(x).
\end{equation}
This simplification is one of the advantages of \textit{balanced} QCoSamps; it may be of great importance to some applications. Note that if we do not need the values of $\lambda_n, \gamma_n$ for the solution of the problem to be solved, there is no need to compute them. The Fourier series value for a given $x$ can be obtained, with the reconstruction process taken purely as multiplication by the reconstruction factor.

\subsection{Quantum programming techniques for construction of QCoSamp operators}\label{sec-quantum-programming-technique}

In this section we describe methods of quantum programming used for the QCoSamp generating operators, where the conventional techniques, presented with notation consistent with the rest of the paper, are placed in \ref{sec-gates} to \ref{sec-oracle}. We also introduce new quantum programming methods and concepts for the purpose of QCoSamp construction: \textit{twice permuted controlled operator}, \textit{ordering coordinates}, \textit{comparison of state}, \textit{constant data encoding}, \textit{forging reference probability}, \textit{quantum sliding window} and \textit{quantum kernel filtering} in the following section.

\subsubsection{Ordering the coordinates}\label{sec-coeff-ordering}

The order of coordinates for the quantum state does not influence the norm of this state but is important from three perspectives: (1) the complexity of the unitary operator generating the state desired, its creation described in details in the next subsection – \ref{sec-cascade-controlled-operator}), (2) the content of each of the states projected to the measurement basis, which is relative to the final form of the probability distribution over the measurement, which determines the resulting computation and (3) last but not least – clarity in the description of states. Therefore, there is a need for a machine, fixing the coordinates' order needed during computation.\\  
The works of Planat \cite{planat_magic_2017_Gadik,planat_magic_2017} show that the application of permutation groups as gates in quantum computers are possible. He uses the analysis of so-called "magic states" (first announced by Bravyi and Kitaev in \cite{bravyi_universal_2005}). However, we can use this to change the order of quantum state coordinates. We will call the element of permutation group used as evolution operator by \textit{ordering operator} $\mathcal{S}$ to distinguish from exchange operators existing in the Fock spaces (announced by Polychronakos in \cite{polychronakos_exchange_1992}), often called a permutation operator (e.g. Basu-Malik in \cite{basu-mallick_spin-dependent_1996}).\\
In the ordering operator matrix in each row and each column there is precisely one appearance of $1$ and $0$ everywhere else. Hence the number of $1$ is equal to the dimension of the matrix. It means that acting on the state does not change the value of coordinates but only their order. An ordering operator's trivial example is the identity; a non-trivial example is the X-Pauli gate or controlled X Pauli gate.\\
Following the example from the permutation group, the ordering operator is denoted with a permutation, in the two-line notation, omitting their identities, in square brackets and then, as usual, the qubits it touches are in round parenthesis. For example:
\begin{align*}
\mathcal{S}
\left [
    \begin{array}{ccccc}
        1 & 3 & 4\\
        3 & 4 & 1
    \end{array}
\right ](q_1, g_2)[a\ket{00}+b\ket{01}+c\ket{10}+d\ket{11}]=
\begin{bmatrix}
    0 & 0 & 1 & 0 \\
    0 & 1 & 0 & 0 \\
    0 & 0 & 0 & 1 \\
    1 & 0 & 0 & 0 
\end{bmatrix}
\begin{bmatrix}
    a\\b\\c\\d
\end{bmatrix}
=
\begin{bmatrix}
    d\\b\\a\\c
\end{bmatrix}
\end{align*}
Meaning, that the operator moves the first coordinate to the third eigenstate, third to fourth and fourth to first eigenstate.\\
The ordering operator is isomorphic between state in the subspace of qubits it touches, preserving the norm of the state from subspace:
\begin{equation}
\norm{\mathcal{S}(q_1,...,q_n)\braket{q_1...q_nq_{n+1}...q_Q}{q_{n+1}...q_Q}}^2=\norm{\braket{q_1...q_nq_{n+1}...q_Q}{q_{n+1}...q_Q}}^2
\end{equation}
It is obvious, because the ordering operator does not change the value of a coordinate and addition in the scalar field is commutative.

\subsubsection{Twice permuted controlled operator}\label{sec-cascade-controlled-operator}

In the previous section \ref{sec-coeff-ordering}, we mention that the ordering operator influences the operator's complexity generating the desired state at the current stage of preparation of the final evolution operator, which comes down to the count of the gates implemented on the specific quantum hardware and used to obtain the selected unitary operator. 
The example of complexity related to the coordinates' order in the case of a $4$ qubits system, in which we must change (e.g., multiply by quantum phase) one, arbitrary chosen coordinate. If it is the last coordinate, for the eigenstate $\ket{11}$, we need one gate only – a controlled phase shift gate. In the case of $\ket{01}$ and $\ket{10}$ being coordinates, we need two gates – a quantum phase shift gate changes the last coefficient as well, which we un-compute with the controlled phase shift gate of the opposite phase. For the $\ket{00}$ coordinate, we cannot change it without changing the order of the coordinates, achieved in this case by a combination of $ X $ Pauli gates.\\ 
To be more formal and precise, the quantum state of $\Phi$ qubits, during evolution, is a tensor product of all qubits. Therefore if we act on the $k$-th qubit gate acting as follows: $a\ket{0}+b\ket{1}$, coordinates having $0$ in the $k$-th position of the corresponding eigenstate are multiplied by $a$ and $1$ – by $b$. But if we still can require, during the generation of QCoSamp operator, to change arbitrary chosen coordinates, we use for this purpose the \emph{twice permuted controlled operator}. \\
First, we observe that $c$ times controlled operators, which let us recall, are responsible for (cluster) entangled states, are made of the identity operator acting on \emph{controlling} qubits and one qubit operators for \emph{controlled} qubit. Such operators change those coordinates with eigenstates having $1$ in the controlling qubit; therefore, the number of changed coordinates is equal to $2^{\Phi-c}$. So if we have a state $\ket{\Phi}=\ket{\varphi_0\dots\varphi_{\Phi-1}}$ and we want to change the $n$-th coordinate, we should proceed as follows:

\begin{enumerate}
    \item Act on state $\ket{\Phi}$ with ordering operator: 
    \begin{align}
    \ket{\Psi_1}=
    \mathcal{S}
    \left [
        \begin{array}{c}
            n\\
            \Phi-1
        \end{array}
    \right ](\varphi_n, \varphi_{\Phi-1})\ket{\Phi}
    \end{align}.
    After this operation, the $n$-th and last coordinates swap.
    \item Create $\Phi-1$ times controlled operator:
    \begin{align}
    c^{\Phi-1}\mathcal{G}=\left [
        \begin{array}{c|c}
            \mathbf{1}^{\otimes(\Phi-1)} & \mathbf{0}\\
            \hline
            \mathbf{0} & \mathcal{G}
        \end{array}
    \right ]
    \end{align}
    
  and act upon it on the just obtained permuted state $\ket{\Psi}$. As a result, we obtain the state where, in general, \emph{two} the last coordinates are changed by the operator $\mathcal{G}$. So if we want to change only one, the first raw coordinate of the operator $\mathcal{G}$ has to be equal to $[1, 0]$. Since it is unitary, the first value of the second row must be $0$ as well. 
   \item Act upon the obtained state with the same permutation operator to reproduce the coordinates' original order. Hence the name of the operator is \emph{twice permuted}.

\end{enumerate}
So the final form of the twice permuted controlled operator has the form as follows:
\begin{align}
    \mathcal{S}
    \left [
        \begin{array}{c}
            n\\
            \Phi-1
        \end{array}
    \right ](\varphi_n, \varphi_{\Phi-1})\circ
    c^{\Phi-1}\mathcal{G}\circ
    \mathcal{S}
    \left [
        \begin{array}{c}
            n\\
            \Phi-1
        \end{array}
    \right ](\varphi_n, \varphi_{\Phi-1})
\end{align}

We can easily generalize this operator to widen the number of coordinates to change by the same value, assuming that its number is equal to 
$K\leq2^{\Phi-1}$
and each of them has to be multiplied by a correct value of $\psi_k$, which means that they do not make the resulting state inconsistent with the postulates of quantum mechanics. In that case, the ordering operator has the form:

    \begin{align}
    \mathcal{S}
    \left [
        \begin{array}{ccc}
            n_1 & \dots & n_k\\
            \Phi-1 &\dots & \Phi-k 
        \end{array}
    \right ](\varphi_n, \varphi_{\Phi-1})
    \end{align}
    and controlled operator:
    \begin{align}
    \left [
        \begin{array}{cccccr}
            1 &\multicolumn{5}{r}{\mathbf{0}}\\
            0&1 &\multicolumn{4}{r}{\mathbf{0}}\\
            \multicolumn{6}{c}{\dots}\\
            \mathbf{0}&\dots&1&\multicolumn{3}{r}{\mathbf{0}}\\
            \multicolumn{3}{l}{\mathbf{0}}& \psi_1&\multicolumn{2}{r}{\mathbf{0}}\\
            \multicolumn{3}{l}{\mathbf{0}}&0& \psi_2&\mathbf{0}\\
            \multicolumn{6}{c}{\dots}\\
            \multicolumn{4}{l}{\mathbf{0}} & 0 &\psi_K\\
        \end{array}
    \right ]
    \end{align}

\subsubsection{Comparison of the states}\label{sec-eq-states}

The part of an algorithms basing on QCoSamp is often the amplitude amplification algorithm \cite{brassard_exact_1997, Grover_algorithm, brassard_quantum_2000}, which its core is to define the two operators: Oracle $U_\omega$ and diffusion operator $U_\varphi$, which are related to the \emph{desired state} $\ket{\omega}$ and the \emph{beginning state} $\ket{\varphi}$ within projection operators: $U_\varphi=2\ket{\varphi}\bra{\varphi}-\mathbf{1}, U_\omega=\mathbf{1}-2\ket{\omega}\bra{\omega}$, see \ref{sec-oracle} for more details. Therefore using this operator comes down to the definition of states $\ket{\varphi}$ and $\ket{\omega}$, which is much simpler than defining the oracle and diffusion operator from scratch. The beginning state is usually a uniform superposition of the initial state. Within quantum phase shifts of coordinates if needed, which does not destroy the superposition's uniformity because its norm is equal to $1$, so does not change the resulting probability. Generally speaking, the desired state depends on the problem to be solved. However, often is considered as the equality of two states, e.g., in the curve fitting task, we compare the value of the FCoSamp function for $x_k$ with reference value $y_k$, so the desired situation is when the state representing FCoSamp and the reference value is the same. Therefore, we need a machine for a comparison of two states.\\
Now we discuss the problem of state comparison in the general case. Let us assume that:

\begin{align}\label{eq-equalization-formulation}
    \ket{\hat{1}}=&\frac{1}{\sqrt{2^{N}}}\sum_{k=0}^{2^N-1}\ket{k}\\
    \ket{W}=&\frac{1}{\sqrt{2^{N}}}\sum_{k=0}^{2^N-1}w_k\ket{k}\\
    \ket{Y}=&\frac{1}{\sqrt{2^{N}}}\sum_{k=0}^{2^N-1}y_k\ket{k}\\
\end{align}
We will call $\ket{W}$ the \textit{argument state}, $\ket{Y}$ \textit{reference state} and $\ket{\hat{1}}$  – \textit{idle state}.  Moreover, let's assume that we want to check if the states $\ket{W}$ and $\ket{Y}$ are the same or not. First we prepare the states:
 \begin{align}\label{eq-equalization-states}
    \ket{\Phi_1}=&\frac{1}{\sqrt{2}}\big(\ket{\hat{1}0}+\ket{W1}\big);\nonumber\\
    \ket{\Phi_2}=&\frac{1}{\sqrt{2}}\big(\ket{\hat{1}0}-\ket{Y1}\big);
 \end{align}
 
The state above is attainable, e.g., by 
$H\ket{0}\otimes\ket{Q}=\ket{Q0}+\ket{Q1}$ 
and un-computation of the coordinates $\ket{Q0}$. After we make a tensor product of such states and the un-computation of the coordinate $\ket{11}$ we obtain the formula as follows:

 \begin{align}\label{eq-equalization-state-building}
     \ket{\Phi_1}\otimes\ket{\Phi_2}=&\frac{1}{2}
     \Big[
        \ket{\hat{1}0\hat{1}0}-
        \ket{\hat{1}0Y1}+
        \ket{W1\hat{1}0}-
        \ket{W1Y1}
     \Big]\stackrel{\tiny{uc(W,Y,\ket{11})}}{\longrightarrow}\nonumber\\
     &\frac{1}{2}
     \Big[
        \ket{\hat{1}0\hat{1}0}-
        \ket{\hat{1}0Y1}+
        \ket{W1\hat{1}0}-
        \ket{\hat{1}1\hat{1}1}
     \Big]\stackrel{\tiny{\mathcal{S}[...]}}{\longrightarrow}\nonumber\\
     &\frac{1}{2}\ket{\hat{1}}
     \Big[
        \ket{\hat{1}00}-
        \ket{Y01}+
        \ket{W10}-
        \ket{\hat{1}11}
     \Big]\nonumber\\
     \ket{\Psi c_1c_2}:=&\ket{
        \hat{1}00}-
        \ket{Y01}+
        \ket{W10}-
        \ket{\hat{1}11}
 \end{align}
 
In the equation above, $uc$ means un-computation, and the $\mathcal{S}[...]$ 
means the ordering operator; the permutation details are omitted for clarity, but they are easy to determine.
Now we will proceed with state $\ket{\Psi c_1 c_2}$, where If we want to check the equality of the state, we can act on the \textit{ancillae} with $H^{\otimes2}$ operator and obtain the state:

 \begin{align}\label{eq-state-equalization-hadamard}
        \big(\ket{W}-\ket{Y} \big)&\ket{00}+\nonumber\\
        \big(2\ket{\hat{1}}+\ket{Y}+\ket{W} \big)&\ket{01}+\nonumber\\
        \big(2\ket{\hat{1}}-\ket{Y}-\ket{W} \big)&\ket{10}+\nonumber\\
        \big(\ket{Y}-\ket{W} \big)&\ket{11}
 \end{align}
 
If we enact the $cX(c_1, c_2)$ and measure the right \textit{ancilla}, the obtaining of $\ket{0}$ 
is impossible in the case where the states are equal. So the probability of obtaining the $\ket{1}$ is the measure of similarity of state – 
if it is equal to $1$, the states are the same; if it is equal to $0.5$, it means that $\ket{Y}=-\ket{W}$, the values between $0.5$ and $1$ means that no one of the two cases appears, but if the result is closer to $1$ the closer the state is balanced and vice-versa.\\
Let us consider that the argument state $\ket{W}$ depends on another state $\ket{p}$ called parameter state. 
We understand that coordinates of $\ket{W}$ are the functions of $\ket{p}$. The task of state comparison is to find the state $\ket{p}$ for which it holds $\ket{W(p)}=\ket{Y}$. 
The above method is not useful for this task because the measurement basis size is equal to $2^{P+2}$. 
Therefore, we use the amplitude amplification algorithm (see \ref{sec-oracle}). We define the beginning and desired state using the equations \ref{eq-equalization-state-building} and \ref{eq-state-equalization-hadamard} , described as:

 \begin{align}\label{eq-equalization-state-phi-omega}
    \ket{P}=&\frac{1}{\sqrt{2}^P}\sum_{p=0}^{2^{P}-1}\ket{p}\nonumber\\
    \ket{\varphi}=&\frac{1}{2}H^{\otimes 2}(c_1, c_2)\ket{P\Psi c_1c_2}=\nonumber\\
        &\frac{1}{2}\sum_{p=0}^{2^P-1}\ket{p}\bigg[
            \big(\ket{W(p)}-\ket{Y} \big)\ket{00}+
            \big(2\ket{\hat{1}}+\ket{Y}+\ket{W(p)} \big)\ket{01}+\nonumber\\
            &\big(2\ket{\hat{1}}-\ket{Y}-\ket{W(p)} \big)\ket{10}+
            \big(\ket{Y}-\ket{W(p)} \big)\ket{11}
        \bigg]\nonumber\\
    \ket{\omega}=
    &\frac{1}{2}\sum_{p=0}^{2^P-1}\ket{p}\bigg[
            0\ket{00}+
            \big(2\ket{\hat{1}}+2\ket{Y} \big)\ket{01}+
            \big(2\ket{\hat{1}}-2\ket{W(p)} \big)\ket{10}+
            0\ket{11}
        \bigg]
 \end{align}
 
The correct state $p$ is be found by virtue of amplification algorithm (see \ref{sec-oracle}) after, at worst $1/\sqrt{2^{P+\Psi+2}}$. 
We have to remember here that there is the state $\ket{\hat{1}}$ 
extracted in the eq.  \ref{eq-equalization-state-building} in the whole quantum state. 
but it is not a part of the amplitude amplification, so we omit them to decrease the number of repetitions. 
Each coordinate has the same value: $1/\sqrt{2}^{N}$ and the states $\ket{\varphi}, \ket{\omega}$ would contain at the beginning, the extra summation over all coordinates, each coordinate would be divided by $\sqrt{2}^N$, after measurement, this gives each coordinate division by $2^N$. If we do not take this $\ket{\hat{1}}$ into account within the measurement basis, the quantities divided by $2^N$ would sum $2^N$ times, returning the original value, so finally, this $\ket{\hat{1}}$ state has no contribution to the final measurement probability density.

\subsubsection{Encoding the constant data}\label{sec-huge-number-encoding}

 By distribution phase encoding we can encode the variable using $N$ qubits. This technique is useful in the case where we want to process such a variable – e.g. we can make uniform superposition on all of qubits encoding it and then use it for quantum summation, amplitude amplification etc. However, in our algorithm there can be values that we do not change – e.g. arguments of the function. This variable we call \textit{constant}. We can encode constants using the operator acting on the uniform superposition:
 \begin{align}\label{eq-constant-encoding-epsilons}
    \mathcal{E}_K\ket{H\ket{0}}^{\otimes K}:=&
    \frac{1}{\sqrt{2}^K}
     \begin{bmatrix}
         e^{i\varepsilon_1}&0&...&0&...&0\\
         0&e^{i\varepsilon_2}&.&0&.&0\\
         \multicolumn{6}{c}{...}\\
         0&0&...&e^{i\varepsilon_k}&...&0\\
         \multicolumn{6}{c}{...}\\
         0&0&...&0&...&e^{i\varepsilon_{2^K}}\\
     \end{bmatrix}
     \begin{bmatrix}
         1\\
         1\\
         ...\\
         1\\
         ...\\
         1\\
     \end{bmatrix}=
     \frac{1}{\sqrt{2}^K}
     \begin{bmatrix}
         e^{i\varepsilon_1}\\
         e^{i\varepsilon_2}\\
         ...\\
         e^{i\varepsilon_k}\\
         ...\\
         e^{i\varepsilon_{2^K}}\\
     \end{bmatrix}=
     \frac{1}{\sqrt{2}^K}
     \ket{\varepsilon}\nonumber\\
     \ket{\varepsilon}=&\sum_{k=0}^{2^K-1}e^{i\varepsilon_k}\ket{k}=\sum_{k=0}^{2^K-1}e^{i*\ket{k}}\ket{k}
 \end{align}
 
The angles $\varepsilon_k$ are the constant values. We say that $*\ket{k}$ is the \textit{angle pointed by eigenstate $\ket{k}$}. We use the notation with star $*$ because the eigenstate $\ket{k}$ is a bit like a pointer in classical programming.\\
Let us suppose that we have the state with the variable $\varphi$ that will be processed: $\frac{1}{2}(\ket{0}+e^{i\varphi}\ket{1})$. If we make a tensor product of this state with the result of a constants encoding, we obtain:

 \begin{align}\label{eq-inject-constant-data}
     \frac{1}{2}(\ket{0}+e^{i\varphi}\ket{1})\otimes \mathcal{E}_K\ket{H\ket{0}}^{\otimes K}=
     \frac{1}{2\sqrt{2}^K}
     \sum_{k=0}^{2^K}
     \big(
       \ket{0k}+e^{i(\varphi+*\ket{k}}\ket{1k}
      \big)
 \end{align}
 
 Therefore, constant values are now part of the computation and values of $\varphi$, optimized according to constraints.\\
 The other benefit of this operator is the possibility of connecting the input data in pairs or, in general, in $n$-tuples. Let's consider $k$ $n-$ tuples: $x^{(1)}, ..., x^{(K)}\in\mathbf{R}^N$. We can use the constant encoding operator on each coordinate of $x^{(k)}$ separately:
 
 \begin{align}
     \ket{\varepsilon}_n=&
     \sum_{k=0}^{2^K-1}e^{ix^{(k)}_n}\ket{k}_n=
     \sum_{k=0}^{2^K-1}e^{i (*\ket{k}_n)}\ket{k}_n\nonumber\\
     *\ket{k}_n=&x_n^{(k)}
 \end{align}
 
For the same coordinate $n$ its eigenstate $\ket{k}_n$ is the same for all $x^{(k)}$. Therefore, it preserves the information, being both the coordinate of the same point and the same $n$-tuple. Further, this eigenstate numbers the tuple.
The straightforward way to generate the operator $\mathcal{E}_K$ or state $\ket{\varepsilon}$ depends on the concrete application and described in detail in the context of QCoSamp operator in sec. \ref{sec-fco}.

\subsubsection{Forging reference probability}\label{sec-comparison-the-probability}

This technique serves to find an operator that creates the state that is quantum sampled to reference probability distribution, which is the solution of the problem, that is why it is called "reference". The idea of this technique is fetched from metal forging: the initial state is "forged" by changing the parameters of the FCoSamp slightly in each repetition of the amplitude amplification algorithm (analogous to the press or hammer hitting the metal), which forms it to shape, similar to the shape defined by the reference, which plays the role of the forging die.\\ 
In the sec. \ref{sec-eq-states} we described that the oracle and diffusion operator can be created by the definition of the desired and beginning state, we discussed the problem in comparison of two states as well, which determine when the states are equal or not and delivers the method of obtaining the level of such similarity (or dissimilarity) in eq. \ref{eq-state-equalization-hadamard} and below. 
In the previous section \ref{sec-huge-number-encoding} we provided the method of injection to the system, the large number of constant data. Hence we can create the state describing the final expected probability distribution, which is the FCoSamp function of the large number of arguments which depend, in the general case, on the set of frequencies and coefficients of FCoSamp, related to the frequencies and coefficients of the Fourier series modelling the problem to be solved.
Since we have the model of the problem described by Fourier series, we can define the relation of its coefficients, arguments, and the solution of the problem, which performed exemplary in the curve fitting problem. This relation will be defined by minimizing the fitting error, which means to find the set of coefficients such that the error is as small as possible. Therefore, in approaching the limit, the error should be equal to zero, even if in the concrete case where zero error is not affordable. 
We can say in that case that our desired state should generate a probability distribution representing the zero error, according to the dictionary meaning of the basis eigenstates obtainable by measurement, in context of the problem to be solved (see sec. \ref{sec-geenral-scheme-of-using-QCoSamp}, pg. \pageref{sec-geenral-scheme-of-using-QCoSamp}). In this section we show how to map this desired (and in fact any) output probability distribution to the state notion and use the algorithm amplification with those mapped states.\\

For that purpose, the output probability distribution over measurement basis eigenstates can be identified by the function $g_p(x)$. We assume that the value of this function depends on the arguments and the set of parameters. 
The parameters of this function are encoded in the state $\ket{P}$ and the input value – on the state $\ket{X}$. 
During computation there arise the state $\ket{p}\ket{W_p(x)}\ket{C}$, which represents the function for the set of the parameters encoded in the state $\ket{p}$; the $\ket{C}$ is the state of \textit{ancillae}.
We prepare the system in such a way that the probability of measuring the state $\ket{0}$ for the one of \textit{ancilla} is equal to $g_p(x)$ for the given $x$. 
Formally, we describe the original state of the system:

 \begin{align}\label{eq-comparison-state}
     \ket{P}\ket{X}\ket{C}=&\frac{1}{\Lambda}\sum_{p=0}^{2^P-1}\sum_{x=0}^{2^X-1}\ket{px}H(c_{C-1})\big(\ket{W_p(x)}\ket{0}+\ket{\hat{1}}\ket{1}\big)=\nonumber\\
     &\frac{1}{\Lambda}\sum_{p=0}^{2^P-1}\sum_{x=0}^{2^X-1}
     \ket{px}\bigg[
        \bigg(\sum_{c=0}^{2^C-2}(w_c(p,x)+1)\ket{c}\bigg)\ket{0}+
        \bigg(\sum_{c=0}^{2^C-2}(w_c(p,x)-1)\ket{c}\bigg)\ket{1}
     \bigg], s.t:\nonumber\\
     \ket{W_p(x)}=&\sum_{c=0}^{2^C-2}w_c(p,x)\ket{c}\nonumber\\
     g_p(x)=&\sum_{c=0}^{2^C-2}(w_c(p,x)+1)(\overline{w_c(p,x)+1})
 \end{align}
 
Comparsion the value of this function for a given $x$ to reference value $y$ during computation, meaning before measurement is important and having many applications task.
For solving it we will use the state comparison technique described in the section \ref{sec-eq-states}, based on the consideration that if $\forall x : g_p(x)=y\Leftrightarrow g_p(x)-y=0 $\\
The state $\ket{W}$ plays the role of the argument state (see eq. \ref{eq-equalization-formulation}). The reference state we define as follows:

    \begin{equation}
        \ket{Y}=\frac{1}{\sqrt{2}^{2^C-1}}\sum_{c=0}^{2^C-2}e^{i cos^{-1}(2y-1)}\ket{c}
    \end{equation}
If we act with Hadamard gate on the state $\frac{1}{\sqrt{2}}\big(\ket{Y}\ket{0}+\ket{\hat{1}}\ket{1}\big)$ and we measure the last \textit{ancilla} we obtain for eigenstate $\ket{0}$:
    \begin{align}
        \frac{1}{4\cdot 2^{2^C-1}}\sum_{c=0}^{2^C-2}
            \bigg(1+e^{i cos^{-1}(2y-1)}\bigg)
            \overline{
                \bigg(1+e^{i cos^{-1}(2y-1)}\bigg)
            }=\nonumber\\
            \frac{1}{4\cdot 2^{2^C-1}}\cdot 2^{2^C-1}\cdot 2\big[
            1+
            cos\big(
                cos^{-1}(2y-1)
            \big)
            \big]=\frac{1}{2}+\frac{2y-1}{2}=y
    \end{align}
    
According to the equation \ref{eq-equalization-state-phi-omega} we define the initial and desired states as follows:

\begin{align}
    \ket{\varphi}(P,C)=
        &\frac{1}{2\Lambda}\sum_{p=0}^{2^P-1}\ket{p}\sum_{x=0}^{2^X-1}\ket{x}\bigg[
            \big(\ket{W_p(x)}-\ket{Y} \big)\ket{00}+
            \big(2\ket{\hat{1}}+\ket{Y}+\ket{W_p(x)} \big)\ket{01}+\nonumber\\
            &\big(2\ket{\hat{1}}-\ket{Y}-\ket{W_p(x)} \big)\ket{10}+
            \big(\ket{Y}-\ket{W_p(x)} \big)\ket{11}
        \bigg]\nonumber\\
        \ket{\omega}(P,C)=&\frac{1}{2\Lambda}\sum_{p=0}^{2^P-1}\ket{p}\sum_{x=0}^{2^X-1}\ket{x}\nonumber\\
        &\bigg[
        \big(2\ket{\hat{1}}+\ket{Y}+\ket{W_p(x)} \big)\ket{01}+
        \big(2\ket{\hat{1}}-\ket{Y}-\ket{W_p(x)} \big)\ket{10}
        \bigg]\label{eq-fi-omega-probab-comparison}\\
    \pm\ket{W_p(x)}\pm\ket{Y}=&\sum_{c=0}^{2^C-2}\big[\pm w_c(p,x)\pm e^{i cos^{-1}(2y-1)}\big]\ket{c}\nonumber\\
    2\ket{\hat{1}}\pm\ket{W_p(x)}\pm\ket{Y}=&\sum_{c=0}^{2^C-2}\big[2\pm w_c(p,x)\pm e^{i cos^{-1}(2y-1)}\big]\ket{c}\label{eq-fi-omega-probab-comparison2}
\end{align}

Since the amplitude amplification algorithm creates the operators from the state $\ket{\varphi}$ and $\ket{\omega}$ the notation $\ket{\varphi}(P,C)$ and $\ket{\omega}(P,C)$ means the same as in the operator cases, implying that the algorithm changes the states $\ket{P}$ and $\ket{C}$ exclusively. It indicates that this algorithm aims to desire states in which coordinates for eigenstates $\ket{00}$ and $\ket{11}$ are zeros \emph{for all} $\ket{x}$ states. If there are many $\ket{x}$ states, this situation could be impossible to achieve. In that case, each repetition of the amplitude amplification algorithm makes the probability for best fitting $p$ greater and other probabilities less so.
The FCoSamp function, as a result, has increasingly refined those eigenstates that represent the best fitting of the argument and reference states. Therefore, if we take a quantum sampling of the output, we obtain the higher probabilities for those parameters $p$ of $g_p$ function that makes this function most similar to the argument and reference data.\\
We can use the technique of constant data encoding (sec. \ref{sec-huge-number-encoding}) to provide to the large amount of argument and reference data. In that case we have to use the $\varepsilon_k=e^{i cos^{-1}(2y_k-1)}$ (eq. \ref{eq-constant-encoding-epsilons}) while coding the reference values. Except that of the equation $\ref{eq-fi-omega-probab-comparison2}$ which has the form:

\begin{align}
    \pm\ket{W_p(x)}\pm\ket{Y}=&\sum_{c=0}^{2^C-2}\big[\pm w_c(p,*\ket{x})\pm *\ket{y}\big]\ket{c}\nonumber\\
    2\ket{\hat{1}}\pm\ket{W_p(x)}\pm\ket{Y}=&\sum_{c=0}^{2^C-2}\big[2\pm w_c(p,*\ket{x})\pm *\ket{y}\big]\ket{c}
\end{align}

we use this technique, remembering, the assumption that $g_p$ is made of $2^{C}-2$ parts, which sum up to $g_p$ for each $x$. 
We check the condition that each part building $g_p$ is equal to $y/2^{C-1}$.
This is the more robust condition and demanded because $g_p(x)=y$ fulfills, but there could be a function in which parts correspond to the non-uniformly divided $y$.

 \section{Theory for constructing quantum cosine sample operator}\label{sec-theory-calculations}
In this section, we will describe the method of Quantum cosine series sampling (QCoSamp), creating the operators which the quantum sampling generates the cosine sampled function (FCoSamp) on, which is mapped (sec. \ref{sec-mapping}, \ref{appx-sec-mapping}) to the sine-cosine Fourier series, which finally is the model of the problem to be solved (sec. \ref{sec-geenral-scheme-of-using-QCoSamp}). We describe here the architecture of QCoSamp operators which is the full binary tree (\cite{binary-tree, full_binary-tree}); introducing the notion of \emph{balanced} QCoSamp operators for which the architecture is a perfect binary tree \cite{perfect_binary-tree} and its advantages are connected with simplified reconstruction of the original function (sec. \ref{sec-reconstruction}).\\
Firstly however, we describe the QCoSamp method for the creation of QCoSamp operators from scratch using \textit{bottom-up} convention:
\begin{enumerate}
    \item At the very bottom there appears one-qubit operator $\mathcal{P}_{n, x, p}$, which we call the QCoSamp \textbf{base of computation} (BC) operator.
  \item Two such operators acting on two-qubit states will produce a \textbf{constant parameters component} (CPC) operator: $\hat{\mathcal{C}}_{n,x,r_n, s_n}$.
  \item The application of \textit{distributed phase encoding} allows for the encoding of the variable $x$ and FCoSamp phase shifts. At this point we create an \textbf{argument and phase steerable} (APS) operator $\mathcal{L}$.
  \item The frequency $n$ is introduced by repeating $n$ times, the encoded argument. The number of repetitions $n$ is encoded using the distribution phase encoding: the \textit{frequency steerable } (RS) operator $\mathcal{F}$. Then it is easy to create a \textbf{fully steerable} (FS) operator $\mathcal{D}$ as the connection APS and RS.
  \item Finally, we define the three building blocks of the QCoSamp operator: 
  \begin{enumerate}
  	\item The \textbf{$n$-th component} CMPN (e.g. CMP1, CMP2) operator $\mathcal{M}_n$, which are the leaves of the tree representing QCoSamp, the inclusion of steering technique of APS and FS to CPC for part or all elements of the component.
  	\item The \textbf{connection} operator which are the nodes of such a tree.
  	\item The \textbf{interference} operator is connected with the root of the QCoSamp tree.
  \end{enumerate}
\end{enumerate}
Using these three building blocks, we present the architecture and the method of building the QCoSamp  $\mathcal{M}$.\\

 \subsection{Base of computation}\label{mmsubsec-base-phase-operator}
 
We start the process of building QCoSamp with just one \textit{ancilla} $\ket{c}$ which we initialize with $\ket{0}$. The QCoSamp base of computation operator has the form: 
\begin{equation}\label{eq-mbo-it1}
    \mathcal{P}_{n, x, p}=HR_x^{\circ n}R_{p}H
\end{equation}

where $x$ is an input value and $R^{\circ n}$ means $n$ times, made of the operator's composition with itself. 
The Hadamard gate produces a superposition of one qubit. Sandwiching of two Hadamard gates and the phase-shift gate allows for the extraction of the phase-shifting and preserves it until measurement. Evolution with this operator will result in the state:

\begin{align}\label{eq-mbo-state-it1}
    \mathcal{P}_{n, x, p}\ket{0} =& HR_x^{\circ n}R_{r_n}H\ket{0}=
        1/2\big[
            \big(
                1+e^{i(nx+p)}
            \big) \ket{0}+
            \big(
                1-e^{i(nx+p)}
            \big) \ket{1}
        \big]
\end{align}

The measurement of the basis $\{\ket{0}, \ket{1}\}$ results in the probability for eigenstate $\ket{0}$ as follows:

\begin{align}\label{eq-mbo-meas-it1}
    \mathfrak{p}\ket{c=0}=\norm{
        \bra{0}\mathcal{P}_{n, x, p}^\dagger\ket{0}
    }^2=
        \frac{
            \big(
                1+e^{i(nx+p)}
            \big)
        }{2}\cdot
        \overline{
            \frac{
                \big(
                    1+e^{i(nx+p)}
                \big)
            }{2}
        }
    =
    \frac{1+cos(nx+p)}{2}
\end{align}

\subsection{Constant parameters component operator}\label{sec-fco}

Specification of the base of computation (35-37) allows for the creation of QCoSamp in the form of the two \textit{ancillae} operators which represents one component of FCoSamp function. For the initial moment, it has all constant parameters (frequency and two phases for it). In the next three subsections we will present the parameter steering techniques, which allow for parameters of this component to be steerable according to needs. The goal is to produce the state that, in the sampling process will generate the function $\nu_n(x)$ for given $r_n, s_n$.\\
For creating said operator, we include the entanglement phenomena. From a mathematical point of view, the entanglement appears when the n-qubits quantum state could not be constructed as the tensor product of these qubits. For that purpose we use the composition of two gates: $cR_\varphi$ – controlled phase shift and $\mathbf{1}\otimes H$ which is one of 2-qubits Hadamard gates. Together with the un-computation technique (see \ref{sec-methods-uncomputation}) they create their state of entanglement.\\
The evolution with BC operator leads, for the first qubit with $p=r_n$ and for second one $p=s_n$, to obtain the state:

\begin{align}\label{eq-state-with-phase}
    \ket{c_1} &=\mathcal{P}_{n, x, r_n}\ket{0}=
    1/\sqrt{2}\big(
        \ket{0}+
        e^{i(nx+r_n)} \ket{1}
        \big)\nonumber\\
     \ket{c_2} &=\mathcal{P}_{n, x, s_n}\ket{0}=
    1/\sqrt{2}\big(
        \ket{0}+
        e^{i(nx+s_n)} \ket{1}
        \big)
\end{align}
The tensor product of such a two qubits states is equal to:
\begin{align*}
\ket{c_1 c_2}=\mathcal{P}_{n, x, r_n} \otimes \mathcal{P}_{n, x, s_n} \ket{00}=
    \frac{1}{2}
    \big(\ket{00}+
        e^{i(nx+s_n)}\ket{01}+
        e^{i(nx+r_n)}\ket{10}+ 
        e^{i(nx+s_n)}\cdot e^{i(nx+r_n)}\ket{11}
    \big)
\end{align*}

If we use the sandwiching method and act with the $\mathbf{1}\otimes H$, the first two coordinates are as expected: $1\pm e^{i(nx+s_n)}$,
however, the last two of them will be disturbed because of the component $e^{i(nx+s_n)}\cdot e^{i(nx+r_n)}=e^{i(2nx+r_n+s_n)}$, 
which is not suitable for any of the FCoSamp elements. 
Therefore we have to un-compute the unwanted quantum phase: $\varphi=(2nx+r_n+s_n)$ with the Hermitian adjoint of the controlled phase operator for $\varphi$. 
At this point, the CPC operator is a  composition of the tensor product for the BC operators and un-computation operator. Where acting on the initial state, $\ket{00}$ runs as follows:

\begin{align}\label{eq-cpc-before-hadamard}
    \hat{\mathcal{C}}_{n,x,r_n, s_n}=
    \big[
        cR_{x}^{\circ n} cR_{r_n} cR_{s_n}
    \big]^\dagger
    \big[
    \mathcal{P}_{n, x, r_n} \otimes 
    \mathcal{P}_{n, x, s_n}
    \big]=
        cR_{-x}^{\circ n} cR_{-r_n} cR_{-s_n} 
    \big[
    \mathcal{P}_{n, x, r_n} \otimes 
    \mathcal{P}_{n, x, s_n}
    \big]\nonumber\\
     \hat{\mathcal{C}}_{n,x,r_n, s_n}\ket{00}=cR_{-x}^{\circ n} cR_{-r_n} cR_{-s_n}\ket{c_1 c_2}=\frac{1}{2}
    \big(\ket{00}+
        e^{i(nx+s_n)}\ket{01}+
        e^{i(nx+r_n)}\ket{10}+ 
        \ket{11}
    \big)
\end{align}

The quantum sampling of the above state produces a uniform probability distribution. 
Therefore, we must perform the Hadamard transform to make a sandwich (see \ref{sec-sandwiching}) to extract the phase shifts of the sampling process. 
If we make a tensor product of two Hadamard gates, the sum of all four coordinates appear, like in the comparison of states technique (see \ref{sec-eq-states}). 
Yet we want to add the coordinates in pairs only: first with the second and third with the fourth. Therefore, we use the tensor product of identity and Hadamard gate for closing the sandwich, as follows:

\begin{equation}
    \mathcal{C}_{n,x,r_n, s_n}= 
    \big[\mathbf{1}\otimes H\big]
    \hat{\mathcal{C}}_{n,x,r_n, s_n}=
    \big[\mathbf{1}\otimes H\big]
    cR_{-x}^{\circ n} cR_{-r_n} cR_{-s_n} 
    \big[
    \mathcal{P}_{n, x, r_n} \otimes 
    \mathcal{P}_{n, x, s_n}
    \big]
\end{equation}

The states after evolution with this operator are equal to:
\begin{align}\label{eq-cpc-after-hadamard}
    &\mathcal{C}_{n,x,r_n, s_n}\ket{00}=\nonumber \\
    &\frac{1}{2\sqrt{2}}
    \big[
        \big(1+e^{i(nx+s_n)}\big)\ket{00}+
        \big(1-e^{i(nx+s_n)}\big)\ket{01}+
        \big(1+e^{i(nx+r_n)}\big)\ket{10}+
        \big(1-e^{i(nx+r_n)}\big)\ket{11}
    \big]=\nonumber \\
    &\frac{1}{2\sqrt{2}}
    \bigg[
        \big(1+e^{i(nx+s_n)}\big)\ket{0}+
        \big(1+e^{i(nx+r_n)}\big)\ket{1}
    \bigg]\ket{0}+\nonumber\\
    &\frac{1}{2\sqrt{2}}
    \big[
            \big(1-e^{i(nx+s_n)}\big)\ket{0}+
            \big(1-e^{i(nx+r_n)}\big)\ket{1}
        \big]\ket{1}
\end{align}

For the third part of this equation, we see that if we measure the \textit{ancilla} and we obtain the state $\ket{0}$, we cannot be sure if the remaining qubit is $\ket{0}$ or $\ket{1}$ but we know the probability of it being one or the other.
In the opposite case (obtaining $\ket{1}$), the situation is similar, but the probability distribution is different. Therefore, the above state is in entanglement (because there is no tensor product based decomposition for it), but it is \textit{not} maximal entanglement, so this is not one of Bell states \cite{bell_einstein_1946}. This is a benefit from a quantum computing's perspective: the right qubit by entanglement stores the rest of the system's information. If the rest of the system contains more than one qubit, we say that this is \textit{cluster entanglement}. 
The measure of such a qubit is, in a sense, the measure of the whole system, which allows limits to the dimension of the measurement basis, which is very significant for quantum sampling because of the smaller dimension of measurement basis the smaller number of samples needed for trustworthy results. Hence the cluster entanglement allows limiting the output measurement basis within the extraction of the desired part of the information stored in the rest of the quantum system. \\
In this case, the measurement is a standard two-dimensional basis, as: 
$\{\ket{0}, \ket{1}\}$ 
and we measure the right qubit $\ket{c_2}$ The probability of obtaining $\ket{0}$ on the right qubit is expressed by the formula:

\begin{align}\label{eq-mu-m-not-controlable}
    \mathfrak{p}\ket{c_2=0}=\norm{\bra{00}\mathcal{C}^\dagger_{n,x,r_n, s_n}\ket{0}}^2= 
    \frac{1}{8}\norm{
        \big(1+e^{i(nx+s_n)}\big)\ket{0}+
        \big(1+e^{i(nx+r_n)}\big)\ket{1}
    }^2
    =\nu_n(x)
\end{align}

We see that constant component \textbf{operator} with argument $n$ produces the $n$-th component of the FCoSamp \textbf{function}: $\nu_n$ (see equation \ref{eq-base-model}).\\
\textbf{Setting up the values} of the FCoSamp function in CPC operator can be done in three ways:

\begin{enumerate}
    \item Directly, by setting up the proper phases of phase shifts gates which the CPC operator relies on.
    \item Directly, using the method of constant data encoding (sec. \ref{sec-huge-number-encoding}). For example, if we want to encode the argument of the FCoSamp, with $2^X$ values we generate the initial state $\ket{Xc_1 c_2}=\ket{0^{\otimes X}00}$ and the CPC operator obtain the form:
    
    \begin{equation}\label{eq-quantum-summation-operator}
    \mathcal{C}_{n,*\ket{x},r_n, s_n}= 
    \big[\mathbf{1}\otimes H\big]
    cR_{-*\ket{x}}^{\circ n} cR_{-r_n} cR_{-s_n} 
    \big[
    \mathcal{P}_{n, *\ket{x}, r_n} \otimes 
    \mathcal{P}_{n, *\ket{x}, s_n}
    \big],
    \end{equation}
    
All quantum phase shift operators that previously have acted with the once defined $x$ phase now act with a different phase, defined for each $\ket{x}$ separately. 
Those phases represent the set of input $2^X$ arguments of the FCoSamp function. Therefore system state is described as follows:

    \begin{align}\label{eq-quantum-summation-acting}
    &\mathcal{C}_{n,x,r_n, s_n}\ket{0^{\otimes X}00}=\nonumber \\
    &\frac{1}{\sqrt{2}^{X+3}}
    \sum_{x=0}^{2^X-1}
    \ket{x}\bigg[
        \big(1+e^{i(n*\ket{x}+s_n)}\big)\ket{0}+
        \big(1+e^{i(n*\ket{x}+r_n)}\big)\ket{1}
    \bigg]\ket{0}+\nonumber\\
    &\frac{1}{\sqrt{2}^{X+3}}
    \sum_{x=0}^{2^X-1}
    \ket{x}\big[
            \big(1-e^{i(n*\ket{x}+s_n)}\big)\ket{0}+
            \big(1-e^{i(n*\ket{x}+r_n)}\big)\ket{1}
        \big]\ket{1}.
    \end{align}
    
    At this point, there arises a question: what to (quantum) sample? Well, the answer is – it depends on the application. For example, if we make a quantum sample in the same way as was described in, e.g., \ref{eq-mu-m-not-controlable} we obtain:
    
    \begin{align}\label{eq-quantum-summation-sampling}
        &\mathfrak{p}\ket{c_2=0}=\norm{\bra{0^{\otimes X}00}\mathcal{C}_{n,x,r_n, s_n}^\dagger\ket{0}}^2=\nonumber\\
        &\frac{1}{2^{X+3}}\sum_{x=0}^{2^X-1}
        \bigg[
        \frac{1+cos(n*\ket{x}+r_n}{L}+
        \frac{1+cos(n*\ket{x}+s_n}{L}
        \bigg]=\sum_{x=0}^{2^X-1}\nu_n(*\ket{x}),
    \end{align}
    
For $L=2^{X+3}$, which means that the resulting quantum sampling for the eigenstate $\ket{0}$ after reconstruction, described in sec. \ref{sec-reconstruction}, multiplication by $L$, is equal to the sum of values of the $n$-th component for all $x$-s introduced to the system and computed in the constant time. 
There was a single pass of the algorithm only, in fact, this is one of the versions of the quantum summation algorithms \cite{heinrich_quantum_2002, heinrich_quantum_2003, heinrich_problem_2003}. Therefore, we see in this example that constant encoding is a potent tool for quantum computing. 
\item Indirectly using the additional initial states for introducing to system the values of elements of FCoSamp function as the single value encoded on those states. This technique will be described in next two subsections: \ref{sec-seo} and \ref{sec-full-steerable-operator}.

\end{enumerate}
 
\subsection{Steering of the argument and phase of the FCoSamp}\label{sec-seo}

So far, we can fix the parameters and arguments of the FCoSamp directly with the phases of the phase shift gate or using the constant value data encoding technique. At this point, we introduce the concept of steerable elements of FCoSamp function, which we understand as those elements at which a value is injected into the quantum system by its initial state. It is a necessary way to create search or optimization algorithms using, e.g., amplitude amplification algorithms.\\
We use a distributed phase encoding \cite{ruiz-perez_quantum_2017, draper_addition_2000, beauregard_quantum_2003} to create the extension of a fixed component operator in which the variable $x$ and the phase shifts $r_n, s_n$ are not fixed but encoded by separate inputs in the initial quantum state. We present in this subsection, the most general situation when all of these value is encoded. Nevertheless, one has to consider which of them should be set up directly (by setting up phases or using the constant values encoding), and which – encoded using this method – is dependent on the application and is described in \ref{sec-selection-of-parameters}.\\
Similar to the constants data encoding, we need, for the purpose of introducing the variable $x$ into the system, additional qubits, which form all together the state $\ket{X}$ consisting of $X$ qubits, thus $2^X$ coordinates, and this number ($2^X$)we call the \emph{resolution} of encoding. However, this time we treat this state in the same way like register in classical programming – we encode a number with it, but in specific way:
\begin{align}\label{eq-value of the x}
    \ket{X}=&\ket{x_0\dots x_{X-1}}\nonumber\\
    x=&-\pi+\sum_{j=0}^{X-1}\frac{\pi x_j}{2^j},
\end{align}

which allows encoding $2^X$ numbers from the range $x\in[-\pi, \pi]$ on $X$ qubits. 
Our goal is to create the same output probability distribution as in the case of CPC (sec. \ref{sec-fco}), but this time dependent on the value encoded in the initial state $\ket{X}$. Thus our initial state is equal to $\ket{Xc}=\ket{x_1...x_X0}$. Now, we can define the argument steering operator for the number $x$ encoded in the state $\ket{X}$:

\begin{equation}\label{eq-argument-steering}
    \mathcal{L}=
    \bigg[
        \prod_{j=0}^{X-1} cR_{\pi/2^{j}}(x_j, c) 
    \bigg]R_{-\pi}(c)H(c)
\end{equation}

The product: $\prod_{j=0}^X cR_{\pi/2^{j}}(x_j, c)$ uses the idea of distributed phase encoding (based on the QFT). However, the distribution of phases is kicked-back, one by one, to the \textit{ancilla}, which is not the canonical version of phase encoding and QFT.\\ 
In the equation \ref{eq-argument-steering} the Hadamard gate acts on the \textit{ancilla} only, because we are encoding one, specific number written in the input register $\ket{X}$. Nevertheless, in a practical application using, e.g., amplitude amplification algorithm, the state $\ket{X}$ does not encode a specific number but set of the numbers, being in the uniform superposition – like in canonical QFT.\\
Let us compute the state after acting with this operator:

\begin{align}\label{eq-aps-state-form}
&\mathcal{L}\ket{X0}=
    \frac{1}{\sqrt{2}}\bigg[
        \prod_{j=0}^{X-1} cR_{\pi/2^{j}}(x_j, c)
    \bigg]
    \big(
        \ket{X0}+e^{-i\pi}\ket{X1}
    \big)=\nonumber\\
    &\frac{1}{\sqrt{2}}
    \bigg[
        \prod_{j=1}^{X-1} cR_{\pi/2^{j}}(x_j, c)
    \bigg]
        \left\{
        \begin{tabular}{l}
             $\big[\ket{0x_1\dots x_{X-1}0}+e^{-i\pi}\ket{0x_1\dots x_{X-1}0}\big]$\\
             $\big[\ket{1x_1\dots x_{X-1}0}+e^{-i\pi}e^{i\pi}\ket{1x_1\dots x_{X-1}0}\big]$
        \end{tabular}
        \right.
    =\nonumber\\
    &\frac{1}{\sqrt{2}}
    \bigg[
        \prod_{j=1}^{X-1} cR_{\pi/2^{j}}(x_j, c)
    \bigg]
    \big[\ket{x_0}\ket{x_1\dots x_{X-1}0}+e^{-i\pi}e^{i\delta(x_0,1)\pi} \ket{x_0}\ket{x_1\dots x_{X-1}0}
    \big].\nonumber\\
    &but: x_j\in \{0, 1\}, so: \delta(x_j,1)=x_j, so:\nonumber\\ 
    &\mathcal{L}\ket{X0}=\frac{1}{\sqrt{2}}
    \bigg[
        \prod_{j=1}^{X-1} cR_{\pi/2^{j}}(x_j, c)
    \bigg]
    \big[\ket{x_0}\ket{x_1\dots x_{X-1}0}+e^{-i\pi}e^{i\pi x_0} \ket{x_0}\ket{x_1\dots x_{X-1}0}
    \big]=\dots\nonumber\\
    &\frac{1}{\sqrt{2}}
    \big[\ket{x_0}\ket{x_1}\dots \ket{x_{X-1}}\ket{0}+e^{-i\pi}e^{i\pi x_0} \ket{x_0}e^{i\pi/2}\ket{x_1}\dots e^{ix_{X-1}\pi/{2^{X-1}}}\ket{x_{X-1}}\ket{0}
    \big]=\nonumber\\
     &\frac{1}{\sqrt{2}}
     \big[
     \ket{x_0x_1\dots x_{X-1}0}+
     e^{i\big(-\pi+ \sum_{j=0}^{X-1}\frac{\pi x_j}{2^j}\big) }\ket{x_0x_1\dots x_{X-1}0}
     \big]\stackrel{eq.\ref{eq-value of the x}}{=}
     \frac{1}{\sqrt{2}}
     \big[\ket{X0}+e^{ix}\ket{X1}\big], 
\end{align}

We can use this same idea to encode FCoSamp function phase shift $p$; in that case, we need to extend the initial quantum state by the state $\ket{P}$ encoding the number $p\in[-\pi, \pi]$. 
As a result, we receive the initial state $\ket{PX0}$; nevertheless, we cannot say that argument and phase steerable operator acting on the initial state $\ket{PX0}$ is simply a composition of $\mathcal{L}(R,c)\ket{PX0}$ and $\mathcal{L}(X,c)\ket{PX0}$, because in that case, the Hadamard gate would act upon the \textit{ancilla} twice. Considering that $H\circ H=\mathbf{1}=H\circ H^\dagger$ in that case, we would always obtain the state $\ket{0}$, but we must act with the Hadamard gate only once; therefore, we define the final form of the APS as follows:

\begin{align}\label{eq-aps-final}
    \mathcal{L}\ket{PX0}=&
    \mathcal{L}_{H^-}(P,c)\ket{PX0} \circ \mathcal{L}(X,c) \ket{PX0}=
    \frac{1}{\sqrt{2}}
    \mathcal{L}_{H^-}(P,c)
    \big[
        \ket{PX0}+e^{ix}\ket{PX1}
    \big]=\nonumber\\
    &\frac{1}{\sqrt{2}}
    \big(
        \ket{PX0}+e^{i(x+r)}\ket{PX1}
    \big]
\end{align}

\subsection{Frequency steerable operator and full steerable operator}\label{sec-full-steerable-operator}

APS produces the state with coordinate $e^{i(x+p)}$. To create the final $\nu_N$ function, we introduce the frequency of FCoSamp, which is the multiplication of $x$ by itself by an integer number $n$, which gives $n$ times $x$. In the context of an exponential function it changes to multiplication of phase shifts: $e^{inx}=\prod_{k=1}^n e^{ix}$. Therefore, the exponentiation of the phase shift operator used: $R_{\varphi}^{\circ n}$ in the BC and CPC operators. Now, our goal is to include the possibility of the setting frequency $n$ as the next initial state $\ket{N}=\ket{n_0...n_{N-1}}$ encoding the number of the Fourier $n$-th component, represented by $\nu_n$. For this purpose, we again use the idea of the distribution phase encoding, but this time we will slightly modify it. Firstly, since $n$ is an integer number, we will encode it according to the formula:

\begin{equation}\label{eq-n-enoding}
    n=\sum_{k=0}^{N-1} n_k2^{k}
\end{equation}. 

Secondly, the change of phase of the \textit{ancilla} has to be dependent on both input states $\ket{X}$ and $\ket{N}$, because we have to repeat the influence of the state $\ket{X}$ on the system $n$-times, therefore we use the \textit{twice controlled phase shift gate} $ccR_\varphi$.
Hence, the initial state for this operation has two additional states and is equal to $\ket{NXc}$, where register $\ket{N}$ encodes the frequency and $\ket{X}$ – the argument, as above; the frequency steerable operator (RS) and is of the form:

\begin{align}\label{eq-data-freq-loading}
    \mathcal{F}=&
        \bigg[
            \prod_{k=0}^{N-1}\prod_{j=0}^{X-1}ccR_{\pi/2^{j}}^{\circ 2^k}(n_k, x_j, c)
        \bigg]
        \bigg[
            \prod_{k=0}^{N-1}cR_{-\pi}^{\circ 2^k}(n_k,c)
        \bigg]
        H(c)=\nonumber\\
        &\prod_{k=1}^{N-1}
        \bigg[
            \bigg(
                \prod_{j=0}^{X-1}ccR_{\pi/2^{j}}^{\circ 2^k}(n_k, x_j, c)
            \bigg)
            cR_{-\pi}^{\circ 2^k}(n_k,c)
        \bigg]H(c)
\end{align}

The equation above is the development of equation \ref{eq-argument-steering}; indeed, if $N=1$ and $n_0$ is equal to $1$, then both operators acting on the system in precisely the same way. In other words, the equation \ref{eq-data-freq-loading} differs from the equation \ref{eq-argument-steering} with three elements: (i) its action repeated $N$ times, (ii) the phase shift operator in each repetition is in the $k$-th power and (iii) to each phase shift operator one extra controlling qubit added.\\
Let us notice that:

\begin{align}\label{eq-R-to-k}
    R^{\circ k}_\varphi\Bigg[\frac{\ket{0}+\ket{1}}{\sqrt{2}}\Bigg]=
    \frac{1}{\sqrt{2}}
    \bigg[
        \ket{0}+
        \prod_{j=0}^{k-1}e^{i\varphi}\ket{1}
    \bigg]=
    \frac{1}{\sqrt{2}}
    \big[
        \ket{0}+
        e^{ik\varphi}\ket{1}
    \big]=
    R_{k\varphi}\Bigg[
        \frac{\ket{0}+\ket{1}}{\sqrt{2}}
    \Bigg],
\end{align}

which is an important formula due to a possible change in the $k$ phase shift gates into $1$ with phase multiplied by $k$. 
Hence we can change the formula of RS operator using much fewer phase shift gates, in the form:

\begin{align}\label{eq-data-freq-loading-saving}
    \mathcal{F}=&
        \bigg[
            \prod_{k=0}^{N-1}\prod_{j=0}^{X-1}ccR_{2^{k-j}\pi}(n_k, x_j, c)
        \bigg]
        \bigg[
            \prod_{k=0}^{N-1}cR_{-2^k\pi}(n_k,c)
        \bigg]
        H(c)=\nonumber\\
        &\prod_{k=1}^{N-1}
        \bigg[
            \bigg(
                \prod_{j=0}^{X-1}ccR_{2^{k-j}\pi}(n_k, x_j, c)
            \bigg)
            cR_{-2^k\pi}(n_k,c)
        \bigg]H(c)
\end{align}

Considering the above, it is easier to compute how this operator will act on the initial state $\ket{NX0}$, using the equation \ref{eq-data-freq-loading} for RS operator:

\begin{align}\label{eq-rs-operator}
    \mathcal{F}\ket{NX0}=&\frac{1}{\sqrt{2}}
        \prod_{k=0}^{N-1}
        \bigg[
            \bigg(
                \prod_{j=0}^{X-1}ccR_{\pi/2^{j}}^{\circ 2^k}(n_k, x_j, c)
            \bigg)
            cR_{-\pi}^{\circ 2^k}(n_k,c)
        \bigg]\big(\ket{NX0}+\ket{NX1}\big)\stackrel{eq. \ref{eq-R-to-k}}{=}\nonumber\\
        &\frac{1}{\sqrt{2}}
        \prod_{k=0}^{N-1}
        \bigg[
            \bigg(
                \prod_{j=0}^{X-1}ccR_{2^{k}\pi/2^j}(n_k, x_j, c)
            \bigg)
            cR_{-2^k\pi}(n_k,c)
            \big[\ket{NX0}+\ket{NX1}\big]
        \bigg]\stackrel{eq. \ref{eq-aps-state-form}, dpe} {=}\nonumber\\
         &\frac{1}{\sqrt{2}}
        \big[
            \ket{NX0}+
            \prod_{k=0}^{N-1}e^{in_k 2^kx}\ket{NX1}
        \big]=
        \frac{1}{\sqrt{2}}\big[
            \ket{NX0}+
            e^{i\big(\sum_{k=0}^{N-1}n_k 2^k\big)x}\ket{NX1}
        \big]\stackrel{eq. \ref{eq-n-enoding}}{=}\nonumber\\
       &\frac{1}{\sqrt{2}}
       \big[
            \ket{NX0}+
            e^{inx}\ket{NX1}
        \big],
\end{align}

where $dpe$ means applying the distributed phase encoding technique. Note, that in essence, the presented RS operator combines encoding the frequency and argument. Where the argument is introduced to the system directly, the formula for this system simplifies to:

\begin{enumerate}
    \item In case of directly introducing one value $x$ by the quantum phase shift formula of APS is as follows:
    \begin{align}
        \mathcal{F}=
        &\prod_{k=1}^{N-1}
        \bigg[
                cR_{2^{k}x}(n_k, c)
        \bigg]H(c),
    \end{align} and the initial state is equal to $\ket{N0}$.
    \item In case of directly introducing many values of $x$ by constant data encoding, formula of APS is as follows:
    \begin{align}
        \mathcal{F}=
        &\prod_{k=1}^{N-1}
        \bigg[
                cR_{2^{k}*\ket{x}}(n_k, c)
        \bigg]H(c),
    \end{align} and the initial state is equal to $\ket{NX0}$.
\end{enumerate}

Now, we can define the fully steerable operator (FS), which allows for the encoding of the initial state $\ket{NPX0}$ of all elements in the FCoSamp function: frequency, argument, and phase shift:

\begin{align}\label{eq-final-data-loader}
    \mathcal{D}(N,P,X,c)=&\mathcal{F}_{H^-}(N,X,c)\circ\mathcal{L}(P,c)\nonumber\\
    \mathcal{D}(N,P,X,c)\ket{NPX0}=&
    \frac{1}{\sqrt{2}}\mathcal{F}_{H^-}(N,X,c)
    \big[
        \ket{NPX0}+e^{ip}\ket{NPX1}
    \big]
    =\nonumber\\
    &\frac{1}{\sqrt{2}}
    \big(
        \ket{NPX0}+e^{i(nx+p)}\ket{NPX1}
    \big)
\end{align}

\subsection{Building blocks of the QCoSamp operator }\label{sec-building-blocks}

From this point, we use the ideas described above to create the QCoSamp operator. We use three building blocks that are connected, forming the full binary tree \cite{binary-tree, full_binary-tree} but its final parameters like height, number of the leaves (components), and the way connection depend on their absolute need. Further, we will describe this architecture in more detail. \\
The building blocks, described as:
\begin{enumerate}
    \item The $n$-th component CMPn (CMP1, CMP2,...) operator, which is marked with a black rectangle with a component (frequency) number in the diagrams, and is a leaf in the QCoSamp tree.
    \item The connection CON[X, Y], operator, where X, Y are CMPn's or other connection operators, which is marked as a circle with a plus inside and are nodes of the tree.
    \item Interference operator which is described as the white rectangle with a parameter inside, defining the size of this operator and in fact the size of the whole QCoSamp operator.
\end{enumerate}

 \subsection{The n-th component operator}
 
The target form of the CMPn operator we strive for is coupled with the form of the $n$-th component of the FCoSamp function $\nu_N$ (eq. \ref{eq-base-model}). The equation \ref{eq-cpc-before-hadamard} shows the state after acting with Hadamard gate, the simplest version of interference operator, which transforms to the state described in the equation  \ref{eq-cpc-after-hadamard}. Finally, after quantum sampling, leads to the $\nu_n$ (eq. \ref{eq-mu-m-not-controlable}). 
In short, we demand that the CMPn operator has the form:

\begin{equation}\label{eq-M-n-final}
    \bigg[
    \frac{1}{2}\ket{NRSX}
    \big(
        \ket{00}+
        e^{i(nx+s)}\ket{01}+
        e^{i(nx+r)}\ket{10}+
        \ket{11}
    \big)
    \bigg],
\end{equation}

If we look at the equations \ref{eq-state-with-phase} and \ref{eq-final-data-loader}, notice that they are very similar one another; the difference lays in the leading state $\ket{NPX}$ before state $\ket{c}=\ket{0}$. Therefore, to obtain the $n$-th component operator, we proceed the same way as before.\\
For creating the component operator for a frequency $n$ we must have: different FCoSamp phase shifts $r_n, s_n$, common frequency $n$ and common argument(s) $x$. Similar to CPC there are two \textit{ancillae}, hence the initial state has a form: $\ket{NRSXc_1c_2}$. Now, let us consider the operator:

\begin{equation}
    \hat{\mathcal{M}}_n=\mathcal{D}(N,R,X,c_1)\circ\mathcal{D}(N,S,X,c_2)
\end{equation}

In the equation \ref{eq-aps-final} for APS, we have to use one operator without a leading Hadamard operation to avoiding canceling it (because of Hermitian self-adjoint). In opposition to this, the cases where FS operators $\mathcal{D}$ act on separate \textit{ancillae}, Hadamard gates acts parallel not one after another; thus they should stay for both operators, changing separated \textit{ancillae} and abandoning the rest of the state $\ket{NRSX}$, as follows:

\begin{align}
    \big[H(c_1)\circ H(c_2)\big]\ket{NRSX00}
    =&\ket{NRSX}\otimes \big(H\ket{0}\big)^{\otimes 2}=\ket{NRSX}\otimes \bigg(\frac{\ket{0}+\ket{1}}{\sqrt{2}}\bigg)^{\otimes 2}=\nonumber\\
    &\frac{1}{2}\ket{NRSX}\big(\ket{00}+\ket{01}+\ket{10}+\ket{11}\big)
\end{align}

The remaining part of the operator $\mathcal{D}(N,Y,Q,c_i)$ works in such a way that if $c_i=1$ if multiplies it's coordinate by $e^{i(nx+y)}$. Therefore the acting of $\Hat{\mathcal{M}}_n$ on initial state proceeds like:

\begin{align}\label{eq-M-hat-actin-on-state}
    \Hat{\mathcal{M}}_n\ket{NRSQ00}=&\frac{1}{2}\ket{NRSQ}
    \big(
        \ket{00}+
        e^{i(nx+s)}\ket{01}+
        e^{i(nx+r)}\ket{10}+
        e^{i(nx+r)}e^{i(nx+s)}\ket{11}
    \big)=\nonumber\\
    &\frac{1}{2}\ket{NRSQ}
    \big(
        \ket{00}+
        e^{i(nx+s)}\ket{01}+
        e^{i(nx+r)}\ket{10}+
        e^{i(2nx+r+s)}\ket{11}
    \big)
\end{align}

Now we have to un-compute the phase-shift, keeping the last coordinate as in the case of CPC (sec. \ref{sec-fco}, eq. \ref{eq-cpc-before-hadamard}), but this time we do not know the values: $n,  r, s, x$, because they are encoded in the state $\ket{NRSX}$.
Therefore, we use the modified idea of inverse QFT algorithm, where modification is based on the fact that we make the given $n$-th component acting on the quantum system dependent on the quantum registers' values that encodes $n, r, s, x$. 
First, we define the un-computation procedure as:

\begin{equation}
    \mathcal{L}^{\dagger}(X, c_1, c_2)=
    \bigg[
        \prod_{j=0}^{X-1} c^3R_{-\pi/2^{j}}(x_j, c_1, c_2) 
    \bigg]cR_{\pi}(c_1, c_2)
\end{equation}

The notion $c^mR_\varphi$ means $m$-time-controlled phase shift gate, changing the coordinate if all of the $m$ qubits it touches are equal to $1$ (\ref{sec-gates}). 
It is noteworthy that $\mathcal{L}^{\dagger}(Y, c_1, c_2)$ is \textbf{not} the Hermitian conjugate of arguments steerable operator (eq. \ref{eq-argument-steering}), 
because that operator acts always on one \textit{ancilla} while this one acts on two \textit{ancillae}. 
So, this is the other operator, serving as the un-computation of interference arising \textbf{because} of the tensor product of two states, which the steerable argument operators act upon in parallel.\\
Similarly to the equations \ref{eq-aps-state-form} and \ref{eq-rs-operator} every time this operator $\mathcal{L}^\dagger$ hits on the $\ket{c_1c_2}=\ket{11}$ eigenstate and $q_i=1$ it multiplies its coordinate by $e^{-i\pi/2^{j-1}}$, hence we obtain:

\begin{equation}
    \mathcal{L}^{\dagger}(X, c_1, c_2)\ket{X00}=\xi_x\ket{X}
        \big(
            \xi_{00}\ket{00}+
            \xi_{01}\ket{01}+
            \xi_{10}\ket{10}+
            e^{-ix}\xi_{11}\ket{11}
        \big)
\end{equation}

where $\xi_x, \xi_{jk}$ means the coordinate that has appropriate, stand by eigenvectors before the operator acts.
Similarly, unwanted frequencies due to the interference of two steerable frequency operators can be un-computed. The operator achieves this goal in two ways depending upon if we have to un-compute the same or different frequencies (based on the gate-saving form of the RS operator -eq. \ref{eq-data-freq-loading-saving}):

\begin{align}\label{eq-data-freq-loading-inv}
    \mathcal{F}^{\dagger}(N, X, c_1, c_2)=&
        \bigg[
            \prod_{j=0}^{X-1}\prod_{k=0}^{N-1}c^3R_{-2\cdot 2^{k-j}\pi}(n_k, x_j, c_1, c_2)
        \bigg]
        \bigg[
            \prod_{k=0}^{N-1}c^2R_{2\cdot 2^k\pi}(n_k,c_1,c_2)
        \bigg]\\
        \mathcal{F}^{\dagger}(N, M, X, c_1, c_2)=&
        \bigg[
            \prod_{j=0}^{X-1}\prod_{k=0}^{N-1}\prod_{l=0}^{M-1}c^4R_{-2^{kl-j}\pi}(n_k,m_l, x_j, c_1, c_2)
        \bigg]
        \bigg[
            \prod_{k=0}^{N-1}\prod_{l=0}^{M-1}c^3R_{2^{kl}\pi}(n_k, m_l,c_1,c_2)
        \bigg]
\end{align}

On the index of operators: $c^3R_{-2\cdot 2^{k-j}\pi}$ and $c^2R_{2\cdot 2^k\pi}$ the values are doubled because in equation \ref{eq-M-hat-actin-on-state} there is a frequency doubled by the state $\ket{11}$ in reference to the states $\ket{01}, \ket{10}$. Every time this operator hits on the state where the last two qubits $\ket{c_1c_2}=\ket{11}$, it multiplies the coordinate by $e^{-2ix}$.
Finally, we can define the un-computation version of the final encoding operator as:

\begin{align}
    \mathcal{D}^{\dagger}(N,R,S,X,c_1, c_2)=\mathcal{F}^{\dagger}(N,X,c_1, c_2)\circ \mathcal{L}^{\dagger}(R, c_1, c_2)\circ \mathcal{L}^{\dagger}(S, c_1, c_2)
\end{align}

The above equation is correct because we use it in phase shifts only – specifically, we do not use Hadamard gates to prevent any unwanted doubling of action of Hermitian self-adjoint operator, the Hadamard gate.\\
Remembering our hitherto consideration, we can see that operator $\mathcal{F}^{\dagger}$ changes the coordinate of each eigenstate, which has $\ket{11}$ at the end, making three multiplications of its coordinate:

\begin{itemize}
    \item by $e^{-ir}$ because of $\mathcal{L}^{\dagger}(R, c_1, c_2)$,
    \item by $e^{-is}$ because of $\mathcal{L}^{\dagger}(R, c_1, c_2)$,
    \item by $e^{-2inx}$ because of $\mathcal{F}^{\dagger}(N,X, c_1, c_2)$,
\end{itemize}

So together it gives the multiplication by $e^{-i(2nx+r+s)}$ – exactly what we desire.\\
Now, using the idea from the CPC operator (sec. \ref{sec-fco}, eq. \ref{eq-cpc-before-hadamard}), we can write down the final form of the $n$-th component operator CMPn and its action upon the initial state:

\begin{align}
    \mathcal{M}_{nH^-}=&
    \mathcal{D}^{\dagger}(N,R,S,X,c_1, c_2)\circ
    \hat{\mathcal{M}}_{nH^-}\label{eq-n-th-component-base-model-operator-kicked-back}\\
    \mathcal{M}_{nH-}\ket{NRSXc_1c_2}=&
    \mathcal{D}^{\dagger}(N,R,S,X,c_1, c_2)\nonumber\\
    &\frac{1}{2}\ket{NRSX}
    \big(
        \ket{00}+
        e^{i(nx+s)}\ket{01}+
        e^{i(nx+r)}\ket{10}+
        e^{i(2nx+r+s)}\ket{11}
    \big)=\nonumber\\
    & \frac{1}{2}\ket{NRSX}
    \big(
        \ket{00}+
        e^{i(nx+s)}\ket{01}+
        e^{i(nx+r)}\ket{10}+
        \ket{11}
    \big),\label{eq-n-th-component-base-model-operator-kicked-back1}
\end{align}

where using the $\mathcal{M}_{nH^-}$ operator means that to obtain the FCoSamp value, we must introduce the interference between the zero and first and between second and third components separately, done with the tensor product of identity and Hadamard gate, which has the form:

\begin{align}
    \left[
    \begin{tabular}{c|c}
     $H$ & $\mathbf{0}$ \\
     \hline
     $\mathbf{0}$ & $H$
    \end{tabular}
    \right],
\end{align}

and is called \emph{interference} operator.\\

Similarly to the fixed component operator and sandwiching it with Hadamard gate the $\mathcal{M}_{nH^-}$ operator should be used for further computation while the version after applying the interference operator, is proper for measurement:

\begin{align}\label{eq-n-th-component-base-model-operator}
    \mathcal{M}_n=&
    \big[\mathbf{1}\otimes H\big](c_1, c_2)\circ
    \mathcal{M}_{nH^-}\nonumber\\
    \mathcal{M}_n\ket{NRSXc_1c_2}=&
    \frac{1}{2}
    \big[
        \mathbf{1}\otimes H\big](c_1, c_2)\frac{1}{2}\ket{NRSX}
    \big(
        \ket{00}+
        e^{i(nx+s)}\ket{01}+
        e^{i(nx+r)}\ket{10}+
        \ket{11}
    \big)
   =\nonumber\\
    &\frac{1}{2\sqrt{2}}\ket{NRSX}
    \big[
        (1+e^{i(nx+s)})\ket{00}+
        (1-e^{i(nx+s)})\ket{01}+\nonumber\\
        &(1+e^{i(nx+r)})\ket{10}+
        (1-e^{i(nx+s)})\ket{11}
    \big]=\nonumber\\
    &\frac{1}{2\sqrt{2}}
    \big[
        \big(1+e^{i(nx+s)})\ket{NRSX0}+(1+e^{i(nx+r)}\big)\ket{NRSX1}
    \big]\ket{0}+\nonumber\\
    &\frac{1}{2\sqrt{2}}
    \big[
        \big(1-e^{i(nx+s)})\ket{NRSX0}+(1-e^{i(nx+r)}\big)\ket{NRSX1}
    \big]\ket{1}
\end{align}
Quantum sampling of the most right qubit $\ket{c_2}$ in two dimensional measurement basis, generates the probability of obtaining the eigenstate $\ket{0}$:

\begin{align}\label{eq-nu-x-n-th-component-final}
    \mathfrak{p}\ket{c_2=0}=&\norm{\mel{NRSX00}{{\mathcal{M}_n}^\dagger }{0}}^2=\nonumber\\
    \frac{1}{8}&\norm{
        \big(1+e^{i(nx+s_n)}\big)\ket{NRSX0}+
        \big(1+e^{i(nx+r_n)}\big)\ket{NRSX1}
    }^2
    =\nu_n(x)
\end{align}

The result seems to be the same as in equation \ref{eq-mu-m-not-controlable}, but now we can control the parameters of $\nu_n$ not with the gate parameters but with the input qubits, which allows to integrate the quantum computers with the classical ones. More importantly, its application is the utilization of the input qubits as a part of quantum algorithms. To this avail, including e.g. the amplitude amplification algorithm for solving the practical problems, described in the context of Fourier analysis, will be discussed in section \ref{sec-results} in detail.

\subsection{The connection operator}\label{sec-two-component-connection}

\begin{figure}[ht!]
    \centering
    \includegraphics[width=0.88\textwidth]{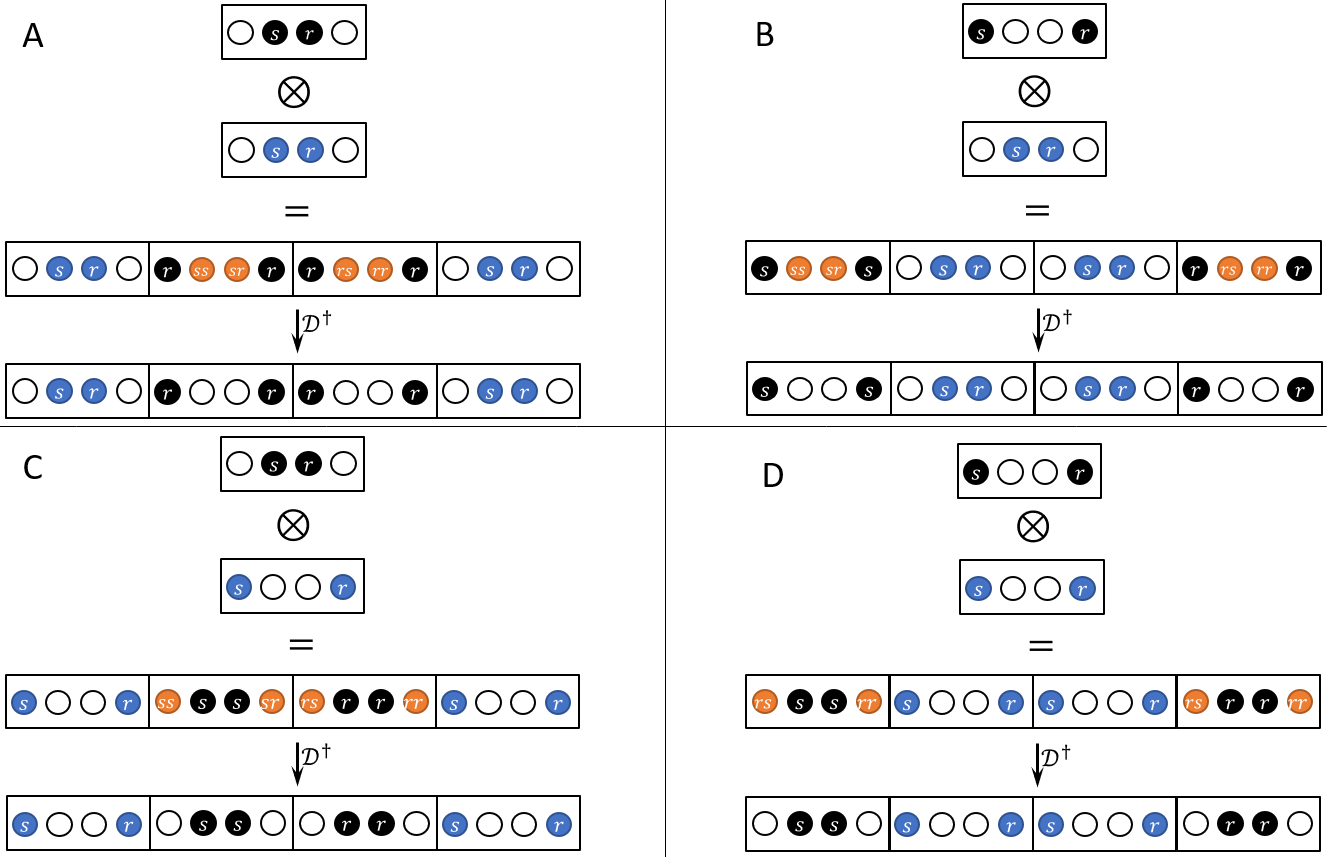}
    \caption{The form of the state vector after connecting two components with various arrangements of non-unit coordinates}
    \label{fig:tensor-product-for-adding-two-states}
\end{figure}

If we create $M$ CMPn's (with full steering components), there are $2^{M(N+R+S+2)+X}$ eigenstates during computation. Most are unwanted because they are multiples of many quantum phase shifts from different components, which is very complicated combinatorically, with complexity growing exponentially with an increasing number of components $M$. The way of dealing with this complexity is ordering the tensor product with a full binary tree structure, which is generally logarithmic and done because the tensor product is associative. Fortunately, if we put the exponential problem to the logarithmic structure, it becomes relatively linear.\\
The essence of a connection operator is the arrangement of the un-computation process for unwanted interference of coordinates, which are now the multiplication of maximum two coordinates with the form of $e^{i(nx+p)}$, e.g., one of the unwanted coordinates for CON(CMP1, CMP2) can be equal to $e^{i(x+r_1)}e^{2x+r_2}$. The maximal number of unwanted interference multiplications is $2$ because we always connect two objects of the previous level. Un-computation works in such a way that at the output, there always are coordinates that are $1$-s or single quantum phase-shifts in the form $e^{i(nx+p)}$. Indeed, the situation is even better – there is always have half of the coordinates equal to $1$ (omitting the normalization questions for now), and half of them in quantum phase shift form. For simplicity, we call the coordinate equal to $1$ unit coordinates, and the quantum phase-shifted – non-unit.\\ 
The input of the connections are always two sibling components $\mathcal{M}_{2kH^-}, \mathcal{M}_{(2k+1)H^-}$ (see equation \ref{eq-M-n-final}, \ref{eq-n-th-component-base-model-operator-kicked-back}, \ref{eq-n-th-component-base-model-operator-kicked-back1}) or previous two (also sibling) connections; that is the reason that QCoSamp can be represented by a full binary tree. Therefore, the tensor product of the connection on the first level (connection of components) will have $16$ coordinates, on the second level $16\cdot16=256$, and in general,  $2^{2^{l+1}}$, where $l$ is level number. This number is divisible by 4; indeed: $\forall l\geq1: 2^{2^{l+1}}/4=2^{2^{l+1}-2}=2^{2(2^l-1)} \wedge 2^l-1>0$. Thus, the tensor product's coordinates made with the connection operator divided into $2^{2^l-1}$ blocks made of $4$ coordinates. We present the exemplary tensor product for the level $1$ (connection of two CMPn's) on the plot (Fig. \ref{fig:tensor-product-for-adding-two-states}A). On this plot, the white circles denote the coordinates equal to $1$ before normalization. The filled circles show the coordinates with phase shifts: $e^i(nx+r), e^i(nx+s)$. The coordinates coming from the left side state (with $n$ frequency) filled black, while the right-side state (with $m\neq n$ frequency) is blue. We divide the resulting tensor product into four-coordinate blocks where the first block comes from the first coordinate of the left-hand state, the second block from the second coordinate, and so on.\\
Looking at the arrangement of the coordinates, within one block after un-computation (see Fig. \ref{fig:tensor-product-for-adding-two-states}.A) we can see there are only two kind of blocks generated: $\circ\bullet\bullet\circ$ and $\bullet\circ\circ\bullet$, where $\circ$ means coordinate equal to $1$ before normalization and $\bullet$ means coordinate equal to $e^{i(nx+r)}$ before normalization. It is a significant observation for further building the QCoSamp operator. That way, we can divide the complicated problem of connection of $M$ components to a set of simpler problems, connecting just two four-coordinate blocks. Furthermore, as we see in figures B, C, and D, the two blocks of types quoted above, always produce a four-block state. Each block of the resulting state is one of the same two types. Therefore, we know the final tensor product of any number of coordinates, constructed by combining two types of blocks (quoting once again): $ \circ\bullet\bullet\circ $ and $\bullet\circ\circ\bullet$. The coordinates will differ in different blocks, and therefore, can describe the unit's and non-unit's coordinate.\\
In conclusion to this subsection, we say that the connection component structures the tensor product of all CMPn into a full binary tree to clarify the un-computation procedure. The significant advantage of such a structure is that the un-computation procedure cancels the unwanted coordinates and makes clones of wanted coordinates from \textbf{unwanted} ones in a controlled way, which allows for the acquisition of a predictable number of copies of each $\nu_n$ function. It decreases the influence of quantum normalization because if we have $K_n$ copies of $\nu_n$, which have a factor of quantum normalization, it means that the new normalization factor is $K_n$ times greater than originally, which decreases the number of demanding repetitions in the quantum sampling procedure. This issue and its influence on the quantum normalization will be discussed in more detail in sections: \ref{sec-base-model-architecture}, \ref{sec-normalization-factor} and \ref{sec-output-correctness}.\\

\subsection{The architecture and building of the QCoSamp operator}\label{sec-base-model-architecture}

\begin{figure}[ht!]
    \centering
    \includegraphics[width=0.78\textwidth]{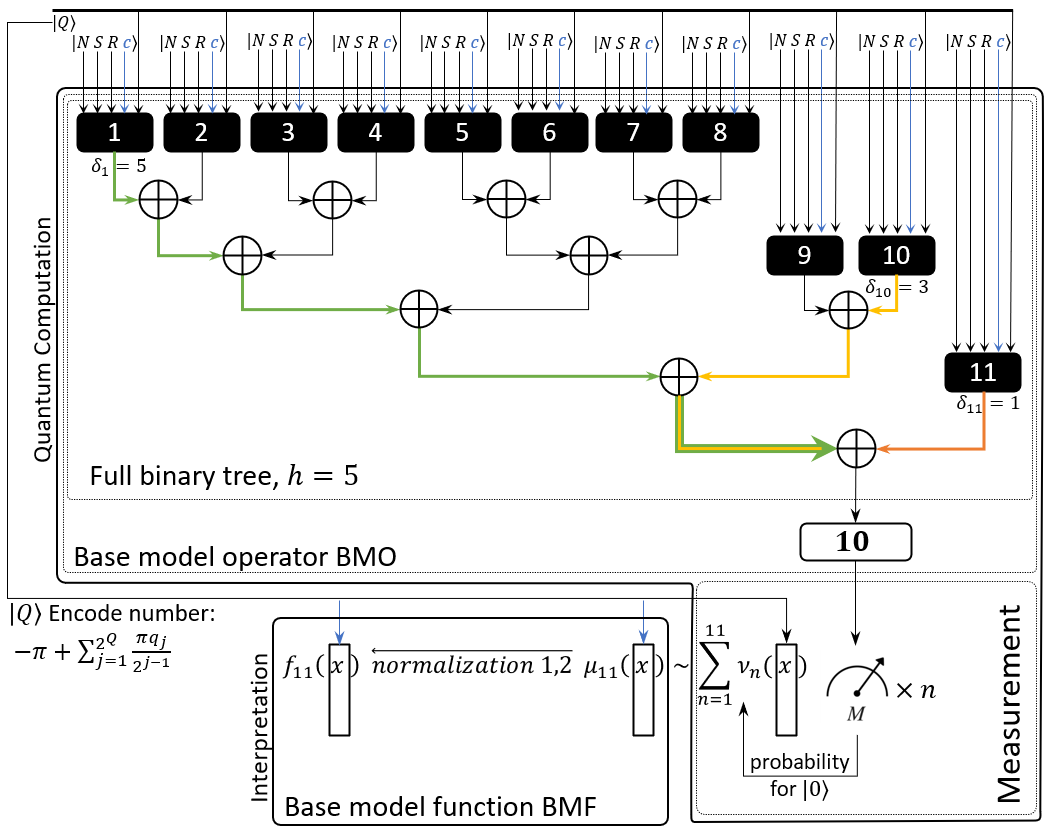}
    \caption{Exemplary architecture of unbalanced QCoSamp for 11 components. Note that a black arrow coming to the component means several qubits designed for the input parameter and the blue arrow described as $\ket{c}$ means \textbf{two} \textit{ancillae}. The full binary tree is made of the components and connection operators. Interference is connected to the root of such a tree. QCoSamp is not balanced because the tree is not perfect. While the components 1-8 has the depth $\delta_n=5$ the components 9 and 10 depth is equal to $3$ and the depth of the last one is equal to $1$. It results in a different quantum normalization factor for the components: 1-8 has $1/136$; 9, 10 has $1/32$ and 11 has $1/8$ normalization factor. After measurement it resulted with: $1/64$ for components 1-8, $1/16$ for 9, 10 and $1/4$ for 11 for measuring $\ket{0}$ and the same values for measuring $\ket{1}$. So they sum up to 1: $8\cdot\frac{2}{64}+2\cdot\frac{2}{16}+1\cdot\frac{2}{4}=1$. The number $2$ in the numerator arise because of two dimensional measurement basis $\ket{0}, \ket{1}$. We can see here the general scheme of quantum computation using QCoSamp.}
    \label{fig:base-model-operator-final}
    
\end{figure}
The architecture of the QCoSamp is forced by its building block structure, from the bottom it is built of $N$ component operators $\mathcal{M}_{nH^-}$ defined in equations \ref{eq-n-th-component-base-model-operator-kicked-back} and \ref{eq-n-th-component-base-model-operator-kicked-back1}. Then, in the first level they are connected in pairs with the two-component connection operator described in sec. \ref{sec-two-component-connection}. Then, in the second level, the connected pairs are connected in pairs once again, and so on. So, the QCoSamps has the full binary tree (see Black \cite{binary-tree}, \cite{full_binary-tree}) architecture. The components are leaves, the depth $\delta_n$ of the leaves is the depth of $n$-th component corresponding to a single leaf; the internal nodes are connection operators. This is the reason that QCoSamp is \textbf{full} binary tree – the connection operators always connect two lower level elements, so there is not possible to have a node exist with one child only. The exemplary two architectures of QCoSamp are presented on the Fig. \ref{fig:base-model-operator-final}.\\
We say that QCoSamp is \textit{balanced} if and only if its structure is a \textit{perfect full binary tree} (see Youming Zhou \cite{perfect_binary-tree}). If the operator is balanced the component count is the power of $2$. The most important property of the balanced QCoSamp is that it produces the same normalization factor for all components, so the procedure of reconstruction from the output QHF to the final Fourier series is simple – we have to use the same equation \ref{eq-fourier-reconstr-for-balanced}, for all FCoSamp phase shifts to compute the coordinates of the series. Where architectures are not balanced the normalization for different components could be different because it depends on the depth of the component in the tree. This complicates the reconstruction, since in the second normalization (see sec. \ref{sec-reconstruction}, eq. \ref{eq-reconstruction-FCoSamp-Fourier}), we must use different normalization factors for different frequencies. However, in some applications it could be an advantage because it weights the influence of the components on the results. For example, the unbalanced architectures of QCoSamps can work similarly to low or high pass filters. This normalization is discussed in detail in sec. \ref{sec-normalization-factor}.\\
The root of the quantum harmonic cosine operator is connected with interference operator that is not an element of the tree and has the very simple form: $\mathbf{1}^{\otimes (N-1)}\otimes H$, where $N$ is the number of components. In the matrix form it has the Hadamard gate on the diagonal and zeros everywhere else. We have noticed that the state before acting on the interference operator is built of blocks consisting of four coordinates of two types: $\circ\bullet\bullet\circ$ and $\bullet\circ\circ\bullet$. On each of such a block will act the $4\times 4$ sub-matrix with Hadamard gate on the diagonal. It results with: $[\circ+\bullet, \circ-\bullet, \bullet+\circ, \bullet-\circ]^T$ for the first type and $[ \bullet+\circ, \bullet-\circ, \circ+\bullet, \circ-\bullet]^T$. Hence, in every second position we obtain the pattern $\bullet+\circ$ which hides one of the formulas: $1+e^{i(nx+w)}$ – where $w$ is $r$ or $s$. On the other hand, every second position corresponds to an eigenstate that has $0$ at the end. Hence if we measure the most right \textit{ancilla}, in a two dimensional basis, we obtain the probability for sampling $\ket{0}$:

\begin{align}\label{eq-final-base-model-function}
    \mathfrak{p}\ket{c_{2N-1}=0}=&
    \norm{\mel{(NRSX00)^N}{{\mathcal{M}}^\dagger }{0}}^2=\nonumber\\
    &\sum_{n=1}^N
        \bigg(
            \frac{1+cos(nx+r_n)}{L_n}+\frac{1+cos(nx+s_n)}{L_n}
        \bigg)\approx
        \mu_N(x)\stackrel{r_{1,2}}{\longrightarrow}f_N(x)
\end{align}

where $\stackrel{r_{1,2}}{\longrightarrow}$ means the first and second reconstruction (see sec.  \ref{sec-mapping}).
If the QCoSamp is balanced (see \ref{eq-base-model}), $L_n=4N$ for all $n$ in opposite case normalization factors could be different for different $n$ – see sec. \ref{sec-normalization-factor} for details\\
 
\subsubsection{The quantum normalization factor}\label{sec-normalization-factor}

The quantum normalization factor arises due to quantum mechanics (see Hayashi et al. \cite{hayashi_introduction_2015}). First, we discuss a balanced QCoSamp. The operator acts on a system made of $N$ components. Each of the components uses $R+S+2$ qubits. Except for that, there are $X$ qubits that encode the arguments (no matter how) shared for all components. Therefore, the system contains $N(R+S+2)+X$ qubits which gives $2^{N(R+S+2)}\cdot 2^X$ eigenstates. When we make uniform superposition on some of the qubits, the normalization factor is equal to:

\begin{align}\label{eq-q-norm-Lambda}
    \Lambda=\frac{1}{\sqrt{2}^{NT}\cdot \sqrt{2}^X\cdot \sqrt{2}},
\end{align}

where $T$ is several qubits engaged in superposition, and the last $\sqrt{2}$ appears due to the interference operator. We write $T$ instead of $R+S+2$ for simplicity, firstly, but for a reason, the qubits count per component depend on the application. The qubits that are not superposed are in the state $\ket{0}$ or $\ket{1}$ with phase shift eventually, so they produce only one eigenstate – the other one has coordinate $0$. Tensor product multiplies the non-zero coordinate count by several non-zero coordinates of the product components. So, the non-superposed qubits do not extend the number of non-zero coordinates. Only non-zero coordinates are normalized. Hence the non-superposed qubits do not influence the normalization factor.\\
On the other hand, because we measure the last \textit{ancilla}, only the state after evolution is divided into halves: one for measure state $\ket{0}$ and the other for $\ket{1}$. Each of these halves has $2^{NT-1}\cdot 2^X$ coordinates. Each component consists of two \textit{ancillae}, so each coordinate of the state for measure state $\ket{0}$ contains the one part of one component $1+e^{i(nx+r_n)}$ or $1+e^{i(nx+s_n)}$. So the number of slots for components is equal to $2^{NT-2}\cdot 2^X$. We create QCoSamp in such a way that all slots are not empty, so that means that each component is in

\begin{align}\label{eq-normalization-tau}
    \tau=\frac{2^{NT-2}\cdot 2^X}{N}
\end{align}

copies. 

The number $\tau$ we call the \textit{appearance factor}. Because we are considering the balanced QCoSamp, the copy count is equal for all components.\\
The measurement generates an extra $2$ for each eigenstate, because $(1+e^{i(nx+r)})\overline{(1+e^{i(nx+r))}}=2(1+cos(nx+r))$. Hence the final normalization factor is equal to:

\begin{align}\label{eq-normalization-L-balanced}
    \frac{1}{L}=\Lambda\overline{\Lambda}\cdot \tau\cdot 2= \frac{2^{NT-2}\cdot 2^Q}{2^{NT}\cdot 2^Q\cdot 2 \cdot N}\cdot 2=
    \frac{2^{-2}}{N}=\frac{1}{4N}.
\end{align}

We see that the normalization factor $L=4N$ in the case of balanced QCoSamp does not depend on the number of superposed qubits but only on the number of components.\\
Now we discuss the non-balanced case, the normalization factor $\Lambda$ as well as the numerator of the $\tau$ remains unchanged because they depend on the qubit number, not on the architecture of QCoSamp, and superposition of states is assumed to be uniform. On the other hand, any full binary tree $T_F$ of height $h$ could be generated from the perfect binary tree $T_P$ of the same height by deleting the left and right children, within whole sub-trees of this internal node in $T_P$, which is the leaf $\mathcal{M}_n\in T_F$ of depth $\delta_n<h$. Each of those sub-trees has $h-\delta_n-1$ depth so they both would generate $2\cdot 2^{h-\delta_n-1}=2^{h-\delta_n}$ extra components. The denominator of $\tau$ for each of them would be equal to $H=2^h$. But now all values of $\tau$ for the extra component are passed to the $\mathcal{M}_n$. So, the denominator of the appearance factor for $n$-th component $\tau_n$ is divided by the appearance of extra components because the whole number of appearances for an existing component must increase by this number and described as follows:

\begin{equation}
     \tau=\frac{2^{NT-2}\cdot 2^X}{2^h/2^{h-\delta_n}}=\frac{2^{NT-2}\cdot 2^X}{2^{\delta_n}}=\frac{2^{NT-2}\cdot 2^X}{M_n},
\end{equation}

where $M_n=2^{\delta_n}$, we can say that the appearance factor of $n-th$ component is equal to the number of coordinates of a whole quantum system divided by the number of nodes of the perfect binary tree. Where the height is equal to the depth of this component in the QCoSamp architecture, thus we can determine the normalization factor for non-balanced QCoSamps:

\begin{align}\label{eq-normalization-L-not-balanced}
    \frac{1}{L_n}=&\Lambda\overline{\Lambda}\cdot \tau_n\cdot 2= \frac{2^{NT-2}\cdot 2^X}{2^{NT}\cdot 2^X\cdot 2 \cdot M_n}\cdot 2=
    \frac{2^{-2}}{M_n}=\frac{1}{4M_n}
\end{align}

where $h$ is the height of the tree representing QCoSamp, and $\delta_n$ is the depth of the $n$-th component. Note that if QCoSamp is balanced, then for all nodes $M_n=N$ and therefore $1/L_n=1/4N$, which is convergent with the equation \ref{eq-normalization-L-balanced}.

\subsubsection{Correctness of the output}\label{sec-output-correctness}

Due to postulates of quantum mechanics, the sum of probabilities for the output eigenstates $\ket{0}$ and $\ket{1}$ must be equal to $1$.
\begin{proof}

Let us denote by $\Delta$ the set of all nodes in the QCoSamp architecture, for which the depth is less than the height, as we see e.g. in equation \ref{eq-n-th-component-base-model-operator}, the coordinates for the last \textit{ancilla} are equal to $1-e^{i(nx+r)}$, therefore the sum component after measurements for $1$ are: $1-cos(nx+r)$, and therefore the sum of mentioned probabilities is equal to following:

\begin{align}
    &\norm{\mel{(NRSQ00)^N}{{\mathcal{M}}^\dagger }{0}}^2+\norm{\mel{(NRSQ00)^N}{{\mathcal{M}}^\dagger }{1}}^2=\nonumber\\
    &\sum_{n=1}^N
        \bigg(
            \frac{1+cos(nx+r_n)}{4M_n}+\frac{1+cos(nx+s_n)}{4M_n}
        \bigg)+
    \sum_{n=1}^N
        \bigg(
            \frac{1-cos(nx+r_n)}{4M_n}+\frac{1-cos(nx+s_n)}{4M_n}
        \bigg)=\nonumber\\
    &\sum_{n=1}^N
        \bigg(
            \frac{2}{4M_n}+\frac{2}{4M_n}
        \bigg)=
        \sum_{n=1}^N \frac{1}{M_n}\stackrel{(?)}{=}1
\end{align}

As mentioned, we generate the full binary tree by deleting consecutive sub-trees from the perfect tree of the same height. Therefore proving the $(?)$ equality from the equation above where we start with the perfect tree and then in each step delete two children of one node until the tree is equal to given one. 
We can use the mathematical induction where the base step is the case with balanced QCoSamp; the inductive step will assume the condition is fulfilled for the QCoSamp with $K$ components with $\delta_k<h$. We prove that the QCoSamp with $K+1$ such components also fulfills the $(1)$ condition if only the deletion does not decrease the tree's height.\\
\textbf{Base step} – QCoSamp is balanced with $N_0$ coordinates.

\begin{proof}
In case of balanced QCoSamp every $M_n=N_0$, so:
\begin{align*}
   \sum_{n=1}^{N_0} \frac{1}{M_n}=\sum_{n=1}^{N_0} \frac{1}{N_0}=1 
\end{align*}
\end{proof}

\textbf{Inductive step}. Let us assume the QCoSamp $B_K$ is not balanced, having $N_k$ components ($N_k<N_0$) and $K$ components that $\delta_k<h$. Assuming that this tree fulfills the condition $(1)$, we have to prove that if we delete two children for one of any internal node of a QCoSamp and the height of the new QCoSamp $B_{K+1}$ will be the same, then the $B_{K+1}$ fulfill the $(1)$ condition as well. Note that the deletion of the given node's two trees is the same operation as the deletion of the node's sub-tree without a given node (which is the root of the sub-tree).

\begin{proof}
Note that the set $\Delta_K=\{M_1,...,M_k\}$ for the tree $B_K$ is the set of those nodes which have the depth equal to tree height. So these are the nodes which we hasn't delete its children for. First we reorganize the sum for the $B_K$:
\begin{align*}
    \sum_{n=1}^{N_K} \frac{1}{M_n}=\sum_{n\notin \Delta_K} \frac{1}{{N_K}} +\sum_{n\in \Delta_K}\frac{1}{2^{\delta_k}}=\frac{N_K-K}{2^h}+\sum_{n\in \Delta_K}\frac{1}{2^{\delta_k}}\stackrel{asmp.}{=}1,
\end{align*}
where $\stackrel{asmp.}{=}$ means that this equality is true by the inductive step assumption. Now we compute the sum for tree $B_{K+1}$. Since we have deleted tree of depth $h-\delta_{k+1}$ without root the number of components decreases and now is equal to $N_K-2^{h-\delta_{k+1}}+1$. Hence the sum for the tree $B_{K+1}$ is as follows:

\begin{align*}
    &\sum_{n=1}^{N_K-2^{h-\delta_{k+1}}+1} \frac{1}{M_n}=
    \sum_{n=1, n\notin \Delta_{K+1}}^{N_K-2^{h-\delta_{k+1}}+1} \frac{1}{2^h}+
    \sum_{n\in \Delta_{K+1}}\frac{1}{\delta_k}=
    \frac{N_K-2^{h-\delta_{k+1}}+1-K-1}{2^h}+
    \sum_{n\in \Delta_{K+1}}\frac{1}{\delta_k}=\\
    &\frac{N_K-K-2^h\cdot 2^{-\delta_{K+1}}}{2^h} +
    \sum_{n\in \Delta_K}\frac{1}{2^{\delta_k}}+\frac{1}{2^{\delta_{k+1}}}=
    \bigg[
    \frac{N_K-K}{2^h}+\sum_{n\in \Delta_K}\frac{1}{2^{\delta_k}}
    \bigg]
    -\frac{2^h}{2^h\cdot 2^{\delta_{K+1}}}+\frac{1}{2^{\delta_{k+1}}}\stackrel{(A)}{=}1
\end{align*}

The equalization $(A)$ we obtain by substituting the sum for the tree $B_K$ from the inductive step assumption.\\
\textbf{Conclusion}: Since both the base case and inductive step have proved true, by mathematical induction every QCoSamp generated from the balanced QCoSamp of the same height, by deleting both children of the components $\delta_n<h$ holds the equation:

\begin{align*}
    \sum_{n=1}^N \frac{1}{M_n}=1.
\end{align*}
\end{proof}

The above corollary, in the context of the full binary tree: "The sum of the inverse of the lengths of all leaves in a full binary tree is equal to one".\\ 
Since we can obtain all QCoSamp as such, we prove that the probability of getting eigenstates $\ket{0}, \ket{1}$ that sum to $1$, we demonstrate the correctness of the output.

\end{proof}

\subsection{Higher dimension case}

Let us consider two QCoSamp $\mathcal{M}^{(1)}, \mathcal{M}^{(2)}$ having the same frequency and a different set of coefficients $r_n^{(1)}, s_n^{(1)}, r_n^{(2)}, s_n^{(2)}$ and separate input encoding different arguments: $\ket{X^{(1)}}, \ket{X^{(2)}}$. We apply the interference operator for each of them separately, we measure the last \textit{ancilla} for the first QCoSamp and the last \textit{ancilla} for the second one together the probability of obtaining the state $\ket{00}$, according to the formula \ref{eq-final-base-model-function} and because the tensor product of the two states creating the QCoSamp's, we obtain two dimensional version of the quantum harmonic cosine function:

\begin{align}
    \sum_{k=1}^N
    \sum_{n=1}^N
        &\bigg(
            \frac{1+cos(kx^{(1)}+r_k^{(1)}x^{(1)})}{4N}+\frac{1+cos(kx^{(1)}+s_k^{(1)})}{4N}
        \bigg)\cdot\nonumber\\
        &\bigg(
            \frac{1+cos(nx^{(2)}+r_n^{(1)}x^{(2)})}{4N}+\frac{1+cos(nx^{(2)}+s_n^{(2)})}{4N}
        \bigg)\approx\nonumber\\
        &\approx\mu_N(x^{(1)}, x^{(2)})
        \stackrel{n 1,2}{\longrightarrow}
        f_N(x^{(1)}, x^{(2)})
\end{align}

This two-dimensional version of the quantum harmonic cosine function can extend to any number of dimensions; in the case of $D$ dimensions, we have to use $D$ QCoSamp's and measure the $D$ last \textit{ancilla} for QCoSamp's together. The probability of state $\ket{0}^{\otimes D}$ will be the value of $D$ dimensional quantum harmonic cosine function, both of the normalizations (sec. \ref{sec-mapping} work in the same manner).

\subsection{Selection of the method of argument and parameters encoding}\label{sec-selection-of-parameters}

We already discussed the methods of encoding the data being arguments and parameters of a quantum algorithm made of QCoSamps: Directly by setting values in FCoSamp phases like in CPC (sec. \ref{sec-fco}) or constant data encoding \ref{sec-huge-number-encoding}) and indirectly using steerable elements APS (sec. \ref{sec-seo}), RC or FC (sec. \ref{sec-full-steerable-operator}). We present the general rules of the selection of the proper method of the argument and parameters encoding, as below:
\begin{enumerate}
    \item Every parameter that a value has to be selected by an algorithm from the set of possible values should be encoded by steerable operators (APS, RC, or FC).
    \item Every parameter for which the whole bunch of uniformly distributed values has to be processed simultaneously should be encoded by steerable operators.
    \item Every parameter that values are fixed for each quantum sampling process (but they may differ between two separate sampling processes), can be encoded directly by quantum phase setting in case of small number of values, or by constant data encoding (sec. \ref{sec-huge-number-encoding}) if this parameter has a relatively large number of values.
\end{enumerate}

 \section{Results}\label{sec-results}
 
This section presents experimental confirmation regarding the correctness of quantum harmonic cosine operators and several prospective applications to signal and image processing. First, we show the experimental result for a one-shot experiment for one ‘fixed values’ component and one for phase encoding. Then we present the random values experiment for one component but with randomized parameters: frequency and phase shifts ($n, r_n, s_n)$.

\subsection{One shot experiment}
\begin{SCfigure}
    \includegraphics[width=0.6\textwidth]{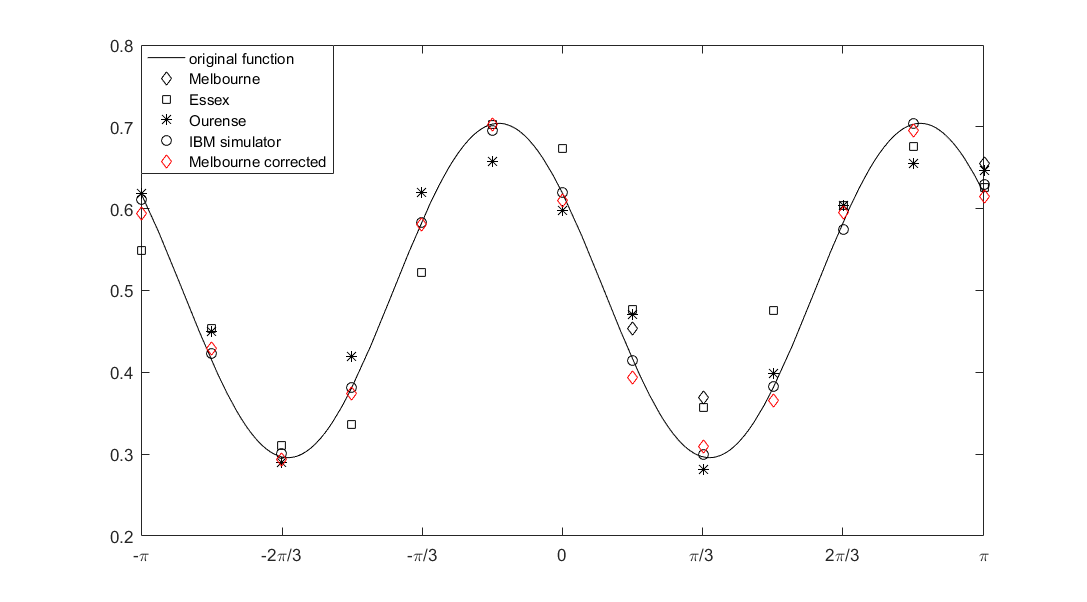}
    \caption{"One shot" experiment for the component $\nu_2(x)$. The original function is plotted with the solid line. The circle markers are results from the simulation, the computation from three quantum computers: Melbourne, Essex and Ourense – before error correction are plotted with black markers. The red diamonds are the results of Melbourne computer after the correction of errors.}
    \label{fig:res-one-component}
\end{SCfigure}
\begin{table}[ht!]
    \centering
    \small
    \begin{tabular}{l|r|r|r|r|r|r|r|r}
        \hline
        &\multicolumn{2}{c|}{read-out error} 
        &\multicolumn{2}{c|}{gate error} 
        &CNOT
        & \multicolumn{3}{c}{computed MSE}\\
        Name 
        &$\scriptscriptstyle{c_0[10^{-2}]}$ &$\scriptscriptstyle{c_1[10^{-2}]}$  
        &$\scriptscriptstyle{c_0[10^{-4}]}$ 
        &$\scriptscriptstyle{c_1[10^{-4}]}$
        &$\scriptscriptstyle{[10^{-2}]}$
        &$\scriptscriptstyle{1024[10^{-4}]}$
        &$\scriptscriptstyle{4096[10^{-4}]}$
        &$\scriptscriptstyle{8192[10^{-4}]}$ \\
        \hline
        Melbourne&1.80& 2.39&4.78 &4.38&1.93&-&-&9.34\\
        Essex &3.00& 3.17 & 3.08 & 5.56&1.09&25.37& 23.52&23.25\\
        Ourense & 1.30 & 2.40& 3.45 &5.82& 0.95&9.65&11.57 &10.32\\
        IBM Simulator &--&--&--&--&--&3.47&0.49&0.32\\
        Melbourne corrected &--&--&--&--&--&-&-&1.53\\
    \end{tabular}
    \caption{The results of the one-shot experiment, column one contains the name of the experiment. First three rows are results of raw data and comes from real quantum computers, shared by IBM, the fourth one is the raw data results from simulation by the same manufacturer; the last is the result is that of the real quantum computer after the correction procedure. Where the next five columns contain the error published after each calibration by the manufacturer. The last three columns contain the Mean Squared Error computed for experiments with 1024, 4096 and 8192 so called "shots" which are a repetition of the same algorithm}
    \label{tab:errors-one-shot}
\end{table}

In the "one-shot" experiment we examined the results for one fixed component operator according to the one component function: $\nu_2(x)=\frac{1+cos(2x-0.2)}{4}+\frac{1+cos(2x+2.1)}{4}$: $n=2, r_n=-0.2, s_c=2.1$. We examined this component with five result-sets. Three of them are the raw results from a realized quantum computer made and shared by IBM; their code-names are Melbourne, Essex, Ourense. The third is raw data from the simulator shared by IBM. The fifth is the results from the Melbourne beck-end after the simple error correction procedure. For Essex, Ourense and the simulator we made three variants of an experiment with 1024, 4096, and 8192 shots, which are the number of repetitions of the same algorithm after which the output probability density is constructed.\\
In figure \ref{fig:res-one-component} there are visible results of this experiment for these five result-sets. On the table \ref{tab:errors-one-shot} there are written three kind of erros: ,,read-out error'', ,, gate error'' and ,,CNOT error'' together with mean square error according to the authentic values of the $\nu_2(x)$ computed with Matlab.  The figure and table shows that the simulator and Melbourne back-end results after the correction procedure show good qualitative agreement with the original plot.

\subsection{Random values experiment}

\begin{figure}[ht]
    \centering
    \begin{tabular}[t]{c|c}
        \includegraphics[width=0.4\textwidth]{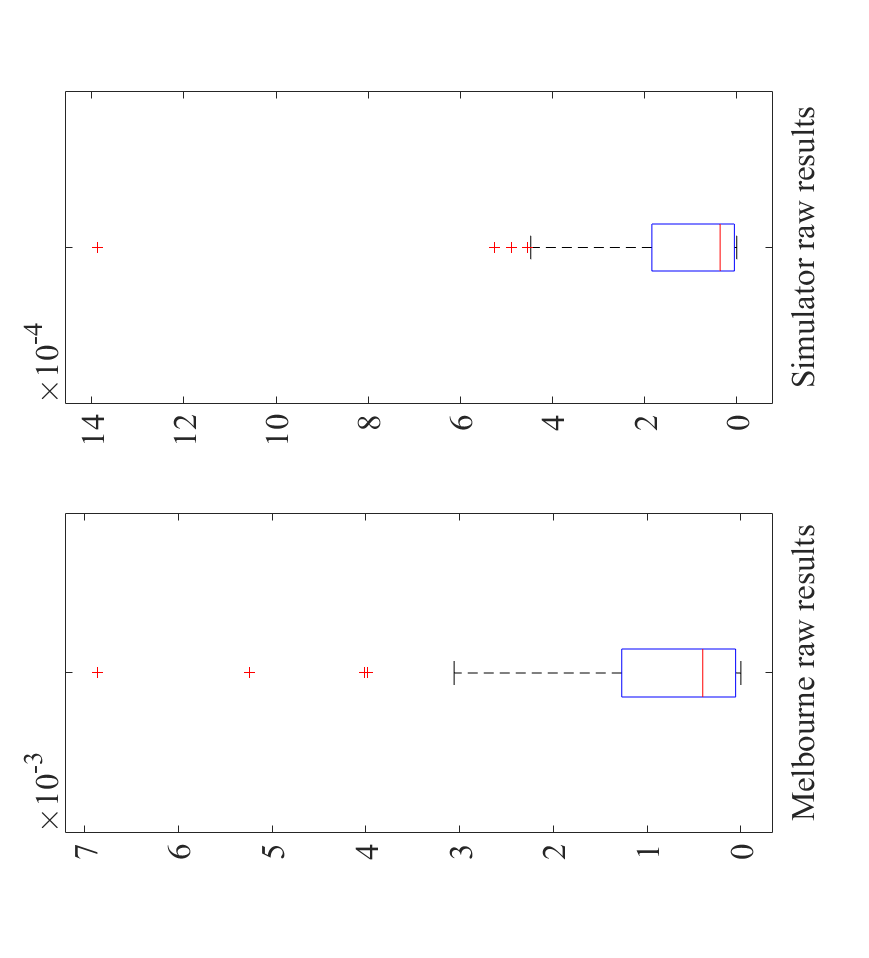} & 
        \begin{tabular}[b]{l|r}
            \hline
             parameter & value \\
             \hline
             Name & Melbourne\\
             $c_0$ read-out error & $1.35\cdot10^{-4}$\\
             $c_1$ read-out error & $1.80\cdot10^{-4}$\\
             $c_0$ single qubit gate error & $3.89\cdot10^{-4}$\\
             $c_1$ single qubit gate error & $10.15\cdot10^{-4}$\\
             $CNOT(c_0, c_1)$ gate error & $2.047\cdot10^{-2}$\\
             \multicolumn{2}{c}{}\\
             \multicolumn{2}{c}{}\\
             \multicolumn{2}{c}{}\\
             \multicolumn{2}{c}{}\\
             \multicolumn{2}{c}{}\\
             \multicolumn{2}{c}{}\\
             \multicolumn{2}{c}{}\\
        \end{tabular}
    \end{tabular}
    \caption{The box plot results in the random variable experiment. The plots show the MSE for the result-sets of Melbourne raw data and Simulator raw data. On the right side, there are parameters of the Melbourne computer during the experiment.}
    \label{fig:random-variable-experiment}
\end{figure}

In the "random variable experiment" we examine the one fixed component function $\nu_n=\frac{1+cos(nx+r_n)}{4}+\frac{1+cos(nx+s_n)}{4}$ with randomized parameters $n, x, r_n, s_n$. In each repetition of the experiment, the new set of such parameters was randomized. We made 500 repetitions on the Melbourne quantum computer and the simulation. Then we computed quartiles of the mean squared errors of this two-result set according to the reference values computed on the classical computer. The results in Fig. \ref{fig:random-variable-experiment} show the left side in the form of a boxplot, and on the right, there are calibration parameters of the Melbourne computer.\\

\begin{table}[ht!]
    \centering
    \begin{tabular}{l|r|r|r}
        \hline
        data set & first quartile & median & third quartile\\
        \hline
        Melbourne raw results & 
        $5.57\cdot 10^{-4}$ & 
        $4.07\cdot 10^{-4}$ &
        $1.27\cdot 10^{-3}$ \\
        Simulator results &
        $5.22\cdot 10^{-6}$ &
        $3.61\cdot 10^{-5}$ &
        $1.84\cdot 10^{-4}$ \\
    \end{tabular}
    \caption{The numerical results of the random values experiment.}
    \label{tab:random-values-experiment}
\end{table}
The numerical values of the mentioned statistics shown in table \ref{tab:random-values-experiment}. We see that the median and third quartile of squared error is better in the order of magnitude in the simulation results. The first quartile is smaller by two orders of magnitude.

\subsection{Application to signal processing – integration}

QCoSamp as a method can be applied as the base of the quantum computation for solving different problems. First, we can use the superposition phenomena for the input qubits $\ket{X}$ encoding the variable $x$. In that case the equations \ref{eq-M-n-final} and \ref{eq-n-th-component-base-model-operator-kicked-back1} of CMPn and becomes:

\begin{equation}\label{eq-M-n-final-var-x}
    \mathcal{M}_{nH^-}\ket{NRSX00}=
    \big[
    \frac{1}{2}\ket{NRS}
    \big(
        \ket{X00}+
        e^{i(nx+s_n)}\ket{X01}+
        e^{i(nx+r_n)}\ket{X10}+
        \ket{X11}
    \big)
    \big]
\end{equation}

The entry $x$ is encoded by the eigenstate $\ket{X}$ using the distributed phase encoding technique. In the equation \ref{eq-M-n-final}, the formula describes the four coordinate for different eigenstates: $$\ket{NRSX00}, \ket{NRSX01}, \ket{NRSX10}, \ket{NRSX11}$$ – the rest of coordinates are zero, one or phase shits (omitting the normalization) so they not influence the final probability, since the values of qubits $\ket{NRSX}$ are fixed and not superposed. Here, we apply the quantum summation technique (see Heinrich works \cite{heinrich_quantum_2002, heinrich_quantum_2003, heinrich_problem_2003} and equations \ref{eq-quantum-summation-operator}, \ref{eq-quantum-summation-acting} and \ref{eq-quantum-summation-sampling}), where the qubits $\ket{X}$ are superposed uniformly, so for each $x$ there are four coordinate appearing: $\ket{X00}, \ket{X01}, \ket{X10}, \ket{X11}$. 
Thus each value of $x$ appears one time in each component, which was proven to be true in the work of Lee and Selby (in \cite{lee_generalised_2016} p. 7 Lemma 1) that the probability of measuring the system in one of eigenstate created by superposition is preserved by controlled transformation. 
After applying the whole architecture of QCoSamp operator  (described in sec \ref{sec-building-blocks}), the most right \textit{ancilla} equal to $\ket{0}$ is the sum of states that can be encoded by $\ket{X}$. Therefore, after quantum sampling and reconstruction, we obtain the sum of the Fourier sine-cosine series' values for all arguments encoded by $\ket{X}$. Hence we can say that this is the value of Riemannian sum with a fixed value of $x_{k+1}-{x_k}=\pi/2^{X}$. Thus, we can obtain:

\begin{align}\label{eq-final-base-model-function-integration}
    \mathfrak{p}\ket{c_{2N-1}=0}=&\norm{\mel{(NRSX00)^K}{{\mathcal{M}}^\dagger }{0}}^2=\nonumber\\
    &\sum_{x\in[-\pi, \pi]}\sum_{n=1}^N
        \bigg(
            \frac{1+cos(nx+r_n)}{4N}+\frac{1+cos(nx+s_n)}{4N}
        \bigg)\approx
        \int_{-\pi}^{\pi}\mu_N(x)dx
\end{align}

Therefore, we can say that the result approximates the integral of the sine-cosine series represented by function $\mu_N(x)$. The accuracy grows exponentially with the number of qubits encoding the $x$ value. However, the most significant fact is that the computation time is constant – we do not need any repetitions of such integration, independently of the size of the $\ket{Q}$. In contrast to classical computation, we can increase the accuracy without penalty by increasing computation time.

\subsection{Case study: curve fitting in signal processing area}\label{sec-app-signal-processing}

The QCoSamp operators are for tasks that are appropriate for determining the Fourier series coefficient and including signal processing; as an example, we use the task of curve fitting. In the input of this task there is a huge number $K$ of pairs: $(x_k,y_k), k\in [0, K-1], x_k\in [0, \pi], y\in[0, 1]$. This curve fitting could be a time-series representing the (noisy) signal. The task is to find the continuous function that has the best fit to the set of points. In terms of Fourier sine-cosine, the task is to approximate the set of given points by a function: $f_N(x)$, for a fixed $N$. It means that we have to find $\lambda_n, \gamma_n$ coefficients such that we can minimize the error between $f_N(x_k)$ and $y_k$. For this purpose we use the forging reference probability technique (sec. \ref{sec-comparison-the-probability}) based on comparison of state (sec. \ref{sec-eq-states}), which uses the amplitude amplification algorithm (see appendix \ref{sec-oracle} after  Grover \cite{Grover_algorithm} and Brassard \cite{brassard_quantum_2000}). This task's solution is described in this section in detail, as an example of the procedure of quantum programming using the Cosine series Quantum Sampling method.\\
In the remaining part of this section we will use notation from sec.\ref{sec-comparison-the-probability}. We will use the balanced QCoSamp. The function $g_p(x)$ is the FCoSamp function, arguments are $x_k$-s encoded by constant data encoding, the parameters which function is dependent on are FCoSamp phase shifts $r_n, s_n$. We encode the frequencies $n$ directly, because the number of them is limited, we use the simplest form – by quantum phase shift setting. Therefore, the QCoSamp state has the form:

\begin{align}
    &\ket{(RS00)^{\otimes N}X}=\ket{R_{(1)}S_{(1)}00\dots R_{(N)}S_{(N)}0X}\stackrel{S}{\longrightarrow}
    \ket{XR_{(1)}S_{(1)}\dots R_{(N)}S_{(N)}0^{\otimes2N}},\nonumber\\ &\text{so according to eq. \ref{eq-equalization-state-building}, \ref{eq-n-th-component-base-model-operator-kicked-back1} and \ref{eq-q-norm-Lambda} and acting with ordering operator for more convenient form:}\nonumber\\
    &QCoSamp\ket{XR_{(1)}S_{(1)}\dots R_{(N)}S_{(N)}0^{\otimes2N}}=\nonumber\\
    &\frac{1}{L}
    \sum_{x=0}^{X-1}\sum_{p=0}^{P-1}\sum_{m=0}^{M-1}
    \ket{xpm}
    \sum_{n=0}^{N-1}
    \bigg[
        e^{i(n*\ket{x}+s_n)}\ket{00}+
        \ket{01}+
        e^{i(n*\ket{x}+r_n)}\ket{10}+
        \ket{11}+
    \bigg]\ket{n}.\label{eq-curve-QCoSamp-basic}\\
    &\ket{P}=\ket{R_{(1)}S_{(1)}\dots R_{(N)}S_{(N)}}\nonumber\\
    &M=\frac{\tau}{2^{P+X}}, (eq. \ref{eq-normalization-tau})\nonumber\\
    &\ket{W_p(x)}=\sum_{n=0}^{N-1}
        \big(
            e^{i(n*\ket{x}+s_n)}\ket{0}+
            e^{i(n*\ket{x}+r_n)}\ket{1}
        \big)\ket{0n}\label{eq-curve-wpx}
\end{align}

In the equation above, we define the working state $\ket{W_p(x)}$ from the forging reference probability technique, where the reference values were encoded using the constant data encoding technique, where the $\varepsilon_k=cos^{-1}(y-1)$. This encoding, together with the fact that $y$ has to be comparable with the FCoSamp, which is the probability distribution, meaning $y\in[0, 1]$. Now we can define the state as follows:

\begin{align}\label{eq-curve-ref}
    \ket{Y}=\frac{1}{\sqrt{2}^{N+1}}\sum_{n=0}^{N-1}
        \big(
            e^{i *\ket{y}}\ket{0}+
            e^{i *\ket{y}}\ket{1}
        \big)
    \ket{0n}
\end{align} 

 And finally we can define the initial and the desired states for the amplitude amplification algorithm, according to eq. \ref{eq-fi-omega-probab-comparison}:
 
 \begin{align}
    \ket{\varphi}=
        &\frac{1}{2L}\sum_{x=0}^{X-1}\sum_{p=0}^{P-1}\sum_{m=0}^{M-1}\ket{xpm}\bigg[
            \big(\ket{W_p(x)}-\ket{Y} \big)\ket{00}+
            \big(2\ket{\hat{1}}+\ket{Y}+\ket{W_p(x)} \big)\ket{01}+\nonumber\\
            &\big(2\ket{\hat{1}}-\ket{Y}-\ket{W_p(x)} \big)\ket{10}+
            \big(\ket{Y}-\ket{W_p(x)} \big)\ket{11}
        \bigg]\nonumber\\
        \ket{\omega}=
        &\frac{1}{2L}\sum_{x=0}^{X-1}\sum_{p=0}^{P-1}\sum_{m=0}^{M-1}\ket{xpm}
        \bigg[
        \big(2\ket{\hat{1}}+\ket{Y}+\ket{W_p(x)} \big)\ket{01}+
        \big(2\ket{\hat{1}}-\ket{Y}-\ket{W_p(x)} \big)\ket{10}
        \bigg]\label{eq-curve-beginning-snd-des-states}
\end{align}

Note that if the state $\ket{Q}=\ket{W_p(x_k)}-\ket{Y}=0$ then the value of FCoSamp function is equal to reference value $y_k$:

\begin{align}\label{eq-nuk=yk}
    \frac{1+cos(cos^{-1}(2y_k-1))}{2}=\frac{1}{2}+\frac{y_k-1}{2}=y_k
\end{align}

Then we just run the amplitude amplification machine. With changing the coordinates for the state $\ket{P}$ in increasing the probability of obtaining those of $\ket{p}$ which make the state of the system as close to the state $\ket{\omega}$ as possible. Obtaining the exact result is, in general case, unavailable because if the number of $x_k, y_k$ is sufficiently large, the sampled FCoSamp function with a far fewer number of components, can approximate the set of points only, minimizing the error between given pairs and the FCoSamp, similar to the curve fitting task with Fourier series on the classical computer. However not exact but best possible result, next to its main task of approximation, will filter the noise from the signal. Except of that the probability of obtaining the states $\ket{Q00}$ and $ \ket{Q11}$ are a measure of the error in the quality of fitting, which we call \emph{quantum similarity measure} (QSM). 
The procedure of obtaining such a measure is as follows:

\begin{enumerate}
    \item Prepare the QCosam according to eq. \ref{eq-curve-QCoSamp-basic}
    \item Define the states $\ket{\varphi}$ and $\ket{\omega}$ and implement sufficient repetitions of the amplitude amplification algorithm, which is described in details in sec. \ref{sec-eq-states} and \ref{sec-comparison-the-probability}.
    \item After the last repetition, apply the interference operator.
    \item Take quantum sampling through the procedure, on a prepared system (pts. 1-3).
\end{enumerate}

The dictionary of the output eigenstates is very simple – each eigenstate consisting of parameter set $P=\\ \{r_1, s_1, \dots, r_n, s_n\}$ with $0$ on the position of the second \textit{ancilla} represents one Fourier series. Therefore the measurement basis, being the subspace of the whole system basis: $\{\ket{xpmc_1c_2n}\}$, consists of state: $\{\ket{\cdot p\cdot c_1\cdot\cdot}\}$, where $\cdot$ mean that this qubit is not measured, which could be written in the extended form, considering that $\ket{p}$ contains information about all FCoSamp phase shifts: $\{\ket{\cdot r_1 s_1 \dots r_N s_N \cdot c_1\cdot \cdot}\}$. Then we apply amplitude amplification through sufficient iterations (see sec. \ref{sec-oracle} for details of "sufficient iterations" meaning) and quantum sampling, where each bin of appearance histogram represents one specific set of values of the parameters $r_n,s_n$. Because of the amplitude amplification procedure, this histogram has picks for those set of parameters which generates the best fit of the FCoSamp function to the input data.\\
We can obtain the fitting error for the chosen set of parameters. For this purpose we prepare the state $\ket{\varphi}$ (eq. \ref{eq-curve-beginning-snd-des-states}) with fixed values of chosen parameter set, but this time we take the quantum sampling procedure in four dimension basis generated of two most right \textit{ancillae} $\ket{\cdot\cdot\cdot c_{1}c_{2} \cdot}$ only. The sum of the values obtained for $\ket{00}$ and $\ket{11}$ eigenstates is the measure of an error. Let us consider the state:

\begin{align}\label{eq-curve-w-y}
    &\ket{W_p(x)}-\ket{Y}=
    \sum_{n=0}^{N-1}
    \bigg[
        \big(
            e^{i(n*\ket{x}+s_n)}-e^{i*\ket{y}}
        \big)\ket{0}+
        \big(
            e^{i(n*\ket{x}+r_n)}-e^{i*\ket{y}}
        \big)\ket{1}
    \bigg]\ket{0n}.\nonumber\\
\end{align}

According to the equations: \ref{eq-curve-QCoSamp-basic}, \ref{eq-curve-wpx} and \ref{eq-curve-beginning-snd-des-states}, the whole state $\ket{\varphi}$, for the found set of parameters $p$ and has the form:

\begin{align}
    &\ket{\varphi}=\ket{Xp0^{\otimes2N}}=
    \frac{1}{2\Lambda}
    \sum_{x=0}^{X-1}\sum_{m=0}^{M-1}
    \ket{xpm}
    \bigg[
            \big(\ket{W_p(x)}-\ket{Y} \big)\ket{00}+
            \big(2\ket{\hat{1}}+\ket{Y}+\ket{W_p(x)} \big)\ket{01}+\nonumber\\
            &\big(2\ket{\hat{1}}-\ket{Y}-\ket{W_p(x)} \big)\ket{10}+
            \big(\ket{Y}-\ket{W_p(x)} \big)\ket{11}
        \bigg]
    .\label{eq-curve-QCoSamp-error-form}
\end{align}

But we remember that the forging procedure goes to zero the coordinate for the state $\ket{xpm00}$; therefore, we suppose that the probability of obtaining such a state is a measure of the fitting error, so let us describe how to compute this probability:

\begin{align}\label{eq-curve-p}
    \mathfrak{p}\ket{c_{2N-2}c_{2N-1} =00}=&
    \big\|
        \braket{00}{\varphi}
    \big\|^2=\frac{1}{16L}
    \sum_{n=0}^{N-1}
        \bigg[
            \big|
                e^{i(n*\ket{x}+s_n)}-e^{i*\ket{y}}  
            \big|^2+
            \big|
                e^{i(n*\ket{x}+r_n)}-e^{i*\ket{y}}
            \big|^2
        \bigg]=\nonumber\\
    &\frac{1}{16L}
    \sum_{n=0}^{N-1}\mathfrak{f}(n*\ket{x}+s_n, *\ket{y})+\mathfrak{f}(n*\ket{x}+r_n, *\ket{y}), where\\
    \mathfrak{f}(x,y)=&2\big(1-cos(x-y)\big).
\end{align}

The quantum normalization factor after measurement L (see sec. \ref{sec-normalization-factor}) is multiplied by $16$ because the factor before measurement $\Lambda$ gives multiplication by $4$. Still, we have quantum normalization equal to $2\Lambda$, which produces a multiplication by $4$.
In the desired state the coordinate for $\ket{00}, \ket{11}$ are equal to $0$, so in that case The function $\mathfrak{f}(x,y)$ has to be equal to $0$, which means that $cos(x-y)=1$, which is fulfilled for $x=y+2k\pi$, but $y\in[0, 1], x\in [-\pi, \pi]$, therefore we can say that for desired state $x=y$. 
Since $y=cos(\hat{y}-1)$, then $x=cos^{-1}(\hat{y}-1)$, which, after applying cosine functions on both side gives $cos(x)=cos(cos^{-1}(\hat{y}+1)$, so  $\frac{1+cos(x_k)}{2}=\frac{\hat{y}}{2}$. By virtue of fact that two fractions $\frac{1+cos(x_k)}{2}$ creates one component of FCoCam $\nu_n$, we can show that:

\begin{align}
    \nu_n(x_k)=y_k\text{ iff }\mathfrak{f}(n*\ket{x}+s_n, *\ket{y})+\mathfrak{f}(n*\ket{x}+r_n, *\ket{y}=0.
\end{align}

The function $\mathfrak{f}$ is never less then $0$ ($cos(x-y)\leq 1 \Longrightarrow 1-cos(x-y)\geq 0$), and the sum of quantities never less then zero is equal to zero if and only if all quantities are equal to zero, therefore we can claim that:

\begin{align}
    \mu_N(x_k)=y_k \Longleftrightarrow \mathfrak{p}\ket{c_{2N-2}c_{2N-1} =00}=
   \big\|
        \braket{00}{\varphi}
    \big\|^2=
    0
    .
\end{align}

Thus, we can say that the value $\mathfrak{p}\ket{c_{2N-2}c_{2N-1} =00}$ is the measure of the error of the quantum fitting algorithm, and we call it, as mentioned, \emph{quantum similarity measure} QSM. 

\begin{SCfigure}
    \includegraphics[width=0.75\textwidth]{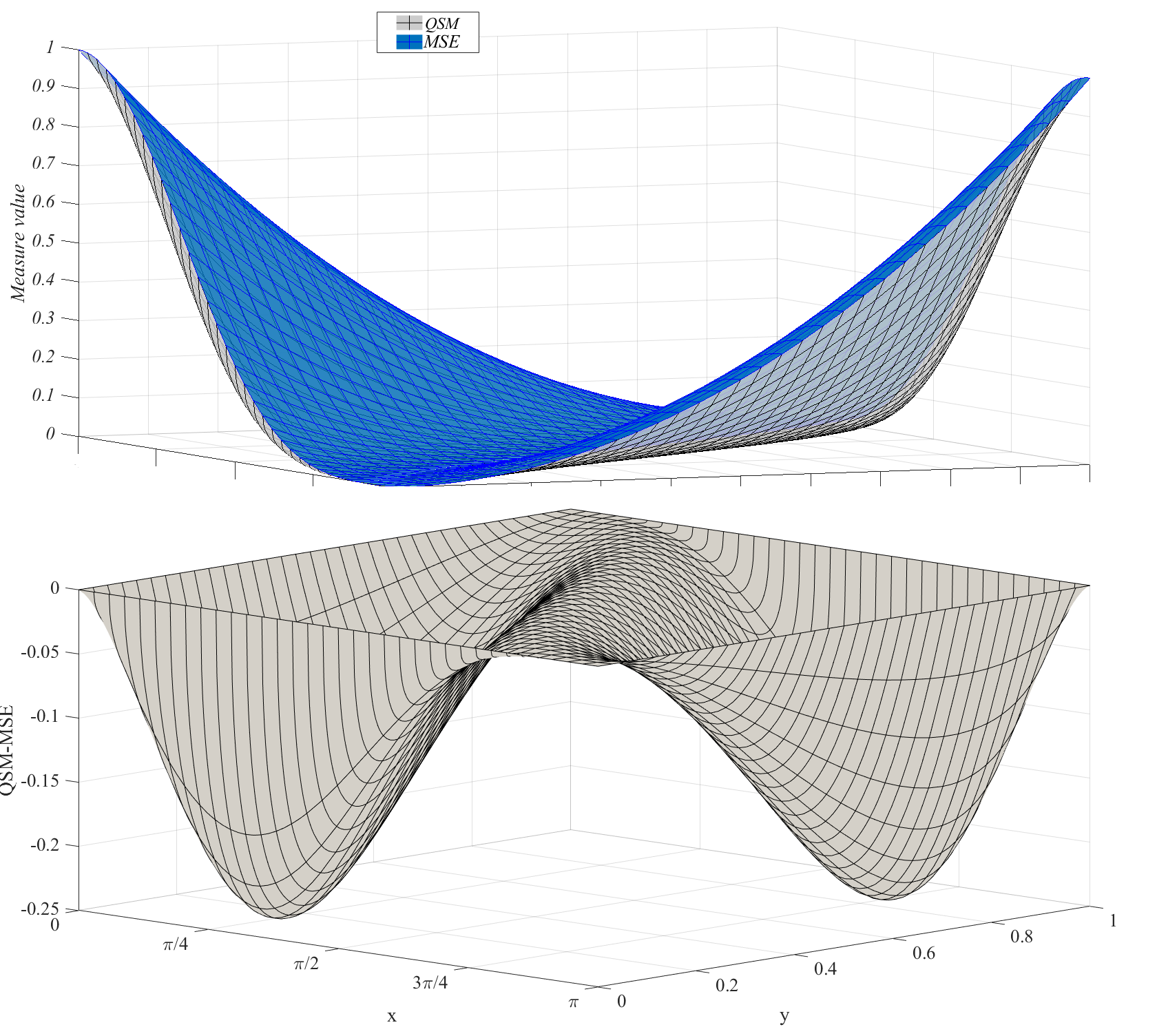}
    \caption{Comparison of the fitting measurement MSE (Mean Square Error), and QSM$^2$. 
There are errors computed for one component for both measures on the top plot: blue is MSE, and grey is QSM$^2$. On the bottom of the plot, the difference QSM$^2$-MSE is shown. Both measures give zero in case of no error. In other cases, the MSE (the blue plot on the top chart) provides a higher result than the QSM$^2$ (gray plot). The difference depends on the values of the function and references, achieving a maximum in two points (bottom plot): $(\pi/4, 0.2), (3\pi/4, 0.8)$ with the maximal difference in the order of $0.25$; we also note that the difference between them is not linear.}
    \label{fig:mse-qsm2-comparison}
\end{SCfigure}

We can compare the QSM$^2$ (squared quantum similarity measure) with the Mean Squared Error MSE, shown on the plots of fig. \ref{fig:mse-qsm2-comparison}.

\subsection{Quantum parallel window and quantum kernel filtering}\label{sec-materials-image-processing}

We use the technique of constant data encoding described in sec. \ref{sec-huge-number-encoding} and the FRQI method \cite{le_flexible_2011, yao_quantum_2017QIP, yan_quantum_2017QIP}  for the image encoding. The image is a composition of pixels consisting of rows and columns. Each pixel is a vector of \textit{channels} – most often $1, 3$ or $4$. We present here the \textit{gray-scale} case ($1$ channel), but extension for the multi-channel case is easy.\\
We map the two dimensional structure of image into the constant encoding operator (sec. \ref{sec-huge-number-encoding}) in such a way that the first $W$ (for image \textit{width}) qubits of the resulting state encode the row number and the remaining $G$ (image \textit{height}) qubits – the column number. The angle of the phase is the intensity of the pixel. For simplicity, we assume that $W, G$ are powers of $2$; if not – there will be some coordinates that are not part of image in the state representing image and they should have the value $0$. So we can write the formula for \textit{$W\times G$ pixels image encoding operator} and in the form of the resulting state is:

 \begin{align}
     \ket{Im}_{W\times G}=\mathcal{I}_{W\times G}\ket{H\ket{0}}^{\otimes W\cdot G}:=\sum_{w=0}^{W-1}\sum_{g=0}^{G-1} (*\ket{wg})\ket{wg} = \sum_{w=0}^{W-1}\sum_{g=0}^{G-1} e^{ip(w,g)}\ket{wg},
 \end{align}
 
where $p(w,g)$ is the intensity of the pixel $(w,g)$.
Using this method, we encode the Full HD image using $21$ qubits and the 16K image (2:1) using $27$ qubits. On classical computers, we need, respectively, $8\cdot2\cdot 10^{6}$ and $8\cdot1,34\cdot 10^{8}$ of bits with $256$ levels of gray. \\
\textbf{The notion of window} is widely used in image processing, e.g. for filtering the image (Jeong et al. in \cite{jeong_medical_2011}) or for feature extraction in object classification task (Viola and Jones in \cite{viola_rapid_2001}, Lienhart and Maydt in \cite{lienhart_extended_2002}). It relies on the extraction from the whole image, a rectangular area, and the operations performed in this area. It sometimes has a name \textit{region of interest (ROI)}. The window can be shifted on the image by $sh_r$ pixels right. After reaching the right border, it comes back to the left border with a $sh_d$ pixels shift down. In each step, the same operation acts on the window. Such a procedure has a name \textit{sliding window}. The computation time on classical computers depends on the size of an image, the window size, shifts right and down, and the window-in operation complexity. The number of repetitions of the window-in operation is equal to $=\big\lceil\frac{I_w-W_w}{sh_r}\big\rceil\cdot \big\lceil\frac{I_h-W_h}{sh_d}\big\rceil$, where $I_w, I_h$ is the image width and height and $W_w, W_h$ is the window width and height. In the quantum computer, we can define the technique of \emph{quantum sliding window}. First, we define the state for pixel $x, y$ that is in the center of the window, and for which the new value is computed, with the formula:

 \begin{equation}
     \ket{W(x,y)}_{W\times G}:=
     \sum_{w=x-\frac{W}{2}}^{x+\frac{W}{2}}
     \sum_{g=y-\frac{G}{2}}^{y+\frac{G}{2}} e^{ip(w,g)}\ket{wg}:= \sum_{w,g\in W_{x,y}} e^{ip(w,g)}\ket{wg}
 \end{equation}
where $W_{x,y}$ means the window with the center $x,y$.
 We can then act with an arbitrary chosen unitary operator $\mathcal{U}$ on the state $\ket{Im}_{W\times G}$ and cut the window with un-computation using twice permuted controlled operator (see sec. \ref{sec-cascade-controlled-operator}) techniques are obtaining the operators $\mathcal{U}_W(x,y)$ acting on the states $\ket{W(x,y)}_{W\times G}$ exclusively. 
The operators $\mathcal{U}_W(x,y)$ can act simultaneously, which creates the \emph{quantum parallel window}, which is the equivalent of a sliding window, wherein the window is not "sliding" but computed in parallel. For optimization, the un-computation procedures are complete once all operators act. Such that the computation time is equal to the complexity of $\mathcal{U}$ only – there will be no repetition because of simultaneous action of all operators.\\
A significant example of an application for the quantum parallel window method is the image's quantum kernel filtering. For this purpose we use the quantum arithmetic \cite{ruiz-perez_quantum_2017, draper_addition_2000, beauregard_quantum_2003} and the mean computing operator $\mathcal{U}_{mean}$as the $\mathcal{U}$. If we compute the mean of all coordinates belonging to the window, as a result for the center placed eigenstate, we obtain:

 \begin{align}\label{eq-U-mean}
    \kappa(x,y)=&\frac{1}{W\cdot G}\bigg(
        \sum_{w,g\in W_{x,y}}*\ket{wg}
    \bigg)\nonumber\\
     h_r=&\frac{sh_r}{2}, h_d=\frac{sh_d}{2}\nonumber\\
     \mathcal{U}_{mean}\ket{Im}_{W\times G}=&
     \sum_{w_c=h_r}^{W-h_r}
     \sum_{g_c=h_d}^{G-h_d}
     e^{i\kappa(w_c,g_c)}
     \ket{w_cg_c}
 \end{align}
 
where the function $\kappa(x,y)$ is the filter's quantum kernel. The new coordinate for each state $\ket{w_cg_c}$ is the exponent of the mean of coordinates from the window surrounding this state. Hence the angle represents the pixel value; the resulting state contains the image after the mean filter application. Using the weighted mean (ibidem), we can compute another kind of kernel filter. \\
\label{pt-complexity-of-qfilters}The advantage of using this method is that the computation time depends on the window operation only – it does not depend on the size of the image and the density of windows computation. If the process on widows also has a constant time (like in the example), we can freely widen the window sizes without influencing computational time.\\

\subsection{Research opportunities in the area of image processing}\label{sec-app-image-processing}

So far, in the subsection \ref{sec-materials-image-processing} we described some techniques useful for image processing, slightly extended by us: the encoding technique based on FRQI, the quantum (sliding) window processing using the un-computation and filtering using the quantum arithmetic. Now we discuss the new areas of investigation and research that are available using the QCoSamp family concept, on the convolution filters and Wavelet-like quantum features based on the idea of Haar-like features:

\begin{enumerate}

\item Two-dimensional QCoSamp with fixed coefficients $r_n, s_n$ applied to the presented quantum sliding window method will result in quantum convolution kernel filters. In equation \ref{eq-U-mean}, we obtain the sum of QCoSamp responses instead of the sum of pixel intensities. Let's assume that $\mu \big(x^{(1)}, x^{(2)}\big)\ket{x^{(1)}x^{(2)}}$ is the coordinate of the eigenstate $\ket{x^{(1)}x^{(2)}}$. Let us consider two dimensional QCoSamp $\mathcal{M}_N$, with the fixed $N, r_n, s_n$. If we act with such an operator on the window $W_{x,y}$ with the center in the pixel $x,y$ we obtain the quantum kernel:
    
    \begin{align}
    \psi(x,y)=&\sum_{w,g\in W_{x,y}}\mathcal{M}_{NH^-}\big(*\ket{w}, *\ket{g}\big)
    \end{align}
    
We can replace it with this kernel, the one from equation \ref{eq-U-mean}, and finally, obtain the state contained in the image after convolution filtering with the kernel designated by the QCoSamp. Also, if QCoSamp $M_N(x^{(1)}, x^{(2)})=0$ for arguments laying on the boundary of $[-\pi, \pi]\times [-\pi, \pi]$ we can say that the QCoSamp is a wavelet (assuming that everywhere else its equal to zero), we obtain the notion of quantum wavelet kernel filter.
\item The idea of Haar-like features is based on the window's division by two areas – black and white; for each of them, and computing the mean of pixel intensity. The black and white part's intensities rate generates a feature vector, used for classification, and specifically for object (including face) detection. This idea can be included in QCoSamp as \textit{quantum wavelet-like features (QWF)}, defined as two-dimension QCoSamp fulfilling the boundary condition from the previous point, that produces a function satisfying the definition of wavelet. We decided to name this features more general (wavelet-like instead of Haar-like features) because the quantum kernels made with such QCoSamp's represent  wavelets in general, not Haar wavelets only. As we have seen above, the computation time is independent of the shape of the QCoSamp, in contrast to the classical ones, where the computational time for complicated wavelets is greater than in simple Haar wavelet case. Having defined QCoSamp's for QWF by fixing frequencies and coefficients $r_n,s_n$, we can use as the kernel operator in the quantum parallel window procedure described above. But this time, we are not interested in the final result from filtering, but the interest is in comparing the pixel results from the given window and the shape of the kernel given by QCoSamp. Applying the QCoSamp to pixels of the window $x,y$ separately, unlike the kernel procedure, but according to the comparison of state technique
 (sec. \ref{sec-eq-states}, eq. \ref{eq-state-equalization-hadamard}, pg. \pageref{eq-state-equalization-hadamard} and description below), coordinate (and probability) of eigenstate $\ket{0}$ obtained by measuring the right \textit{ancilla} of the state $\ket{xyc_1c_2}$ is a measure of similarity of kernel and the window. If they are identical – the probability is equal to $\mathfrak{p}\ket{xyc_1c_2=\cdot\cdot\cdot 0}=0$, (so $\mathfrak{p}\ket{xyc_1c_2=\cdot\cdot\cdot 1}=1$) while if they are exactly opposite each other to $\mathfrak{p}\ket{xyc_1c_2=\cdot\cdot\cdot 0}=\mathfrak{p}\ket{xyc_1c_2=\cdot\cdot\cdot 1}=0.5$. Omitting the remaining part of the system for simplicity and introduced using the same method as in the curve fitting case study (sec. \ref{sec-app-signal-processing}, eq. \ref{eq-curve-wpx} and \ref{eq-curve-ref}), we write the simplified formula for the states, which generates the measure of similarity of the pixel intensity distribution through the window and the shape of the kernel. Described as follows:

    \begin{align}\label{eq-image-proc-x-y}
        \ket{\Theta c_1c_2}=&
        \sum_{w, g\in W_{xy}}\bigg[
        \big(
            \mu(w,g)-*\ket{xy}
        \big)\ket{wg00}+
        \big(
            2+\mu(w,g)+*\ket{xy}
        \big)\ket{wg01}+\nonumber\\
        &\big(
            *\ket{xy}-\mu(w,g)
        \big)\ket{wg10}+
        \big(
            2-*\ket{xy}-\mu(w,g)
        \big)\ket{wg11}
        \bigg]\nonumber\\
        \vartheta(W_{x,y})=&\mathfrak{p}\ket{c_1=0}=2\big|\mu(w,g)-*\ket{xy}\big|^2,
    \end{align}
this is equal to $0$ in the case where all of the pixels are fitted to the FCoSamp function perfectly. It is closer to $1/2$ (with accuracy to the normalization) if the dissimilarity increases.
At this point, we would like to highlight two of the possible applications of QWF:
    \begin{enumerate}
    	
        \item The canonical approach uses the similarity measure as the feature vector for pixel $x,y$, and uses the $\ket{\vartheta(W_{x,y})}$ in further quantum processing. On the other hand, we can measure the values $\vartheta(W_{x,y})$ and use it as the features in further, additional classical processing, making the so-called \textit{quantum-classical hybrid} algorithm.
        \item Direct object detection using the two-dimensional QWF version, discussed in section \ref{sec-app-signal-processing}, with the quantum curve fitting algorithm, we make a training phase using the set of images containing one class of object (e.g., dog, cat, cow, etc.). For different classes, we should obtain different shapes of QCoSamp. Then, on the set of the images unseen for the system, we check the similarity $\vartheta(Im)$. The class of the QCoSamp with the best similarity measure value for the image, defines the object class for the given image, with a threshold of minimal similarity accepted for "unknown" objects. This way, we create a separate quantum model for each class.
    \end{enumerate}
\end{enumerate}

 \section{Discussion \& conclusions}\label{sec-discussion}

Based on the experiments made on real quantum computers, we can observe that:
\begin{enumerate}
	
\item The best results (smallest MSE) is for the simulation, which is not surprising as simulation does not factor for errors but is an idealized result – see \cite{noauthor_ibm_nodate}.
\item The correction procedure affects the results, such that they are similar to that in the idealized case; for this specific case, it improves the result by a factor of six.
\item The read-out error of the qubit measured has a decisive influence on the final computational error. Measured $c_1$ qubit error for Essex computer has ca $1.3$ times greater error than Melbourne and Ourense. Since the mean gate error is similar on these three computers, we see that read-out error causes over two times greater final error. This dependency is not precise because several other factors influence the result, but the trend is evident. 
\item In the case where we cannot observe the CNOT gate error's influence on the final result. 
\item Increasing the number of shots has a noticeable impact on the simulated results only. Increasing the number of shots from 1024 to 4096 ($4$ times) decreases the MSE by $7$ times. However, the further increase to $8192$ decreases the error by a factor $1.5$ only. 
\item Increasing the number of shots in the case of real back-ends has no noticeable influence on the final error. We speculate that the influence of quantum errors is greater than the statistical effect of repetitions.
    
\end{enumerate}

We can see that some equations are repeated in different variants in the paper, like the equation of the $n$-th component on the two variants before and after applying interference: eq. \ref{eq-inject-constant-data}, \ref{eq-mbo-state-it1}, \ref{eq-cpc-after-hadamard}, \ref{eq-quantum-summation-acting}, \ref{eq-aps-final}, \ref{eq-M-n-final}, \ref{eq-n-th-component-base-model-operator-kicked-back1}, \ref{eq-nu-x-n-th-component-final}; the equation of the state comparison: eq. \ref{eq-state-equalization-hadamard}, \ref{eq-comparison-state}, \ref{eq-fi-omega-probab-comparison}, \ref{eq-curve-w-y}, \ref{eq-curve-p} and \ref{eq-image-proc-x-y} etc. shows that we use the same machinery to solve different problems, which is mentioned as the main motivation of this paper.\\
The quantum sampling method used here to extract results itself is significant, even ignoring the architecture of QCoSamp, the importance is described further. While deterministic quantum computation must obtain one state, the amplitude amplification procedure must extract just one state, which is the problem's solution. In the quantum sampling model of computation, such a procedure aims to filter the worst cases, reflected in the narrowing peaks (representing possible solutions) in the sampled state's appearance histogram. This enables a reduction in the number of eigenstates from a measurement basis is affordable in practice, to make the sampling procedure available to perform in the context of requiring a sampling count. Therefore, the quantum sampling procedure limits the needed repetition count of the amplitude amplification procedure.\\ 
Generalizing, the proposed method (QCoSamp) is the method of generation for the quantum evolution operator. The quantum sampling procedure performed in measurement and the eigenstate proper, for the problem to be solved and, after the reconstruction, generates the value of Fourier sine-cosine series for the argument and parameters encoded on several input qubits. Therefore, we can say that we proposed solving the problems described by the Fourier series on the quantum computer.\\
Furthermore, we gave the experimental proof of concept for this idea by checking the simulator's basic building block computed on three quantum computers. The simulator experiments proved that the accuracy of computation improves with an increasing number of measurements. The mean square error quartile analysis shows that the results obtained with $8192$ repetitions will achieve accuracy from $2.2\cdot 10^{-3}$ to $1.3\cdot 10^{-2}$ with a probable error $6\cdot 10^{-3}$. Looking at the decreasing trend, we can expect that the error will decrease with an increasing number of repetitions. Still, we were not able to check it experimentally due to the limitations of the available hardware. In real quantum computers, the error is more significant, by about one order of magnitude. These two facts show that our method is theoretically correct and implementable on real quantum computers. There are currently limitations due to the errors caused by quantum decoherence and the small number of available qubits. Along with reducing errors and increasing the qubits quantity, the proposed method will be more practical to use.\\

\subsection{The meaning of the QCoSamp architecture}

Following the process of generating the QCoSamp, we will come to the state which can be constructed by a diagonal operator with values $1, e^{i\varphi_k}$ on the diagonal multiplied by $1^{\otimes h}H$, where $h$ is sufficient for the size of the diagonal operator and depends on the number of components, which seems to be quite simple evolution operator. Nevertheless, the full description and considerations about the internal structure of QCoSamp are relevant for the following reasons:
\begin{enumerate}
    \item The architecture of QCoSamp allows the implementation of this operator using the standard gate set available on existing quantum computers.
    \item The QCoSamp method describes how to use the QCoSamp operator as a part of the wide class of algorithms based on the model described by the Fourier sine-cosine series. There is a possibility, for the basic building blocks – CMPn, introducing the frequencies, arguments, and series coefficient in three ways: 
    \begin{enumerate}
        \item Directly, by fixing the quantum phases (sec. \ref{sec-fco}) representing such an element.
        \item Directly, by constant data encoding (sec. \ref{sec-huge-number-encoding}), which allows encoding of the large amount $D$ of data, using a logarithmic, smaller number of qubits – $\lceil\log_2(D)\rceil$, introduce to the system, objects that demand a lot of resources. Such as images, video sequences, and large databases of content from the area of big data, etc. 
        \item Indirectly, using the steerable parameters (sec. \ref{sec-seo}, \ref{sec-full-steerable-operator}), which allow the creation of the uniform superposition of the $2^Q$ values of such parameters, perform transformations on all of those values in parallel and extract the results by amplitude amplification algorithm or similar.
    \end{enumerate}
    It enables the quantum program to flexibly manipulate the qubits' arrangement in the initial state and interpretation of them at the Fourier series' abstraction levels. The problem represented by this series is not at the low level of quantum bits, which makes the process of input state preparation much less complicated. In other words, knowing the meaning of series parameters (arguments, coefficients, frequencies) in the context of the problem, one can immediately and almost mechanically translate them to the consecutive sub-sequences of qubits in the initial state according to the representation of Fourier series by QCoSamp (sec. \ref{sec-mapping}, \ref{appx-sec-mapping}) its reconstruction (sec. \ref{sec-reconstruction}) and the rules of parametric encoding selection (sec. \ref{sec-geenral-scheme-of-using-QCoSamp}). 
    \item It discovers the complex, closed in the form of the full binary tree, structure of the the QCoSamp hidden behind a simple outer form. Thus, allowing intentional manipulation of this operator to solve complex tasks may not be noticeable by just observing this simplified diagonal matrix form of the operator. 
    This structure (sec. \ref{sec-base-model-architecture}), similarly to encoding selection rules described above, takes the evolution operator defining the process to a higher level of abstraction because knowing the counts of series and the encoding of parameters, which methods are already selected (and are described above). Defined precisely by the QCoSamp method, where the parameters of the operator components (leaves) should be steerable and for which of them should be fixed (sec. \ref{sec-building-blocks}), so we know immediately (sec. \ref{mmsubsec-base-phase-operator}, \ref{sec-fco}, \ref{sec-seo}) how to make components from gates and because of the connection (sec \ref{sec-two-component-connection}) and interference operator are mechanically applicable. Finally, we know how to obtain an operator representing this series. 
    \item Except that this structure clarifies the un-computation procedure, implementing QCoSamp quantum algorithms is much more uncomplicated; influencing the procedure by decreasing the number of demanded repetitions in quantum sampling is of greater importance.
    
\end{enumerate}

QCoSamp enables creating a quantum algorithm based on the problem's modeling to be solved by Fourier sine-cosine series, taking this process to a higher level of abstraction in a similar way to higher-level programming languages and frameworks. It takes quantum programming to a higher level of abstraction, lowering the level of required skills in quantum information theory and quantum physics for people who would like to use quantum computation to solve its practical problems. Bringing quantum programming closer to specialists makes it more applicable to practical activity, especially in signal and image processing, but certainly not exclusively in these areas.

\subsection{Examples of the future research and development opportunities}

As described in sec. \ref{sec-materials-image-processing}, we show the general scheme of quantum kernel filtering. Nevertheless, in image processing, there are many types of filters for image sharpening, edge detection, Gaussian blur, differential filters, etc. So there is a vast area for creating and developing the quantum version of such filters and finding new ones. Except that, because the computation time does not depend (pg. \pageref{pt-complexity-of-qfilters}) on the size of the window and generally is constant, there arises a possibility in checking the impact of larger sized windows upon the images.\\
In the area of Quantum Wavelet Features (sec. \ref{sec-app-image-processing}) which, theoretically promising, needs confirming in both simulation and real quantum computers. Which demand the development of a framework for real solutions (like QISKit or F\#) and research in both approaches: (a) creating a quantum-classical algorithm of object recognition, where the descriptors (feature vectors) of objects are fetched from the quantum computer using QCoSamp and classification done by classical computing and (b) where both – feature extraction and classification, run on a quantum computer. The first approach is currently more accessible due to the limitations of qubit count in real quantum computers. Further research could be in the direction of the shapes of the QCoSamp with the best object distinguishing power. Other issues to check the influence of the parameter sharing property of QCoSamp on this power, which we understand has the same parameters in both dimensions by the components. Suppose this property holds the distinguishing power of a sufficient level. In that case, qubits are saved, because sharing parameters by a two-dimensional component of QCoSamp needs only the two additional \textit{ancillae} in the one-dimensional case, because the initial parameter qubits can be shared in two dimensions as well, while not sharing one demand doubling the qubit count.\\
Another issue arises in connection with the tree structure of QCoSamp operators; hidden in its arrangement is made of component QMPn's, and the connection operators. The idea that stays behind this structure stays on the observation that if the tensor product is associative, we can place the parenthesis in the manner we want to. Each parenthesis contains no parenthesis inside is the connection of the component and forms the first level; parenthesis includes the first level objects and forms the second level of the tree, and so on. In the case of the QCoSamp, we connect two objects of the previous level; therefore, we obtain the full binary tree. The goal guiding us in that activity was making the very complicated internal structure of QCoSamp far simpler, and what we achieved. Still, it turns out that this structure of QCoSamp limits the number of necessary un-computations, on average are fourfold. Like a connection of more than two objects at each level, the other structures, changing the number of connected objects on different levels, etc. can limit such demands even more. If so, the tree structuring of the tensor product would be a useful tool for optimizing a quantum circuit. Still, it demands much more in-depth theoretical research and practical experiments.\\

\subsection{Conclusion}

This paper presents a new family of the quantum evolution operators – QCoSamp, used for modeling problems with single- or multidimensional Fourier sine-cosine series on quantum computers. We show the details of the internal structure, architecture, and the way of building such operators. We also present experimental results involving their implementations on real quantum computers or quantum simulators. We have developed several new quantum programming techniques specifically for those operators and create several algorithms for practical application. We present its importance from the perspective of the level of abstraction in quantum programming.\\
The proposed family of QCoSamps operators opens new investigation possibilities in the domain of quantum signal and image processing. In the future, we plan to develop the proposed operators to create a framework for implementing this notion in one of the quantum programming languages (QISKit or F\#). We also plan to investigate the areas of research mentioned above.


 
\appendix
\section{Mapping between quantum cosine sampled function and Fourier sine-cosine series}\label{appx-sec-mapping}

In this section, we construct the mapping between the sine-cosine Fourier series (eq. \ref{eq-fsn-approx}) and the FCoSamp function (eq. \ref{eq-base-model}) for which, the approximation is obtained by a quantum sampling of the QCoSamp operator.\\
First, we reorganize the FCoSamp as follows:

\begin{align}\label{eq-base-model-sine-cosine}
    \mu_N(x)= &
    \frac{1}{L}
    \sum_{n=1}^N
        \big(
            2+cos(nx+r_n)+cos(nx+s_n)
        \big)=\nonumber\\
    &\frac{1}{2}+\frac{1}{L}
    \sum_{n=1}^N
        \big(
            cos(r_n)cos(nx)-sin(r_n)sin(nx)+
            cos(s_n)cos(nx)-sin(s_n)sin(nx)
        \big)=\nonumber\\
     &\frac{1}{2}+\frac{1}{L}
     \sum_{n=1}^N
        \big[
            [cos(r_n)+cos(s_n)]cos(nx)+
            [-sin(r_n)-sin(s_n)]sin(nx)
        \big]
\end{align}

As usual, we consider that the value $1/2$ represents the $0$-s component of the series where $\lambda_0=1/2$.
The above form of $\mu_N$ is similar to $f_N$, and we can see that $r_n, s_n$ is in the place of coefficients. 
However, there is one crucial difference: while in Fourier transform coefficients of one frequency depend on the approximated function itself, it does not depend on another. In the FCoSamp, the coefficients for the same frequency is dependent on one another, and it is formed by trigonometric functions as follows:
\begin{align*}
    \lambda_n &= cos(r_n)+cos(s_n)\\
    \gamma_n &= -sin(r_n)-sin(s_n),
\end{align*}
which is the mapping from equation \ref{eq-mapping-FCoSamp-Fourier}.  
However, the question remains, can we find for every $\lambda_n, \gamma_n$ the coefficient $r_n, s_n$. Because if not, some of the Fourier series would not have the representation by FCoSamp. \\
Therefore we have to prove that for every pair of phases in FCoSamp, there exists a pair of two Fourier coefficients and vice versa for the same frequency (component number) $n$. For that reason, we will justify the below observations made for the same frequency $n$:

\begin{enumerate}
    \item The set $ \{(r_n, s_n) | r_n, s_n \in [-\pi, \pi] \wedge s_n\neq r_n-k\pi\}$ is a double cover of the filled open circle without the center with radius $2$ on the 2D Cartesian space: $\lambda^2+\gamma^2<2 \wedge (\lambda, \gamma)\neq (0,0)$
    \item The set $ \{(r_n, s_n) | r_n, s_n \in [-\pi, \pi] \wedge s_n= r_n\}$ is a cover of the edge of such a circle: $\lambda^2+\gamma^2=2$
    \item The set $ \{(r_n, s_n) | r_n, s_n \in [-\pi, \pi] \wedge s_n= r_n-\pi\}$ maps to a point $(0,0)$.
\end{enumerate}

\begin{SCfigure}
    \includegraphics[width=0.8\textwidth]{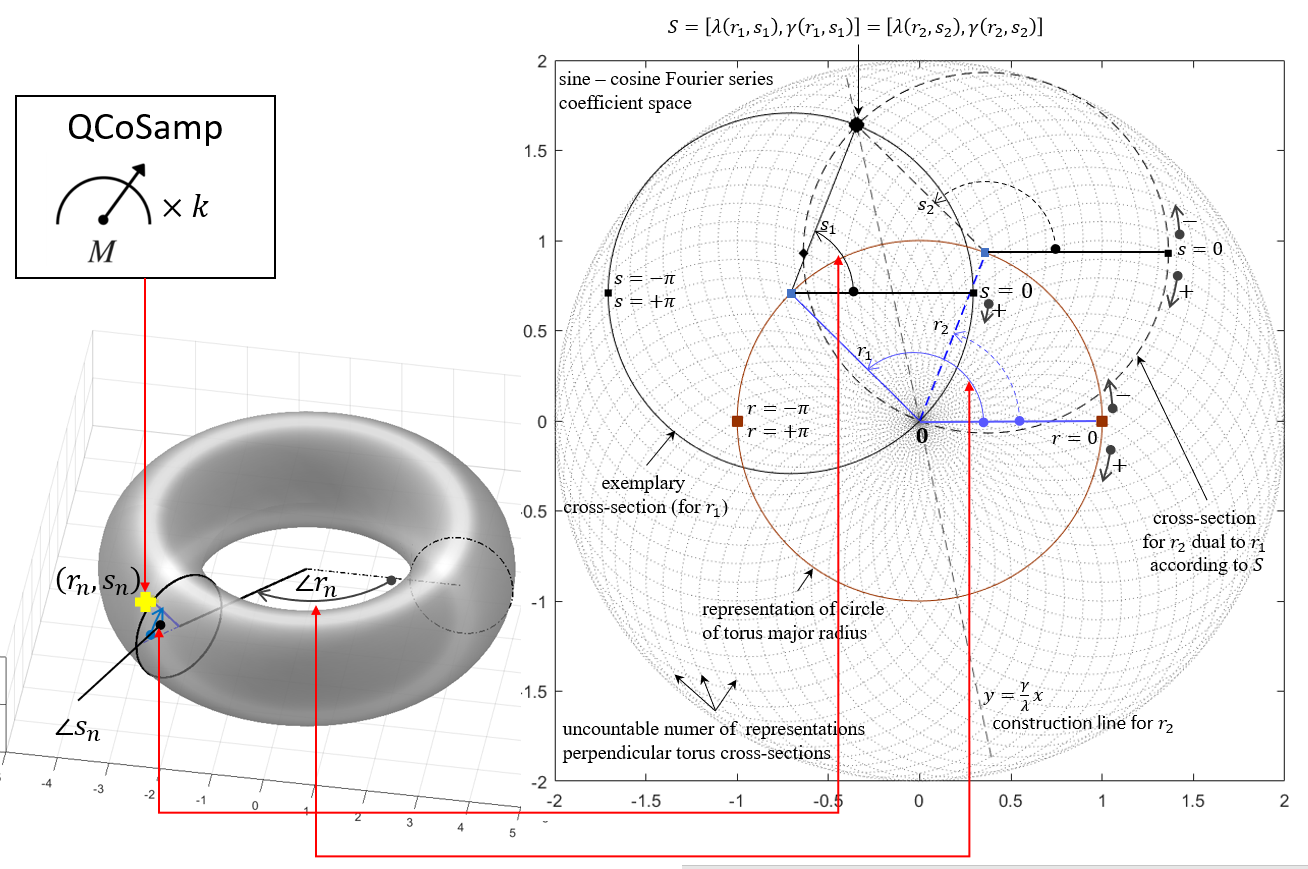}
    \caption{Mapping from the base model coefficient space to the Fourier series coefficient space. On the left picture, there is the torus formed by quantum harmonic cosine function phase shifts $r_n, s_n$. There is an exemplary mapping of one base model coefficient pair to one pair of Fourier series coefficient in the right-hand image.}
    \label{fig:torus}
\end{SCfigure}

The equation \ref{eq-mapping-FCoSamp-Fourier} gives the mapping from FCoSamp to the Fourier series. The phase shifts $r_n, s_n\in[-\pi, \pi]$ due to the periodicity of the sine and cosine functions. Therefore we can say that they lie on the torus (see Fig. \ref{fig:torus}). Now, if we fix the $r_n$ then $\{(\lambda_n, \gamma_n)\}$ forms a circle with the center in the point $(cos(r_n), -sin(r_n))$ and the radius $1$. But centers of those circles create the circle, with the center in the origin and radius $1$, as shown in Fig. \ref{fig:torus} right. The brown circle is made of the $r_n$ coefficients considering that it is the parametric variable for parametric equation of the circle: $x=cos(r_n), y=-sin(r_n)$. Therefore we can say that $r_n$ represents the angle. The zero angle, is set horizontally to the right (thicker blue line with a red square at the end). Angles increase to $\pi$ clockwise and decrease to $-\pi$ counter-clockwise. Therefore the major circle of the torus is mapped to this brown circle. If we have this coefficient set (e.q. to $r_1$ – thinner blue line) then $s_n$ treated as angle and the same reference angle and direction of growth. Therefore each section of the torus parallel to major radius and uniquely defined by it, is mapped to exactly one circle of the unit radius that is tangential to the point $(0,0)$. The inverse is true as well, certainly: any circle of a radius $1$ tangential to a point $(0,0)$ is mapped to exactly this section of the torus, defined by the angle the radius of the brown circle ending with the center of the circle mapping. If we take any point belonging to a circle of radius $2$ or its interior, it will lay on one such a circle. Therefore if we have a pair $(r_n, s_n)$ we can find the Fourier coefficient by drawing the radius defined by $r_n$ and then on its end – the radius defined by $s_n$. And vice-versa: if we dispose of a Fourier series coefficients $(\lambda, \gamma)$ we can find the $r_n, s_n$ coefficients by drawing a line from that point to the brown circle. The angle between this line and the line $s=0$ (black line on the picture) is the value of $s_n$. The angle of the radius going to the point of this intersection and the reference $r=0$ is the value of $r_n$.\\
In fact, we can find two FCoSamp coefficients pairs for one pair of Fourier series coefficients for most cases. We can see it clearly in the image. Two lines are connecting the point $(\lambda, \gamma)$ with the brown circle. The construction line $y=\frac{\lambda}{\gamma}x$ splits into two halves the rhombus, which has the vertices in the $\mathbf{0}$ point, the $(\lambda, \gamma)$ point, and two points on the brown circle distant from this point by 1. The construction line angle is equal to $\alpha =arctan(\gamma/\lambda)$; we have to add the value $-\pi$ for the upper half part of the space or $-2\pi$ for the bottom one due to the differences of the reference frames. Since this line divides the angles through which it pass by half $r_2=(\alpha-(r_1-\alpha)=2\alpha -r_1)$. So this is the proof of the observation $1$. For the points laying on the border (observation $2$) of the $\gamma^2+\lambda^2$ there is only one pair of $r_n = s_n$ coefficients. For the point $(0,0)$ each pair $(r_n,s_n): s_n=r_n-\pi$ is mapped – which proves the observation $3$.

\section{Gates used for the implementation of QCoSamp}\label{sec-gates}

DiVincenzo describes all gates that we will use in current work in \cite{gates_divincenzo_quantum_1998}. Below we present the gate list that we use in our work:
 \begin{itemize}
     \item $
            H=\frac{1}{\sqrt{2}}
            \begin{bmatrix}
            1&1\\
            1&-1
            \end{bmatrix}
            $ – Hadamard gate, which creates the superposition of the one qubit while used for not superposed qubit and can be used for generating the interference of superposed qubits.
    \item $
            R_\varphi=
            \begin{bmatrix}
            1&0\\
            0&e^{i\varphi}
            \end{bmatrix}
            $ – quantum phase shift gate which multiplies the second coordinate of qubit by the angle $\varphi$.
    \item $
            X=
            \begin{bmatrix}
            0&1\\
            1&0
            \end{bmatrix}
            $ – NOT gate, X-Pauli gate changes the coordinates between each other
    \item $
            \mathbf{1}=
            \begin{bmatrix}
            1&0\\
            0&1
            \end{bmatrix}
            $ – identity gate
    \item $
            cG=
            \begin{bmatrix}
            1&0&0&0\\
            0&1&0&0\\
            0&0&g_{11}&g_{12}\\
            0&0&g_{21}&g_{22}\\
            \end{bmatrix}
            $ controlled version of one-qubit gate $G$. 
            
It participates in the entanglement between two qubits creating, but the created entanglement is not necessarily maximal. This gate is conditional – it changes the second qubit's coordinates if the first one is equal to $\ket{1}$.
 The maximum entanglement (\cite{bell_einstein_1946, Higuchi-entanglement-2000}) appears in the case when the state is one of the Bell states, achieved by using a controlled X Pauli gate acting on the state $H\ket{0}\otimes \ket{0}$.
We can express the entanglement as the violation of the Bell's inequality, which occurs in the case of Bell states indeed and also many other issues – for all those states that are inseparable in the sense of the tensor product, and those states are also entangled, but not maximally. 
There is a well-known story about Alice, Bob, and entanglement, which is proper for maximal entanglement but not always for not maximal one. Let's consider the state:    
$$\ket{E}=\frac{1}{2}\big(\ket{00}+e^{ix}\ket{01}+e^{iy}\ket{10}+\ket{11}\big).$$ 
The Alice-Bob story does not work in this case because the probability of measuring each of the eigenstates is uniform and equal to $1/4$, so if Alice takes her qubit and measures it, there is still fifty-fifty chance for Bob to measure $\ket{0}$ or $\ket{1}$.
But on the other hand, if we assume, that this state is separable, which mean that there exists two one-qubit states: $\ket{\Phi}=[\varphi_1, \varphi_2]^T, \ket{\Psi}=[\psi_1, \psi_2]$ fulfilling the equation: $\ket{\Phi}\otimes\ket{\Psi}=\ket{E}$ it lead us to 4 equations:
            \begin{align*}
            \begin{tabular}{cccc}
                $2\varphi_1\psi_1=1 $& 
                $2\varphi_1\psi_2=e^{ix}$ &
                $2\varphi_2\psi_1=1 $& 
                $2\varphi_2\psi_2=e^{ix}, \Leftrightarrow$ \\
                $\varphi_1=\frac{1}{2\psi_1}$ & & & $\varphi_2=\frac{1}{2\psi_2}, \Leftrightarrow$\\ 
                & $\psi_2=e^{ix}\psi_1$ & & $\psi_1=e^{iy}\psi_2$ \\
                & $\psi_2=e^{ix}e^{iy}\psi_2, \Leftrightarrow$ &&\\
                & $e^{ix+y}=1,$ &&
            \end{tabular}
            \end{align*}

which lead us to the conclusion that the state $\ket{E}$ is separable by tensor product if and only if $x=-y$, 
which means that in other cases, the entangled state exists, but the Alice-Bob story has no sense in all the cases; most of these states "doesn't look like" entangled, but it is; however it is not maximally entangled.
Even more – the output probability is the same for all $x, y$, so the (not maximally) entangled and not entangled states produce the same output probability distribution over the measurement basis! There are also the states that "looks like" entangled, that are proper for Alice-Bob story, but not maximally entangled, exemplary:

$$\ket{E}=\frac{1}{2}\big(a\ket{00}+b\ket{11}\big), \wedge \|a\|^2, \|b\|^2\neq1,$$ in which case if Alice measures the $\ket{0}$ state, Bob has to measure $\ket{1}$ and opposite, but the entanglement of the state is not maximal, because output probability is not uniform. Concluding, not maximal entanglement is not determinable by the Alice-Bob story.

    \item $c^nG(x, c_1,\dots,c_n)$ – $n$ times controlled gate, where $x$ is the \emph{controlled} qubit and $c_1,\dots,c_n$ are \emph{controlling} qubits,  which is represented  by $2^{n+1}\times 2^{n+1}$ being equal to identity except of the $2\times 2$ matrix in the bottom – right corner being the $G$ gate.
    
This gate acts on the \textbf{controlled} qubit with operator $\mathcal{G}$ if and only if all \textbf{controlling} qubits are all equal to $1$. This gate, together with Hadamard gate create the (cluster) entanglement, which is not always maximal entanglement as well.
 \end{itemize}

\section{Sandwiching the phase}\label{sec-sandwiching}

Let us consider the state $H\ket{0}=\frac{1}{\sqrt{2}}\ket{0}-\frac{1}{\sqrt{2}}\ket{1}$. If we act on this state with the phase shift gate we obtain: $\frac{1}{\sqrt{2}}\ket{0}-\frac{1}{\sqrt{2}}e^{i\varphi}\ket{1}$. We get a probability, measuring $\ket{1}$: $\frac{1}{\sqrt{2}}e^{i\varphi}\overline{\frac{1}{\sqrt{2}}e^{i\varphi}}=\frac{1}{2}e^{i\varphi}e^{-i\varphi}=\frac{1}{2}$, and the same probability for measuring $\ket{0}$ Tthe result is equal like we never used a gate on the state. Hence we have to use e.g. Hadamard gate to pull out the phase shift 

\begin{equation*}
    H\bigg[ 
        \frac{1}{\sqrt{2}}\ket{0}-\frac{1}{\sqrt{2}}e^{i\varphi}\ket{1}
    \bigg]=
    \bigg(
        \frac{1}{2}+\frac{1}{2}e^{i\varphi}
    \bigg)\ket{0}-
    \bigg(
        \frac{1}{2}-\frac{1}{2}e^{i\varphi}
    \bigg)\ket{1}
\end{equation*}

Now, we obtain probability of measuring $\ket{1}$ equal to: 
$$
    \bigg(
        \frac{1}{2}-\frac{1}{2}e^{i\varphi}
    \bigg)
    \overline{
        \bigg(
            \frac{1}{2}-\frac{1}{2}e^{i\varphi}
        \bigg)
    }=
    \bigg(
        \frac{1}{4}-\frac{1}{4}\big(e^{i\varphi}+e^{-i\varphi}\big)+\frac{1}{4}
    \bigg)=\frac{1}{2}-\frac{cos(\varphi)}{2}
$$
and for measuring the $\ket{0}$:
$
    \frac{1}{2}+\frac{cos(\varphi)}{2}
$
Hence, we see that the fifty-fifty probability now is disturbed by the phase shift's interference by half of the cosine of the phase shift angle. Putting all gates together, we obtain:

\begin{equation}
H\circ R_\varphi \circ H \ket{0}=
 \frac{1}{2}
 \big[
    \big(
        1+e^{i\varphi}
    \big)\ket{0}-
    \big(
        1-e^{i\varphi}
    \big)\ket{1}
 \big]
\end{equation}

We will use similar equations many times in this work; therefore, we described it in detail. The name "sandwiching" comes from the fact that there is a Hadamard gate before and after phase shift – so it creates a sandwich of these gates. Nevertheless, notice that the first and last Hadamard gates play another role. Simultaneously, the first one makes a superposition. The last one introduces interference to the system, which extracts the phase to the measurement, which is very similar to the technique described in the next appendix, but it works for one qubit state.

\section{Phase kick-back}

The phase kick-back (or phase estimation) algorithm, as Cleve says in \cite{cleve_quantum_1998} (pp. 2-3) is the simulation of the Mach-Zehnder interferometer but is also a crucial technique for many quantum algorithms and is used many times in our work. Similarly to the previous subsection, we use sandwiching by two Hadamard gates. Still, this time we do not use the phase shift gate between them, but we consider that there exists auxiliary qubit $\ket{u}$ and the operator $U$ acting this qubit in such a way that it changes the global phase of them: $U\ket{u}=e^{i\varphi}\ket{u}$. We can consider the $U$ operator as the kernel part of our computation, and the the $\ket{u}$ could be the state made of many qubits. In that case the phase shift by angle $\varphi$ is the interesting part of our computation containing the results. Similarly to the previous subsection, the probability of measuring any of eigenstate of $u$ does not have the information about phase because $e^{i\varphi}\overline{e^{i\varphi}}=1$. If we use the Hadamard gate on the state, we obtain: $HU\ket{u}=e^{i\varphi}H\ket{u}$, which means that phase shift will not has the influence on the resulting probability from the same reason.\\
The Mach-Zehnder interferometer idea measures the difference between the phase of the light coming through the examined phase object and the not disturbed beam. Done by splitting the light beam before it hits the examined object using a half-plate beam splitter. Then one of the split beams goes through the item, and its phase is slightly changed while the other is just reflected in the mirror, so its phase doesn't change. Then the two beams are connected in another half-plate beam splitter after reflecting the examination beam. In the quantum algorithm, the operator $U$ plays the role of the examined phase object, the $\ket{u}$ qubit is the examination beam, and the Hadamard gates are the half-plates. In the interferometer, we have two light beams. Therefore we guess that in our quantum system, we need an extra qubit $\ket{c}$, which will play the role of reference (not disturbed) beam. In the interferometer, we have the same light beam splitter up by half-plate; therefore, they have orthogonal polarization, which connect them in one information system. It means that we know the interference pattern without a distortion made by the examined object. The deflection of this pattern avails to compute the phase shift generated by the object. In our case, the connection between $\ket{u}$ and $\ket{c}$ is made by the controlled version of the operator:

\begin{align}
cU\ket{cu}=&
\begin{bmatrix}
1&0&0&0\\
0&1&0&0\\
0&0&U_{11}&U_{12}\\
0&0&U_{21}&U_{22}\\
\end{bmatrix}
\begin{bmatrix}
c_1u_1\\
c_1u_2\\
c_2u_1\\
c_2u_2\\
\end{bmatrix}=
\begin{bmatrix}
c_1u_1\\
c_1u_2\\
c_2(u_1U_{11}+u_2U_{12})\\
c_2(u_1U_{21}+u_2U_{22})\\
\end{bmatrix}, so:\nonumber\\
cU\ket{cu}=&
\big(
    c_1u_1\ket{0}+c_2(u_1U_{11}+u_2U_{12})\ket{1}
\big)\ket{0}+
\big(
    c_1u_2\ket{0}+c_2(u_1U_{21}+u_2U_{22})\ket{1}
\big)\ket{1}=\nonumber\\
=&
\big(
    c_1u_1\ket{0}+c_2u_1e^{i\varphi}\ket{1}
\big)\ket{0}+
\big(
    c_1u_2\ket{0}+c_2u_2e^{i\varphi}\ket{1}
\big)\ket{1}=\nonumber\\
=&
\big(
    c_1\ket{0}+c_2e^{i\varphi}\ket{1}
\big)\otimes
\big(
    u_1\ket{0}+u_2\ket{1}
\big)=\big(
    c_1\ket{0}+c_2e^{i\varphi}\ket{1}
\big)\ket{u}
\end{align}

We see that the usage of controlled $U$ operator on the two-qubit state (with ancilla) without affecting the $\ket{u}$ qubit it phase shifts the second coordinate of ancilla only. Therefore we say that 
the phase is "kicked-back" to the ancilla state, hence the name of this technique. Note that phase in now local – affecting only one coordinate of a qubit, not the whole qubit.
\\
By the above algorithm, the phase estimation was written in the form:

\begin{align}\label{eq-phase-kick-back-general}
    &H(c)\circ cU \circ H(c)\ket{0}\ket{u}=
    \frac{1}{\sqrt{2}}
    H(c) cU \big(\ket{0}-\ket{1}\big)\ket{u}=\nonumber\\
    &=\frac{1}{\sqrt{2}}H(c)
    \big(
        \ket{0}-e^{i\varphi}\ket{1}
    \big)
    \big(
        u_1\ket{0}+u_2\ket{1}
    \big)=\nonumber\\
    &=\frac{1}{2}
    \big(
        \big(1-e^{i\varphi}\big)\ket{0}+
        \big(1+e^{i\varphi}\big)\ket{1}
    \big)
    \big(
        u_1\ket{0}+u_2\ket{1}
    \big)=\nonumber\\
    &=\frac{1}{2}\big[
        u_1\big(1-e^{i\varphi}\big)\ket{00}+
        u_2\big(1-e^{i\varphi}\big)\ket{01}+
        u_1\big(1+e^{i\varphi}\big)\ket{10}+
        u_2\big(1+e^{i\varphi}\big)\ket{11}
    \big]
\end{align}
Now, we obtain the probabilities for measurement auxiliary qubit in state:
\begin{align}
    \ket{0}:&\frac{u_1\overline{u_1}+u_2\overline{u_2}}{2}-
    \frac{
    \big(
        u_1\overline{u_1}+u_2\overline{u_2}
    \big)
        cos(\varphi)
    }{2} \stackrel{(*)}{=}
    \frac{1}{2}-\frac{cos(\varphi)}{2}=\frac{1-cos(\varphi)}{2}
    \nonumber\\
    \ket{1}:&\frac{u_1\overline{u_1}+u_2\overline{u_2}}{2}+
    \frac{
    \big(
        u_1\overline{u_1}+u_2\overline{u_2}
    \big)
        cos(\varphi)
    }{2} \stackrel{(*)}{=}
    \frac{1}{2}+\frac{cos(\varphi)}{2}=\frac{1+cos(\varphi)}{2}
\end{align}

The $\stackrel{(*)}{=}$ are true because of the first postulate of quantum mechanics, precisely: $u_1\overline{u_1}+u_2\overline{u_2}=1$ because these are coordinate of one, $\ket{u}$ state. We see above that the phase kick-back results in measurement with the distortion of the probability of measuring states (without this operation, the probabilities of measuring $\ket{0}$ and $\ket{1}$ would be the same – $1/2, 1/2$). However, the state $\ket{u}$ does not influence measurement. We can say it is transparent for kicking-back operation. Let us note, that we can measure the phase shift produced by an operator $U$ acting on one qubit state $\ket{u}$ by measuring state $\ket{c}$ because of the entanglement (not maximal in that case) and interference phenomena.

\section{Quantum Fourier Transform and distributed phase encoding}\label{sec-distributed-phase-encoding}
Ruiz in \cite{ruiz-perez_quantum_2017} looks at the QFT like on the transformation of basis change. In that case we can say that Quantum Fourier Transform $\mathcal{Q}$ maps the eigenstate $\ket{\psi}$, according to formula:
\begin{equation}\label{eq-qft-basic}
    \mathcal{Q}\ket{\psi}=\frac{1}{\sqrt{2^\Psi}}\sum_{\varphi=0}^{2^\Psi-1}e^{\frac{2\pi i \psi k}{2^{\Psi}}}\ket{\phi}
\end{equation}

The ket-states $\ket{\phi}$ appearing on the right side of this formula is $2^\Psi$ eigenstates, which are the basis after the change. 
We can write each of such a state in the form: $\ket{\phi}=\ket{\varphi_1...\varphi_\Psi}$, 
which is, in fact, the tensor product of $\Psi$ one qubit eigenstates. 
Therefore each of such eigenstates $\ket{\varphi_j}=\ket{0}$ or $\ket{\varphi_j}=\ket{1}$.
We assume that $\varphi_k=0 \Longleftrightarrow \ket{\varphi_k}=\ket{0}$ and $\varphi_k=1 \Longleftrightarrow \ket{\varphi_k}=\ket{1}$. 
Hence, we see that the eigenstate $\ket{\Phi}$ and understood as the binary representation of number $\varphi$ according to the formula:

\begin{equation}\label{eq-binary-representation-of-eigen-state}
    \varphi=\sum_{k=0}^{2^\Psi-1}\varphi_{k+1}2^k
\end{equation}
For the same reason, the number $\psi$ is a value of $\ket{\psi}$ treated as the binary representation.Cleve at al. declare in 
\cite{cleve_quantum_1998} (pp 8-9) the state \ref{eq-qft-basic} is un-entangled and could be factorized by already well known by us state:

\begin{align}
    &\frac{1}{\sqrt{2^\Psi}}\sum_{\varphi=0}^{2^\Psi-1}e^{\frac{2\pi i \psi k}{2^{\Psi}}}\ket{\varphi}=
    \frac{1}{\sqrt{2}}
    \bigg(\ket{0}+e^{\frac{2\pi i \psi}{2} }\ket{1}\bigg)\otimes
    \frac{1}{\sqrt{2}}
    \bigg(\ket{0}+e^{\frac{2\pi i \psi}{4} }\ket{1}\bigg)\otimes...
    \frac{1}{\sqrt{2}}
    \bigg(\ket{0}+e^{\frac{2\pi i \psi}{2^\Psi} }\ket{1}\bigg)=\nonumber\\
    &= \frac{1}{\sqrt{2}}
    \big(
    \ket{0...00}+
    e^{\frac{2\pi i \psi}{2^\Psi}}\ket{0...01}+
    e^{\frac{2\pi i \psi}{2^{\Psi-1}}}\ket{0...10}+
    e^{
        \frac{2\pi i \psi}{2^{\Psi-1}}+
        \frac{2\pi i \psi}{2^\Psi}
    }\ket{0...11}+...+
     e^{\frac{2\pi i \psi}{2}+
        \frac{2\pi i \psi}{4}+...
        \frac{2\pi i \psi}{2^\Psi}
    }\ket{1...11}
    \big)=\nonumber\\
    &=\frac{1}{\sqrt{2^\Psi}}
    \sum_{k=0}^{2^\Psi-1}e^{2\pi i \psi \sum_{k_j}k_j2^{-k_j}}\ket{k}=
    \frac{1}{\sqrt{2^\Psi}}
    \sum_{k=0}^{2^\Psi-1}e^{2\pi i \psi 0.(k+1)}\ket{k}
\end{align}
The notation $0.k$, introduced by Draper in \cite{draper_addition_2000} is a binary fraction and 
\begin{equation}
    0.k=0.k_1k_2...k_K=\sum_{j=1}^K \frac{k_j}{2^j}
\end{equation}
So finally the QFT is defined as:
\begin{equation}\label{eq-qft-final}
     \mathcal{Q}\ket{\psi}=\frac{1}{\sqrt{2^\Psi}}
    \sum_{k=0}^{2^\Psi-1}e^{2\pi i \psi 0.(k+1)}\ket{k}
\end{equation}

Because all coordinates are the phase shift by angle $2\pi \psi 0.(k+1)$, the probabilities of measuring the eigenstates are uniform, precisely equal to $1/2^\Psi$. Therefore it is not an algorithm for computing the Fourier transform for input state, but the series exists during the evolution only. This algorithm is useful as a part of other algorithms – e.g., Shor \cite{shor_algorithms_1994} used it for factoring the integers. \\
Another QFT usage is connected with encoding numbers described by Ruiz after Draper and Bauregard, in the previously mentioned papers. Ruiz proposed the method of encoding numbers using the phase distribution to the eigenstate made by QFT. We can assume that each eigenstate encodes one binary fraction. According to the equation \ref{eq-qft-final} for each such a binary fraction the corresponding coordinate is equal to $e^{2\pi i \psi 0.(k+1)}$. Because QFT is understood as the change of basis, we can say that the value of $\psi$ is encoded on a new basis. This approach is convenient for quantum arithmetic (Ruiz, Draper, Bauregard, ibidem); however, we have slightly changed this technique mixing it with phase kick-back, which allows us to separate the encoding and computation part of the operator. This modification is one of the groundwork of the proposed method. It is described in details in the   \ref{sec-theory-calculations} "\nameref{sec-theory-calculations}".

\section{Un-computation}\label{sec-methods-uncomputation}

The un-computation described in the perspective of reversing computation in general by Aaronson et al. in \cite{aaronson_classification_2015} and was originally introduced by Benett in \cite{bennett_logical_1973}. He noticed that reversible computers could produce unwanted information mixed with the original one in the intermediate stage of computation. Therefore there is a need to erase this unwanted information before we notice the results. The process is possible because of the assumed reversibility of computation. That process of erasing unwanted information is called \textit{un-computation}. To be more precise, after Aaronson, we can divide the computation into two functions: $f(x)$ that is the goal and wanted, and $garb(x)$, which is unwanted. Because of the reversibility there exists function $garb^{-1}(x)$ which un-computes the useless data after storing the $f(x)$ in the safe place (registers).\\
Quantum computing is, in fact, reversible computing because evolution operator $\mathcal{U}$ of the system has to be unitary, which means that $\mathcal{U}^\dagger\mathcal{U}\ket{x}=\ket{x}$. If the gate is by addition, like most basic gates, Hermitian then $\mathcal{H}\mathcal{H}\ket{x}=\ket{x}$. \\
In quantum computation, the unwanted part arises due to the interference phenomena. This phenomenon is one of the grounds being the power of quantum calculations, but their results often could influence the measurement destroying the computations. For example having two qubits: $\ket{\Phi}=\frac{1}{\sqrt{2}}\big(\ket{0}+e^{i\phi}\ket{1}\big)$ and $\ket{\Psi}=\frac{1}{\sqrt{2}}\big(\ket{0}+e^{i\psi}\ket{1}\big)$ we would like to check if the phases are equal. The tensor product of two states produces:

\begin{align}\label{eq-uncomputation-example}
    \ket{\Phi}\otimes\ket{\Psi}=
    \frac{1}{2}
    \big(
        \ket{00}+
        e^{i\psi}\ket{01}+
        e^{i\phi}\ket{10}+
        e^{i(\phi+\psi)}\ket{11}
    \big)
\end{align}
If there would be $1$ by eigenstate $\ket{11}$, we could act with a two-qubit Hadamard gate and obtain:
\begin{align}\label{eq-uncomputation-example-step-2}
    \frac{1}{2} 
    \big[
        \big(2+e^{i\phi}+e^{i\psi}\big)\ket{00}+
        \big(e^{i\phi}-e^{i\psi}\big)\ket{01}+
        \big(e^{i\psi}-e^{i\phi}\big)\ket{10}
        \big(2-e^{i\phi}-e^{i\psi}\big)\ket{11}
    \big]
\end{align}

If phases are equal, the measure of the eigenstates $\ket{01}, \ket{10}$ is impossible. Therefore we extract the information about the equality of the phases.\\
Nevertheless, the coordinate by the state $\ket{11}$ is not equal to $1$ and will not create the situation that for each $\phi=\psi$ measuring any state will be impossible. So, this is the situation that we have the unwanted product of interference. The un-computation operator has to work only for the coordinate $\ket{11}$. The original operator changing the phase of eigenstate $\ket{11}$ only is the controlled phase shift operator: $cR_{\phi+\psi}$. The un-computation operator is equal to its Hermitian: $cR_{\phi+\psi}^\dagger=cR_{-\phi-\psi}$. Acting with this operator on the state from the equation \ref{eq-uncomputation-example} will produce the desired state from which we can achieve the equation \ref{eq-uncomputation-example-step-2}.

\section{Oracle, amplitude amplification and diffusion operator}\label{sec-oracle}
The Oracle is commonly (e.g. by Cleve in \cite{cleve_quantum_1998} or Brassard in \cite{brassard_quantum_2000}) defined as the function: 
\begin{equation}
    \mathbf{\chi}_{\Omega}(\xi)=
     \left \{
        \begin{array}{ll}
        0     & \forall \omega\in \Omega: \xi\neq \omega  \\
        1     & \exists \omega \in \Omega: \xi=\omega
        \end{array}
    \right.=
    \delta_{y_1,\xi}\cdot ... \cdot \delta_{y_N,\xi}
\end{equation}

The values $\omega_1, ... , \omega_N$ are the $N$ values that we call "good" values for which the Oracle returns \textit{true}; for other values it returns \textit{false}. This is the most general definition for the case when there is more than one "good" value. The operator representing such an Oracle has the form:

\begin{equation}
    \mathcal{U}_\omega\ket{x}=
    \left \{
        \begin{array}{rl}
        -\ket{x} & \chi_\omega(x)=1\\
        \ket{x} & \chi_\omega(x)=0
        \end{array}
    \right.=(-1)^{\chi_\omega(x)}\ket{x}
\end{equation}

The notation $\mathcal{U}_\omega$ means that the oracle operator defines the state $\ket{\omega}$, which should be found by the algorithm. We call this state \textit{desired} state.\\ 
As we see, the oracle operator changes the state for the opposite in one case only: when the state it acts is equal to one of the desired states. Grover \cite{Grover_algorithm} and Brassard \cite{brassard_quantum_2000} proposed the diffusion operator $\mathcal{U}_\varphi$ that is defined by the state $\ket{\varphi}$ that is uniform superposition of the all eigenstates as the Hausholder transformation \cite{hausholder1958}:

\begin{align}
    \ket{\varphi}=&\frac{1}{\sqrt{2^K}}\sum_{k=0}^{K-1}\ket{k}\nonumber\\
    \mathcal{U}_\varphi=&2\ket{\varphi}\bra{\varphi}-\mathbf{1}
\end{align}

It means that the Grover diffusion operator is the reflection of the state $\ket{\varphi}$ through the state $\ket{s'}$ orthogonal to the state $\ket{\omega}$ laying on the same plane (in Hilbert space) as the $\ket{\omega}$. Therefore it is very convenient to define the oracle operator in the same way, as the reflection through the state $\ket{\varphi}$:

\begin{equation}
    \mathcal{U}_\omega=\mathbf{1}-2\ket{\omega}\bra{\omega}
\end{equation}
Now, building the amplitude amplification algorithm is relatively easy:
\begin{enumerate}
    \item Define the state of the quantum system $\ket{\varphi}$.
    \item Define the state which system should be reached $\ket{\omega}$.
    \item Repeat the application of $\mathcal{U}_\varphi\mathcal{U}_\omega$ $N$ times.
\end{enumerate}

Grover specified the repetition count to $1/\sqrt{2^\Phi}$ in one "good" element case. For many "good" elements it is equal to $1/\sqrt{a}$, where $a=\bra{s'}\ket{s'}$. Since the state $\ket{s'}$ is generally unknown there is \textit{amplitude estimation} algorithm for estimating the $a$ value given by Brassard in \cite{brassard_quantum_2000}. These numbers of repetitions are given to assume that we make the computation once, and we would like to have the solution. If we make the number of experiments to obtain the probability distribution, there is possible to end the computation before reaching this number. At the start of the algorithm, all states have the same probability. Each repetition increases the probability of obtaining "good" states and decreases the "bad" ones. The probability of obtaining the states, given by Brassard (ibidem p. 7) is equal to:

\begin{equation}
    sin^2((m+1)\theta)
\end{equation}
where $\theta$ is equal $asin(\frac{1}{\sqrt{2^\Phi}})$ for one "good" case and $asin(\frac{1}{\sqrt{a}})$ for many "good" case; $m$ is the number of repetitions of the algorithm.
\bibliographystyle{ieeetr}
\bibliography{ms.bib}
\end{document}